\documentclass{aa}  

\usepackage{graphicx}
\usepackage{txfonts}
\usepackage{chemformula}
\usepackage{placeins}
\usepackage{tabularx}
\usepackage{multirow}
\usepackage{booktabs}
\usepackage{lscape}
\usepackage[]{hyperref}
\hypersetup{
    colorlinks = true,
    linkcolor = {blue}, 
    citecolor = {blue},
    urlcolor = {blue}
}
\usepackage{xspace}

\newcommand{\stark}{\texttt{stark}\xspace}
\newcommand{\hans}{\texttt{HANSOLO}\xspace}
\newcommand{\boldchange}[1]{{#1}}

\begin{document}

\title{JWST reveals a rapid and strong day side variability of 55\,Cancri\,e}


 \author{
         J.~A.~Patel\inst{1}\thanks{\email{jayshil.patel@astro.su.se}}$^{\href{https://orcid.org/0000-0001-5644-6624}{\includegraphics[scale=0.5]{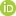}}}$,
         A.~Brandeker\inst{1},
         D.~Kitzmann\inst{2,3}$^{\href{https://orcid.org/0000-0003-4269-3311}{\includegraphics[scale=0.5]{Figures/orcid.jpg}}}$,
         D.~J.~M.~Petit~dit~de~la~Roche\inst{4},
         A.~Bello-Arufe\inst{5},
         K.~Heng\inst{6,7,8,9},
         E.~Meier Valdés\inst{3},
         C.~M.~Persson\inst{10},
         M.~Zhang\inst{11},
         B.-O.~Demory\inst{3,2},
         V.~Bourrier\inst{4},
         A.~Deline\inst{4},
         D.~Ehrenreich\inst{4,12},
         M.~Fridlund\inst{13,10},
         R.~Hu\inst{5,14}$^{\href{https://orcid.org/0000-0003-2215-8485}{\includegraphics[scale=0.5]{Figures/orcid.jpg}}}$,
         M.~Lendl\inst{4},
         A.~V.~Oza\inst{14,5},
         Y.~Alibert\inst{2,3},
         M.~J.~Hooton\inst{15}
        }
\authorrunning{Patel et al.}

   \institute{
             \label{inst:1} Department of Astronomy, Stockholm University, AlbaNova University Center, 10691 Stockholm, Sweden \and
             \label{inst:2} Weltraumforschung und Planetologie, Physikalisches Institut, Universität Bern, Gesellschaftsstrasse 6, 3012 Bern, Switzerland \and
             \label{inst:3} Center for Space and Habitability, Universität Bern, Gesellschaftsstrasse 6, 3012 Bern, Switzerland \and
             \label{inst:4} Observatoire astronomique de l'Université de Genève, Chemin Pegasi 51, 1290 Versoix, Switzerland \and
             \label{inst:5} Jet Propulsion Laboratory, California Institute of Technology, Pasadena, CA 91011, USA \and
             \label{inst:6} Faculty of Physics, Ludwig Maximilian University, Scheinerstrasse 1, D-81679, Munich, Bavaria, Germany \and
             \label{inst:7} ARTORG Center for Biomedical Engineering Research, University of Bern, Murtenstrasse 50, CH-3008, Bern, Switzerland \and
             \label{inst:8} University College London, Department of Physics \& Astronomy, Gower St, London, WC1E 6BT, United Kingdom \and
             \label{inst:9} University of Warwick, Department of Physics, Astronomy \& Astrophysics Group, Coventry CV4 7AL, United Kingdom \and
             \label{inst:10} Department of Space, Earth and Environment, Chalmers University of Technology, Onsala Space Observatory, 439 92 Onsala, Sweden \and
             \label{inst:11} Department of Astronomy and Astrophysics, The University of Chicago, Chicago, IL 60637, USA \and
             \label{inst:12} Centre Vie dans l'Univers, Facult\'e des sciences de l'Universit\'e de Gen\`eve, Quai Ernest-Ansermet 30, 1205 Geneva, Switzerland \and
             \label{inst:13} Leiden Observatory, University of Leiden, PO Box 9513, 2300 RA Leiden, The Netherlands \and
             \label{inst:14} Division of Geological and Planetary Sciences, California Institute of Technology, Pasadena, CA 91125, USA \and
             \label{inst:15} Cavendish Laboratory, JJ Thomson Avenue, Cambridge CB3 0HE, UK 
             }

\date{Received XX; accepted XX}

 
  \abstract
   {The nature of the close-in rocky planet 55\,Cnc\,e is puzzling despite having been observed extensively. Its optical and infrared occultation depths show temporal variability, in addition to a phase curve variability observed in the optical.} 
   {We wish to explore the possibility that the variability originates from the planet being in a 3:2 spin-orbit resonance, thus showing different sides during occultations. 
   We proposed and were awarded Cycle 1 time at the \textit{James Webb} Space Telescope (JWST) to test this hypothesis.}
   {JWST/NIRCam observed five occultations (secondary eclipses), of which four were observed within a week, of the planet simultaneously at 2.1 and 4.5\,$\mu$m. While the former gives band-integrated photometry, the latter provides a spectrum between 3.9--5.0\,$\mu$m.}
   {We find that the occultation depths in both bandpasses are highly variable and change between a non-detection ($-5 \pm 6$\,ppm and $7 \pm 9$\,ppm) to $96\pm8$\,ppm and $119^{+34}_{-19}$\,ppm at 2.1\,$\mu$m and 4.5\,$\mu$m, respectively.
   Interestingly, the variations in both bandpasses are not correlated and do not support the 3:2 spin-orbit resonance explanation. The measured brightness temperature at 4.5\,$\mu$m varies between 873--2256\,K and is lower than the expected dayside temperature of bare rock with no heat re-distribution (2500\,K) which is indicative of an atmosphere. Our atmospheric retrieval analysis of occultation depth spectra at 4.5\,$\mu$m finds that different visits statistically favour various atmospheric scenarios including a thin outgassed CO/CO$_2$ atmosphere and a silicate rock vapour atmosphere. Some visits even support a flat line model.}
   {The observed variability could be explained by stochastic outgassing of CO/\ch{CO2}, which is also hinted by retrievals. Alternatively, the variability, observed at both 2.1 and 4.5\,$\mu$m, could be the result of a circumstellar patchy dust torus generated by volcanism on the planet. }


   \maketitle
%

\section{Introduction}\label{sec:intro}

Ultra-short-period planets (USPs) provide a unique opportunity to study planets in extreme environments that have no counterparts in our Solar system \citep[see][for a review]{2018NewAR..83...37W}. Many USPs are consistent with a bare rock composition, while some of them might have a secondary metal-rich atmosphere or a disintegrating surface \boldchange{\citep[e.g.,][]{2012A&A...545L...5B, 2019Natur.573...87K, 2022A&A...664A..79Z}}. Being in an orbit around the nearby ($d=12.6$ pc), bright naked eye star 55\,Cancri ($V=5.95$ mag), 55\,Cancri\,e (hereafter 55\,Cnc\,e) is one of the best targets for investigating the nature of a USP. Out of the five known planets in the system, planet e is the only one transiting the star.

55\,Cnc\,e was discovered by \citet{2004ApJ...614L..81M} with an orbital period of $\sim 2.8$\,d, which was later found to be an alias of the true $0.74$\,d period \citep{2010ApJ...722..937D}. This was confirmed by the detection of planetary transits in the optical and infrared (IR) independently \citep{2011ApJ...737L..18W, 2011A&A...533A.114D}, enabling its radius measurement. Together with mass estimates derived from radial velocity measurements, the earlier works attempted to constrain the internal structure of the planet and found that the planetary density was consistent with either a purely rocky planet, a rocky planet with a thick super-critical water envelope, or carbon-rich interior with no envelope \citep{2011A&A...533A.114D, 2012A&A...539A..28G, 2012ApJ...759L..40M}. More recently, \citet{2018A&A...619A...1B} refined planetary mass ($8.3\,M_\oplus$) and radius ($1.88\,R_\oplus$) using radial velocity data and HST/STIS transit observations. Their internal structure modelling, based on these updated mass-radius measurements, suggests a rocky planet surrounded by a heavyweight (high mean molecular weight) atmosphere. A low mean molecular weight, or lightweight, atmosphere on the planet is not possible because of intense radiation from its host star. \boldchange{Atmospheric escape simulations also imply that lightweight atmospheres (made of H, He) would not survive on 55\,Cnc\,e for a long time period \citep[e.g.,][]{2012A&A...539A..28G, 2016A&A...586A..75S, 2018A&A...619A...1B, 2021AJ....161..181Z}.} Other attempts to model the internal structure of the planet \citep[e.g.,][]{2017A&A...597A..38D, 2017MNRAS.472..245L, 2018ApJ...860..122C} indicate a rocky interior with a gas or water envelope.

Soon after its discovery, \citet{2012ApJ...751L..28D} used \textit{Spitzer} to detect thermal emission from 55\,Cnc\,e and determined its day-side temperature to be around 2300\,K. \citet{2016Natur.532..207D} constructed a temperature map of the planet using \textit{Spitzer}/IRAC phase curve measurements at 4.5\,$\mu$m. They calculated the average day-side temperature to be around 2350\,K with a maximum $\sim 2700$\,K. Curiously, the hottest location of the planet was found to be shifted by $\sim 41 ^ \circ$ to the east compared to the sub-stellar point, indicating a strong heat re-distribution. On the other hand, the day-night temperature difference was found to be as large as 1300\,K, a sign of inefficient heat transport to the night side. These conflicting results led \citet{2016Natur.532..207D} to speculate that perhaps efficient heat transport is only happening on the day side of the planet by a thick atmosphere, or alternatively, a molten lava flow is responsible for the heat transport. The inefficiency of energy transport to the night side could be due to gases becoming cold enough to condense. Similarly, a lava stream could be hindered by the surface solidifying at the night side. \citet{2017AJ....154..232A} re-analysed the phase-curve data and confirmed the findings of \citet{2016Natur.532..207D}. Their physical model of the phase curve allowed them to show that the radiative and advective time scales must be of the same order to reproduce the observed phase curve. This disfavours the lava ocean scenario since a lava flow would have a too large advective time scale \citep[e.g.,][]{2016ApJ...828...80K} to be an efficient heat transporter \citep[however interior dynamics models of the planet, in some cases, exhibits a mantle super-plume away from the sub-stellar point, which can potentially interact with the lava ocean and increase its temperature at the location of the plume, mimicking hot-spot offset;][]{2023A&A...678A..29M}. \citet{2017AJ....154..232A} further propose that a CO or N$_2$ dominated atmosphere on the day side could explain the phase curve. This claim was corroborated by a 3D global circulation model climate model by \citet{2017ApJ...849..152H} that could potentially describe the observations, assuming a H$_2$ + N$_2$ dominated atmosphere with a trace source of opacity at 4.5\,$\mu$m (such as CO$_2$ or H$_2$O), coupled with the with a presence of night-side clouds. A recent re-reduction and re-analysis of the \textit{Spitzer} phase curve by \citet{2022AJ....164..204M} yielded an even larger day-night temperature difference with a smaller phase offset, more consistent with a poor heat transport typically found on USPs.

The heavyweight atmosphere on the planet, which was implied by the \textit{Spitzer} phase curve, climate modelling and mass-radius constraints, is challenging to detect. Numerous observations have tried but failed to detect any atmosphere on the planet
The singular claim of detection of gas on 55\,Cnc\,e comes from \citet{2016ApJ...820...99T} who identified HCN in the atmosphere using HST/WFC3 transit observations. However, subsequent observations using high-resolution spectroscopy from the ground could not reproduce the detection of HCN \citep{2021AJ....161..209D}. Furthermore, the transit observation of 55\,Cnc\,e in the Ly\,$\alpha$ band by \citet{2012A&A...547A..18E} resulted in a non-detection, suggesting the absence of an extended H upper atmosphere. This was supported by the non-detection of He in the upper atmosphere by \citet{2021AJ....161..181Z}. A lack of H and He in the atmosphere could mean that both gases escaped if they were initially accreted from the disk. In addition to this, several studies attempted but could not detect other atmospheric species such as H$_2$O, TiO, NH$_3$, C$_2$H$_2$, Fe, Ca, Mg, K, Na, H \boldchange{\citep{2016A&A...593A.129R, 2017AJ....153..268E, 2020AJ....160..101J, 2020MNRAS.498.4222T, 2021AJ....161..209D, 2022MNRAS.513.1544K, 2023AJ....166..155R}}. These non-detections mean that either those species are absent from the atmosphere or only present at very low volume mixing ratios if the mean molecular weight of the atmosphere is not too high to be detected by the transit observations. Another possibility is that the atmosphere of the planet is cloudy \citep{2017MNRAS.472..447M}.

The IR observations of 55\,Cnc\,e in emission posed another challenge for understanding the behaviour of the planet. \citet{2016MNRAS.455.2018D} monitored the occultation depths of 55\,Cnc\,e with \textit{Spitzer} at 4.5\,$\mu$m during 2012--2013 and found a variable occultation depth ranging from 47\,ppm to 176\,ppm. This translates into a corresponding change of brightness temperature of 1370\,K to 2530\,K. \boldchange{Variability was also observed in the optical bandpass by MOST (Microvariability and Oscillations of STars) that discovered significant changes in phase curves over several seasons \citep{2011ApJ...737L..18W, 2014IAUS..293...52D, 2019A&A...631A.129S}.} While the optical observations with MOST found a significant phase curve amplitude, the secondary occultation remained undetected. More recently, CHEOPS (CHaracterising ExOPlanet Satellite) extensively observed 55\,Cnc \citep{2021A&A...653A.173M, 2023A&A...669A..64D, 2023A&A...677A.112M} in the optical (G band) and confirmed significant variability not only in phase amplitude but also in phase offset and occultation depth, where the occultation depths at some epochs were consistent with zero. 
TESS (the Transiting Exoplanet Survey Satellite) also observed 55\,Cnc and found a hint of weak variability in occultation depths over three observing sectors \citep{2022A&A...663A..95M}. In contrast to the variability of the occultation depths, no optical or IR variability has been observed in the transit depths \citep[e.g.][]{2023A&A...677A.112M, 2018A&A...619A...1B}.

Multiple studies in the literature propose various hypotheses to explain the observed variability of the occultation depth of 55\,Cnc\,e in the optical and IR. \citet{2016MNRAS.455.2018D} suggested that plumes from volcanic outgassing on the dayside could explain the observed variability in emission. Assuming an Earth-like composition for the interior, it can release gases such as CO or CO$_2$ that are a significant source of opacity around 4.5\,$\mu$m. Gas plumes evolving at different atmospheric pressure levels could be inferred as varying temperatures during occultation observations in the IR. Given that the variability was observed throughout the optical and IR,
it was suggested by \citet{2021A&A...653A.173M} that a circumstellar inhomogeneous dusty torus could provide a variable source of opacity. \citet{2023A&A...677A.112M} studied the dusty torus scenario in detail and concluded that such a torus made up of certain species of a narrow range of particle sizes could indeed reproduce the level of observed variability in the optical.  However, a dusty torus should extent out to its Hill sphere and, if opaque, is inconsistent with the observed transit depths \citep{Heng2023}.  \cite{Heng2023} argued that a thin, transient outgassed atmosphere is consistent not only with the observed optical and infrared occultation depths, but also provides a plausible explanation for their variability.  \boldchange{\citet{2024ApJ...963..157T} demonstrated that CO-\ch{CO2} atmospheres are outgassed under a broad range of conditions (surface pressures, oxygen fugacity, temperatures).} 


Since 55\,Cnc\,e is in a very close-in orbit around its host star, \citet{2020A&A...633A..48F} showed that the planet's orbit is inside the stellar Alfv\'{e}n surface. This means that star-planet interactions (SPI) are plausible for the system, potentially causing variability-inducing star spots. \citet{2018A&A...615A.117B} proposed coronal rain, a kind of SPI, as a reason for the variability in chromospheric lines that they observed with HST \citep[see also][]{2019A&A...631A.129S}. 
\citet{2021A&A...653A.173M} ruled out star spot creation by the planet as a plausible mechanism to explain the optical variability observed by CHEOPS but this does not prohibit other possible forms of SPIs, such as coronal rain.

Although multiple hypotheses have been provided to describe the thermal phase curve and variability from 55\,Cnc\,e, each has difficulties in fully explaining all observed features. The observations with the \textit{James Webb} Space Telescope (JWST), presented here, were in part motivated by exploring \boldchange{an} alternative hypothesis that the planet rotates at an asynchronous rate to its orbit, potentially explaining both the hot-spot shift into the afternoon and the rapid orbit-to-orbit variability.
The idea and the observations motivated by it are presented in Section \ref{sec:3:2&Obs} followed by results in Section \ref{sec:res}. We show the results from atmospheric retrieval analysis in Section \ref{subsec:retrieval}. Finally, we interpret the results from our observations and present our conclusions in Section \ref{sec:interpretations} and \ref{sec:conclusions}, respectively. Details of the data analysis methods used are put into Appendix~\ref{app:a}.

\begin{table*}
    \centering
    \caption{\boldchange{Observation log and wide band occultation depths}}
    \begin{tabular}{ccccccccc}
        \hline
        \hline
        \noalign{\smallskip}
        Visit & Prog.\ & Start date & End date & Parity & Occultation & Occultation & Brightness & Brightness \\
         & ID & & & & depth at & depth at & temp. at & temp. at \\
          & & & & & 2.1\,$\mu$m (ppm) & 4.5\,$\mu$m (ppm) & 2.1\,$\mu$m (K) & 4.5\,$\mu$m (K) \\
        \noalign{\smallskip}
        \hline
        \hline
        \noalign{\smallskip}
        1 & 2084 & 2022-11-18 14:40:17 & 2022-11-18 19:15:53 & even & {\small $47.4 ^{+21.0} _{-15.5}$} & {\small $7.0 ^{+8.8} _{-8.8}$} & {\small $2417 ^{+335} _{-287}$} & {\small $873 ^{+167} _{-187}$} \\
        \noalign{\smallskip}
        2 & 2084 & 2022-11-20 19:43:08 & 2022-11-21 00:18:44 & odd & {\small $-5.1 ^{+5.5} _{-6.0}$} & {\small $65.2 ^{+22.3} _{-42.2}$} & {\small $1247 ^{+190} _{-245}$} & {\small $1716 ^{+230} _{-315}$} \\
        \noalign{\smallskip}
        3 & 2084 & 2022-11-23 00:43:57 & 2022-11-23 05:19:33 & even & {\small $37.3 ^{+4.7} _{-4.6}$} & {\small $101.4 ^{+17.1} _{-32.4}$} & {\small $2234 ^{+86} _{-88}$} & {\small $2078 ^{+172} _{-342}$} \\
        \noalign{\smallskip}
        4 & 1952 & 2022-11-24 11:38:15 & 2022-11-24 17:28:41 & even & {\small $36.8 ^{+27.7} _{-32.9}$} & {\small $119.2 ^{+34.0} _{-19.0}$} & {\small $2302 ^{+413} _{-807}$} & {\small $2256 ^{+330} _{-188}$} \\
        \noalign{\smallskip}
        5 & 2084 & 2023-04-24 11:57:03 & 2023-04-24 16:32:36 & odd & {\small $95.9 ^{+8.1} _{-7.9}$} & {\small $95.4 ^{+13.5} _{-16.8}$} & {\small $3138 ^{+107} _{-107}$} & {\small $2016 ^{+137} _{-179}$} \\
        \noalign{\smallskip}
        \hline
    \end{tabular}
    \label{tab:obs_log_res}
\end{table*}

\section{Asynchronous rotation scenario for 55\,Cnc\,e, Observations and Methods}\label{sec:3:2&Obs}

\subsection{55\,Cnc\,e in 3:2 spin-orbit resonance}\label{subsec:3:2Theory}
The planet 55\,Cnc\,e orbits its host star in about 17.7\,h with a semi-major axis of 0.015\,AU \citep{2018A&A...619A...1B}. When a planet is orbiting this close to its host star, it is usually assumed to be in a tidally locked synchronous spin-orbit configuration because of strong tidal forces. However, if the planet is part of a multi-planetary system, gravitational interactions with the other planets can perturb the planet from its synchronous 1:1 spin-orbit configuration.
\citet{2012MNRAS.427.2239R} simulated the tidal evolution of the orbit of 55\,Cnc\,e and showed that there is a reasonable likelihood for the planet to be trapped in an asynchronous spin-orbit resonance, with the 3:2 spin-orbit resonance being the most likely after 1:1 synchronous rotation. Asynchronous rotation can thus not be ruled out for 55\,Cnc\,e.
The consequence is that the planet would show different faces to the star during the orbit. This in turn means that the hottest point on the planet would not necessarily be the sub-stellar point. Just as on Earth the hottest time of the day is in the afternoon and not at noon, so could thermal inertia on 55\,Cnc\,e shift its hottest spot to the afternoon (east). The thermal inertia could, like on Earth, be provided by the atmosphere. In the case of a bare rock, thermal inertia could be provided by the heating, melting and evaporation of the rock in the morning with subsequent condensation and crystallisation in the afternoon. 
Quantitative models of these scenarios are sensitive to the detailed assumptions of the mass and composition of the atmosphere \boldchange{that, in turn, depends on the material equation of state.}
Using simplified models, \citet{2019ESS.....431107B} showed that the observations up until then could indeed be explained by using reasonable assumptions on the physical properties of the planet, meaning that the asynchronous rotation scenario could not be excluded.

Assuming that the planet is rotating asynchronously in the most probable 3:2 spin-orbit resonance, the planet will show the same face only at every second occultation instead of showing the same face every time. That means the two opposite sides will be seen during consecutive occultations. If there are semi-stable surface features, e.g., due to volcanic activity, on different sides of the planet they will show up differently during alternate occultations. In this case, the observed occultation depths would be expected to highly correlate with the occultation number over a short period, while this correlation could be broken over a longer time scale due to surface changes. The variability in occultation depths observed by \citet{2016MNRAS.455.2018D} can then be attributed simply to the planet showing different faces during occultations. Notably, \citet{2018AJ....155..221T}, who confirmed the \textit{Spitzer} variability of occultation depths, found the variability to be well fitted by a sinusoidal with a period as short as 2 days, but discarded this solution as being unphysical. However, if the planet is indeed in a 3:2 spin-orbit resonance, it is expected that the period of variability should be equivalent to the synodic period ($\sim 35.5$ hr), close to the period of 2 days.

To further test this intriguing hypothesis of asynchronous rotation and simultaneously sensitively measure potential atmospheric signatures, we designed an observation programme for JWST as detailed in the next section.

\subsection{Observations}\label{subsec:obs}

If the planet is indeed in a 3:2 spin-orbit resonance, it will show two opposite sides in consecutive occultations. Assuming the planetary surface to evolve slowly, we would then expect every second consecutive occultation to be strongly correlated. Enumerating the occultations by orbit number, we thus requested two ``odd'' and two ``even'' occultations within a short time-constrained span of two weeks, to rule out significant surface evolution within that time. Since 55\,Cnc is a very bright IR target ($K = 4$\,mag), avoiding saturation while observing it with JWST is challenging. From pre-launch estimates, our options were essentially limited to a grism time-series mode of the Near Infrared Camera (NIRCam).
The proposal was awarded time in JWST Cycle~1 as GO\,2084 \citep{2021jwst.prop.2084B}. The observation log is provided in Table~\ref{tab:obs_log_res}. Due to technical difficulties, only three occultations of the programme were observed within the time constraint of two weeks; the fourth was postponed until five months later. Fortunately, a different programme 
\citep[GO\,1952,][]{2021jwst.prop.1952H} that also targeted 55\,Cnc had an occultation observed in the same instrument mode and within the same first week \citep{2024arXiv240504744H}. In the following, we thus present an analysis of all five visits.

NIRCam offers simultaneous observations in short-wave (SW) and long-wave (LW) channels at 0.6--2.3\,$\mu$m and 2.4--5.0\,$\mu$m, respectively. The SW channel allows the use of a weak lens with a filter providing photometric monitoring of the target, while the LW channel provides a spectroscopic mode using a grism and a filter. Our observations in the LW channel used the F444W filter with GRISMR element and RAPID readout mode. On the other hand, the WLP4/F212N2 weak lens/filter with RAPID readout mode was used in the SW channel. Both channels employed the SUBGRISM64 subarray that has 2048 columns and 64 rows. This gave us spectroscopic data between 3.9--5\,$\mu$m (centred at around 4.5\,$\mu$m) in the LW channel (or, 4.5\,$\mu$m channel) and one single photometric data point in a narrow-band (2.3\%) bandpass at 2.12\,$\mu$m from the SW channel (also referred to as the 2.1\,$\mu$m channel). Given the brightness of the host star, we chose two groups per integration with a total integration time of about 1.03\,s. 

We used five independent pipelines to reduce and analyse the spectroscopic data at 4.5\,$\mu$m and two different pipelines to analyse the short-wave photometric data. The details of these methods are described in Appendix~\ref{app:a}.

\subsection{Retrieval model and atmospheric scenarios}\label{subsec:retrieval_setup}

We chose two representative independent reductions of occultation depth spectra, from \texttt{stark} and \texttt{HANSOLO} pipelines, to perform atmospheric retrieval. Both reductions differ in their treatment of correlated noise and thus produce slightly different results, which was the reason for choosing two different reductions for retrieval (see Appendix~\ref{app:a} for more details).

To interpret the observational data, we used the open-source \texttt{HELIOS-r2} atmospheric retrieval code \citep{Kitzmann2020ApJ...890..174K}, which uses the nested sampling algorithm \citep{Skilling2004AIPC..735..395S} implemented in the \texttt{MultiNest} library \citep{Feroz2008MNRAS.384..449F}. For the atmospheric characterisation, we tested four different models with a varying level of complexity. The simplest model tries to fit the observational data with a flat line, while the second one assumes the planet to emit like a pure blackbody of temperature $T_\mathrm{bb}$. \boldchange{Since observations by, for example, \citet{2012A&A...547A..18E} and \citet{2021AJ....161..181Z} rule out the presence of a thick primordial hydrogen-helium atmosphere, a 
potential atmosphere has to be secondary in nature. There are two essential pathways to create a secondary atmosphere for a hot planet such as 55\,Cnc\,e. The atmosphere can either be dominated by outgassing from the planetary interior \citep[e.g.,][]{2024ApJ...963..157T} or be created through evaporation of mantle material, or a combination thereof. Thus, for the two atmospheric scenarios, we assumed a secondary atmosphere with outgassed carbon monoxide (CO)/carbon dioxide (\ch{CO2}) \citep[e.g.][]{Heng2023} or an atmosphere produced by an evaporating mantle with a bulk silicate earth composition that is composed of silicon oxide (SiO), silicon dioxide (\ch{SiO2}), and magnesium oxide (MgO) \citep{2022A&A...661A.126Z}.}

Nested sampling allows Occam's Razor \boldchange{\citep{Ockham1495quaestiones}} to be enforced via the calculation of the Bayesian evidence (or marginalised likelihood function)\boldchange{, see \citet{2008ConPh..49...71T, Trotta2017arXiv170101467T}}. 
In practice, this allows us to favour simpler explanations for some of the data (e.g.\ flat line or blackbody function). To provide good constraints on the Bayesian evidence values, \boldchange{within MultiNest we used 5000 live points \citep{Feroz2008MNRAS.384..449F}} for each retrieval calculation. Increasing this value further did not alter the resulting evidence values to a significant degree.

The atmosphere was considered to be isothermal with the surface pressure $p_\mathrm{surf}$ as a free parameter in the retrieval model. The atmosphere and surface were allowed to have their own distinct temperatures, $T_\mathrm{atm}$ and $T_\mathrm{surf}$, respectively. 

The cross sections of CO, \ch{CO2}, SiO, \ch{SiO2}, and MgO were taken from \citet{Li2015ApJS..216...15L}, \citet{Yurchenko2020MNRAS.496.5282Y}, \citet{Yurchenko2022MNRAS.510..903Y}, \citet{Owens2020MNRAS.495.1927O}, and \citet{Li2019MNRAS.486.2351L}, respectively. All temperature and pressure-dependent cross sections were calculated with the open-source opacity calculator \texttt{HELIOS-K} \citep{Grimm2015ApJ...808..182G, Grimm2021ApJS..253...30G}.

\begin{table}
  \caption{Summary of retrieval parameters and prior distributions used for the retrieval models.}  
  \label{table:retrieval_parameter}      
  \centering                                     
  \begin{tabular}{lcc}         
  \hline\hline                       
  Parameter  & \multicolumn{2}{c}{Prior}         \\
             & Type                      & Value \\
  \hline
  \textit{Flat line} & & \\
  Occultation depth & uniform & $0$\,ppm -- $200$\,ppm\\
  \hline
  \textit{Blackbody} & & \\
  $d_{wl}$ & Gaussian & see Table~\ref{tab:obs_log_res} \\
  $R_p/R_*$  & Gaussian & $0.0182 \pm 0.0002$ \\
  $T_\mathrm{bb}$ & uniform & $300$\,K -- $3000$\,K\\
  \hline
  \textit{Atmosphere} & & \\
   $d_\mathrm{wl}$ & Gaussian & see Table~\ref{tab:obs_log_res} \\
   $R_p/R_*$  & Gaussian & $0.0182 \pm 0.0002$ \\
   $p_\mathrm{surf}$  & log-uniform & $10^{-10}$\,bar -- $500$\,bar \\
   $T_\mathrm{surf}$  & uniform     & 300\,K -- 3000\,K \\
   $T_\mathrm{atm}$   & uniform     & 300\,K -- 3000\,K\\
   $\xi_j$            & uniform     & $10^{-10} \leq x_j \leq 1$ \\
  \hline
  \end{tabular}
\end{table}

\begin{figure*}
    \centering
    \includegraphics[width=\textwidth]{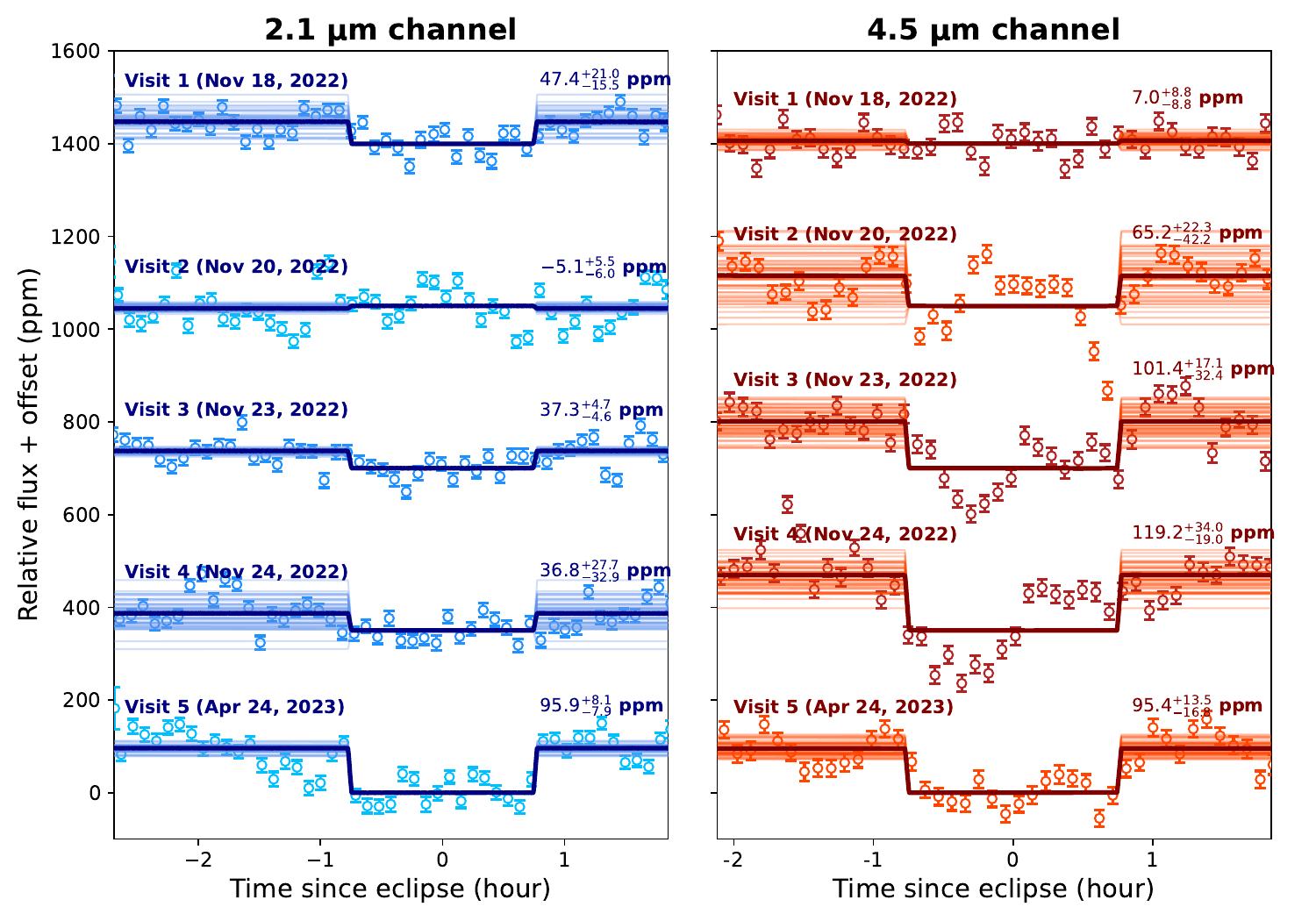}
    \caption{Detrended occultation light curves from the short-wave photometric channel (2.1\,$\mu$m, left panel) and white-light light curves from the long-wave channel (4.5\,$\mu$m, right panel). Only binned data points are shown here. The darker and lighter shades of the points depict even and odd orbital number parity, respectively. The dates and occultation depth (median and 68-percentile confidence intervals) of the visits are indicated above each plot. The best-fitted models and models computed from randomly selected posteriors to show the model uncertainties are plotted with thick and thin lines.}
    \label{fig:detrended-lcs}
\end{figure*}

\begin{figure}
    \centering
    \includegraphics[width=\columnwidth]{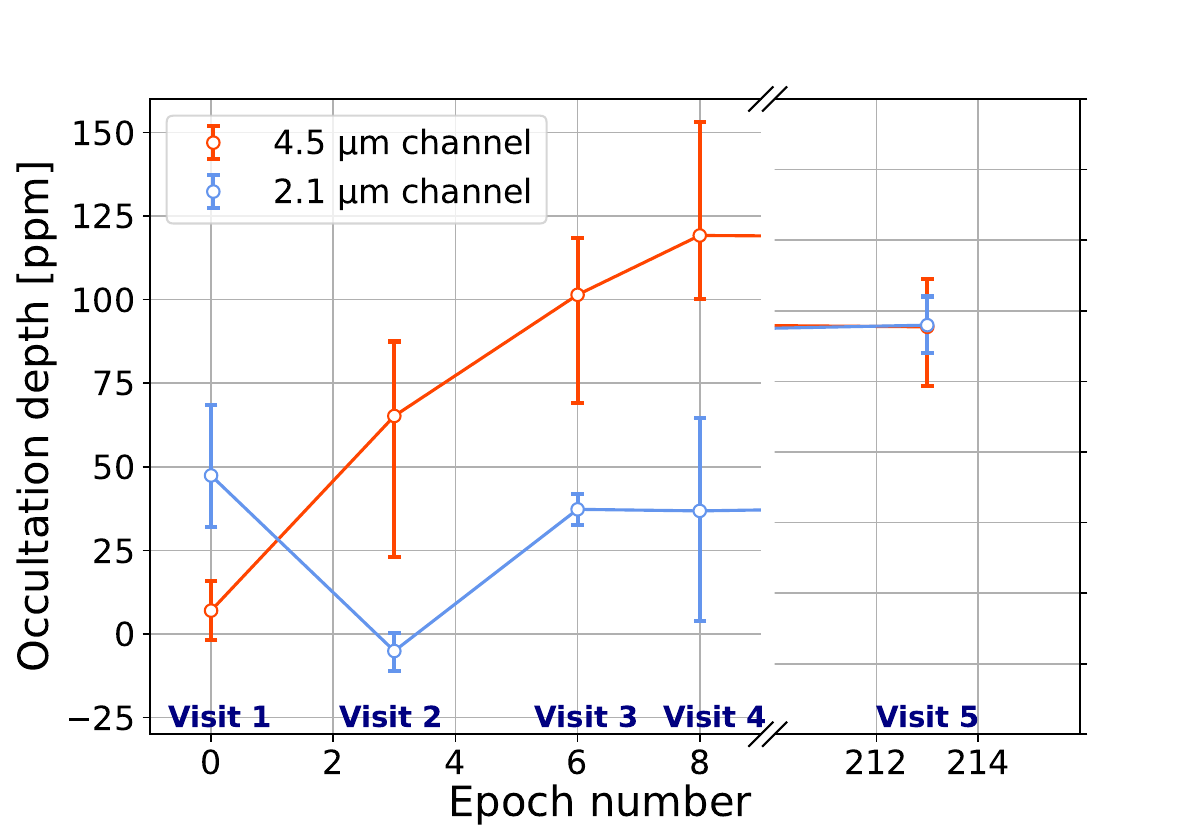}
    \caption{Observed wide band occultation depths in long-wave (in orange) and in short-wave (in blue) channels. The depths are plotted as a function of epoch number starting from the first visit.}
    \label{fig:sw_lw_ecl_dep_comp}
\end{figure}

The atmospheric composition in the retrieval model was described through a centred-log-ratio prior which allows a more optimised sampling of the parameter space when the dominant background gas is not known \citep{Benneke2012ApJ...753..100B}. For a given mixture of $n$ gases, the centred-log-ratio conversion (clr) for the mixing ration $x_j$ of a given molecule $j$ in the mixture is given by 
\begin{equation}
    \xi_j = \mathrm{clr}(x_j) = \ln \frac{x_j}{g(\mathbf x)} \ ,
\end{equation}
where $g(\mathbf x)$ is the geometric mean of all mixing ratios $\mathbf{x}$:
\begin{equation}
    g(\mathbf{x}) = \left( \prod_{j=1}^{n} x_j \right)^{1/n} \ .
\end{equation}
Due to the constraint that 
\begin{equation}
    \sum_{j=1}^n x_j = 1 \quad \text{or}  \quad \sum_{j=1}^n \xi_j = 0 \ ,
\end{equation}
only $n-1$ free parameters are needed in the retrieval. We used uniform priors to produce $\xi_j$ values subject to the constraints that $\min\left(\mathbf{x}\right) = 10^{-10}$ and $\max\left(\mathbf{x}\right) = 1$, see \citet{Benneke2012ApJ...753..100B} for details. We note that the prior boundaries for $\xi_j$ depend on the number of molecules in the retrieval and the chosen value of the smallest allowed mixing ratio.

For the retrieval of the data from the \stark reduction, we performed the calculations on the relative occultation depths.  Thus, for these calculations, we needed to add an additional free parameter to the retrieval: the white-light occultation depth $d_\mathrm{wl}$. For these, we used Gaussian priors with the values provided in Table~\ref{tab:obs_log_res}. Since \texttt{HANSOLO} reduction provides absolute occultation depths this additional parameter was not needed. Additionally, we binned the data provided by \stark which uses the instrument's native resolution to about 30 spectral bins.

All retrieval parameters for the different models are summarised in Table~\ref{table:retrieval_parameter}. The empirically calibrated stellar spectrum of 55\,Cnc from \citet{2012A&A...545A..97C} was used to transform the emission spectra calculated by the retrieval model to wavelength-dependent occultation depths.

\section{Results}\label{sec:res}

\subsection{Wide band occultation depths}

\begin{table*}
  \caption{Overview of retrieval results for the \stark and \hans reductions. Boldface indicates the statistically preferred models.}  
  \label{table:retrieval_results}      
  \centering                                     
  \begin{tabular}{lcccccccccc}         
  \hline\hline                       
  Model & \multicolumn{2}{c}{Visit 1} & \multicolumn{2}{c}{Visit 2} & \multicolumn{2}{c}{Visit 3} & \multicolumn{2}{c}{Visit 4} & \multicolumn{2}{c}{Visit 5} \\
        & $\ln \mathcal Z$ & $B$ & $\ln \mathcal Z$ & $B$ & $\ln \mathcal Z$ & $B$ & $\ln \mathcal Z$ & $B$ & $\ln \mathcal Z$ & $B$ \\
  \hline
  \textbf{\texttt{stark}} & & & & & & & & & &\\                                  
  Flat line           & $-169.98$          & $e^{32.3}$ & $-148.08$          & $e^{14.1}$   & $-154.10$          & $e^{16.2}$ & $-147.70$          & $e^{11.8}$ &  $\mathbf{-135.03}$ & - \\
  Blackbody           & $-159.53$          & $e^{21.8}$ & $-134.26$          & 1.3          & $-154.33$          & $e^{16.4}$ & $\mathbf{-135.90}$ & -          &  $-140.10$          & 159.9 \\
  CO, \ch{CO2}        & $\mathbf{-137.72}$ & -          & $\mathbf{-133.96}$ & -            & $-147.66$          & $e^{9.8}$  & $-135.96$          & 1.1        &  $-137.48$          & 11.6 \\
  SiO, \ch{SiO2}, MgO & $-139.56$          & 6.3        & $-135.17$          & 3.4          & $\mathbf{-137.90}$ & -          & $-136.71$          & 2.3        &  $-141.01$          & $e^{6.0}$  \\
  \hline
  \textbf{\texttt{HANSOLO}} & & & & & & & & & &\\ 
  Flat line           & $-115.19$          & $9.5$ & $-109.72$          & 12.2       & $-143.00$          & 27.2 & $-129.95$          & 1.7 &  $-134.64$         & 1.3 \\
  Blackbody           & $\mathbf{-112.94}$ & -     & $\mathbf{-107.22}$ & -          & $\mathbf{-139.68}$ & -    & $\mathbf{-129.41}$ & -   & $\mathbf{-134.41}$ & -       \\
  CO, \ch{CO2}        & $-113.66$          & 2.1   & $-108.23$          & 2.7        & $-139.72$          & 1.0  & $-130.10$          & 2.0 & $-134.97$          & 1.8    \\
  SiO, \ch{SiO2}, MgO & $-114.06$          & 3.0   & $-108.35$          & 3.1        & $-140.39$          & 2.2  & $-130.36$          & 2.6 & $-135.43$          & 2.5    \\
  \hline
  \end{tabular}
\end{table*}

We used six pipelines to reduce and fit our JWST/NIRCam dataset. The methods are described in detail in Appendix~\ref{app:a}. Here we present results from our primary analysis from the \texttt{stark} pipeline (Appendix~\ref{app:stark}). A summary of our results, along with the observation log, is tabulated in Table~\ref{tab:obs_log_res}.

Our main finding is the strong variability in occultation depths. The white-light occultation depths (computed by fitting an occultation model to the band-averaged occultation time series) at 4.5\,$\mu$m are highly variable even during the short time scale of a week (Table~\ref{tab:obs_log_res}). During the time span of 6 days (8 planetary orbits), the measured occultation depths at 4.5\,$\mu$m continuously increased from basically non-detection in Visit 1 ($7\pm9$\,ppm) to $119^{+34}_{-19}$\,ppm in Visit 4. The occultation depth from our final visit (Visit 5), observed 5 months after the other visits, is $\sim 95\pm16$\,ppm and consistent with the depths from Visits 3 and 4 but differs significantly from the depths from Visit 1 and 2. Fig.~\ref{fig:sw_lw_ecl_dep_comp} shows occultation depths as a function of time, illustrating this point. The best-fitted occultation models along with the de-trended data are shown in Fig.~\ref{fig:detrended-lcs} for all visits.

We used an empirically calibrated stellar spectrum of 55\,Cnc from \citet{2012A&A...545A..97C}, stellar and planetary parameters from \citet{2018A&A...619A...1B} and the NIRCam response function\footnote{\url{http://svo2.cab.inta-csic.es/theory/fps/}} to compute brightness temperatures using the measured white-light occultation depths at 4.5\,$\mu$m. As shown in Table~\ref{tab:obs_log_res}, the brightness temperature changes significantly from 873\,K to 2256\,K within a week. Notably, the brightness temperature almost doubled from Visit 1 to 2, i.e., only after three planetary orbits.

Similarly, the 2.1\,$\mu$m channel occultation depths are also variable. Within a week, the 2.1\,$\mu$m occultation depths remained almost constant at around 40\,ppm for Visit 1, 3 and 4, while we found a non-detection of occultation for Visit 2 which was observed between Visit 1 and 3 (see, Fig.~\ref{fig:sw_lw_ecl_dep_comp}).
However, the final observation that was taken 5 months later (Visit 5) shows a significantly higher occultation depth of $96 \pm 8$\,ppm, which is almost equal to the depth observed at 4.5\,$\mu$m in the same epoch.
\boldchange{The corresponding brightness temperatures varies significantly between 1247\,K and 3138\,K (see, Table~\ref{tab:obs_log_res}).}
Interestingly, there is no correlation between the occultation depth variability observed at 2.1 and 4.5\,$\mu$m (Fig.~\ref{fig:sw_lw_ecl_dep_comp}). Fig.~\ref{fig:detrended-lcs} present the de-trended SW data with best-fitted models.

The variability, plotted in Fig.~\ref{fig:sw_lw_ecl_dep_comp}, is clearly not correlated with the parity of the orbit number. Occultation depths are also variable between occultations from orbits of the same parity, e.g., in even (Visit 1, 3, 4) or odd (Visit 2, 5) visits. The rapid variability thus cannot be explained by simply alternating between two sides of the planet. This does not rule out that the planet rotates asynchronously but means that an explanation for the rapid variability has to be found elsewhere.

All visits showed various degrees of significant correlated noise of unknown origins, in both the 2.1 and 4.5\,$\mu$m channels. The leftover correlated noise can be seen in Fig.~\ref{fig:detrended-lcs} and are also quantified in the Allan deviation plots in Fig.~\ref{fig:allan_deviation}. We perform an injection-retrieval test to estimate proper uncertainties on occultation depths in the presence of correlated noise (see, Sec.~\ref{app:stark_long}). We report uncertainties from this analysis in Table~\ref{tab:obs_log_res}. We, however, found that various methods to account for correlated noise could somewhat change the results of occultation depths and emission spectra (see, Appendix~\ref{app:a} for more details).


\subsection{Occultation depth spectra and atmospheric retrieval}\label{subsec:retrieval}

We computed the relative occultation depth spectra as outlined in Appendix~\ref{app:stark} using the \texttt{stark} reduction and absolute occultation depth spectra from \texttt{HANSOLO} pipeline as described in Appendix~\ref{app:hansolo}. \boldchange{Since different methods to handle the correlated noise could lead to different results, we chose to perform atmospheric retrieval analysis on results from two pipelines, \stark and \hans, which use two representative techniques to deal with the correlated noise (see, Appendix~\ref{app:a} for details).} The occultation depth spectra, shown in Fig.~\ref{fig:spec_comp}, are also variable from visit to visit and do not show any consistent spectral features.

\begin{figure*}
    \centering
    \includegraphics[width=0.161\textwidth]{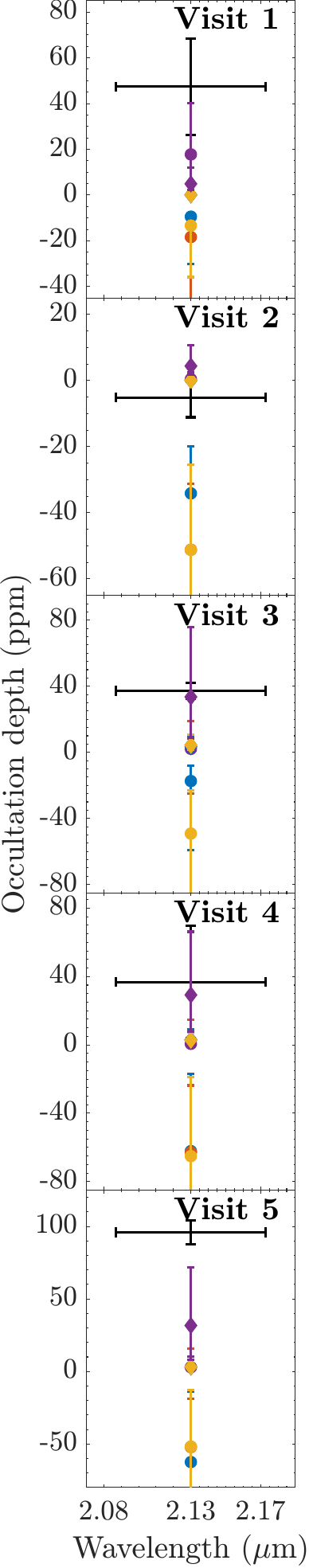}\includegraphics[width=0.41\textwidth]{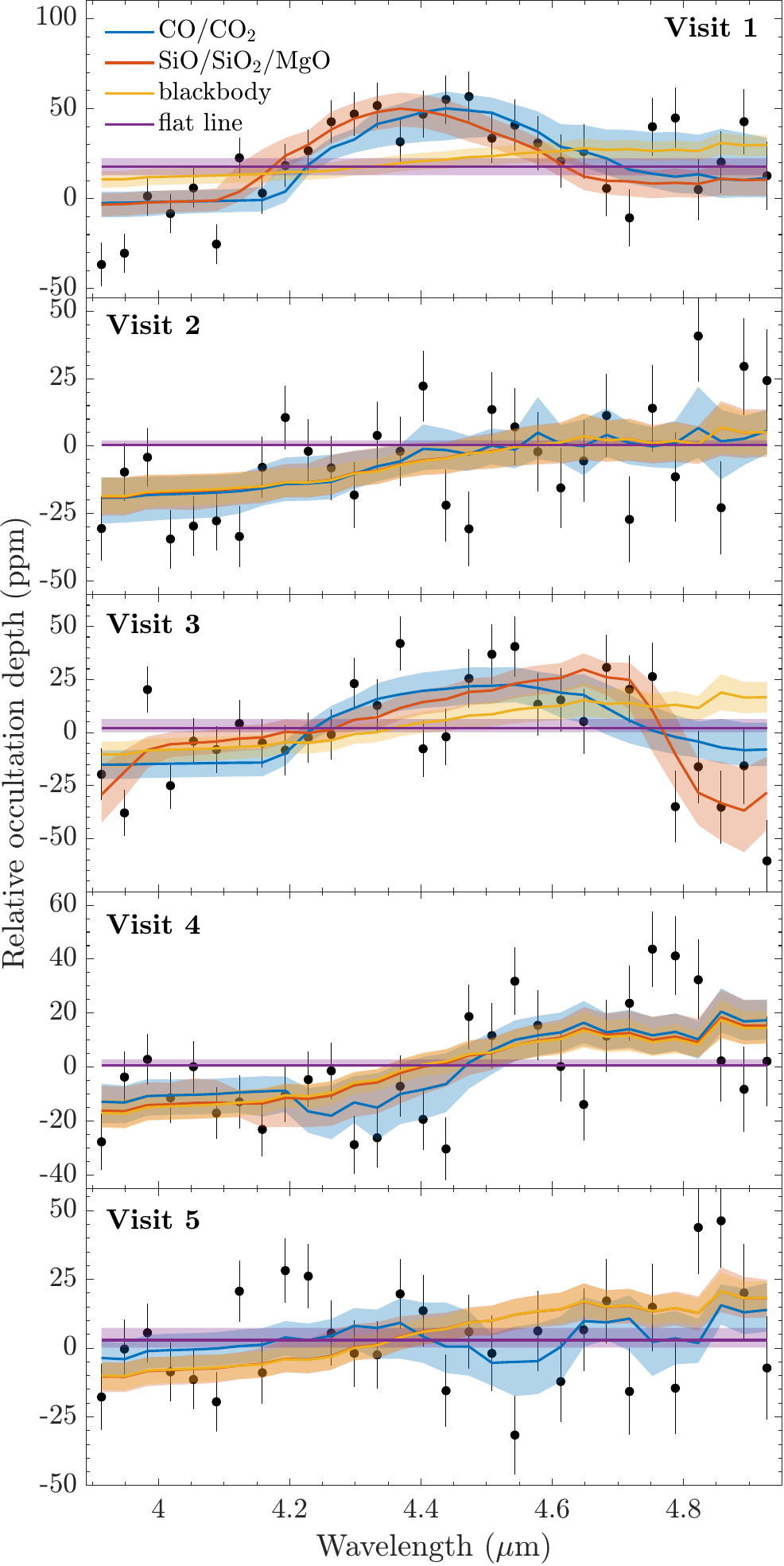}\hspace{0.1cm}\includegraphics[width=0.412\textwidth]{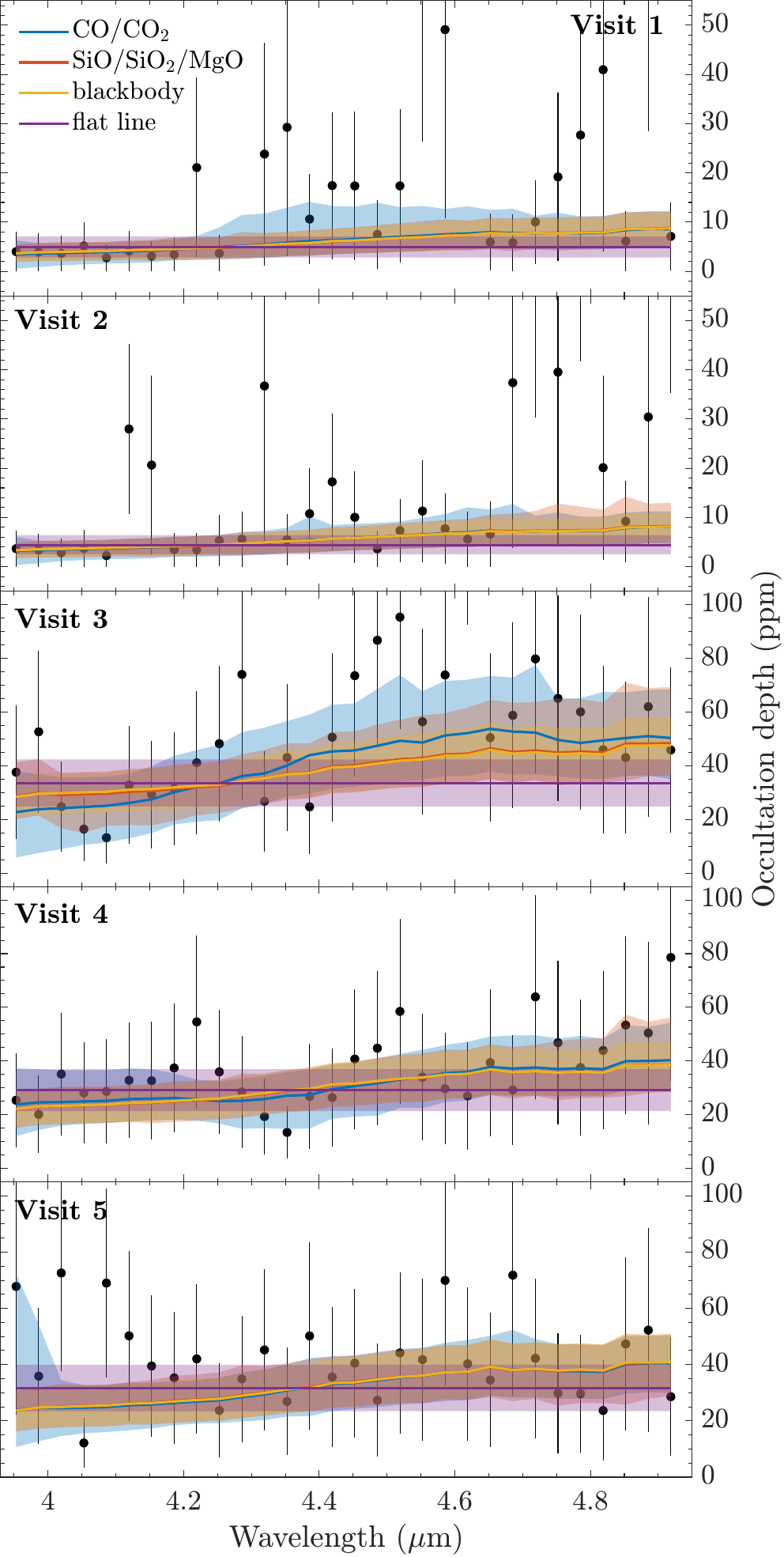}
    \caption{Posterior spectra for all model scenarios and visits. The left column shows predicted occultation depths in the shortwave channel. The black data points indicate the observed value, while diamonds represent the retrieval results for the \hans reduction and squares refer to the outcome for the \stark. The vertical error bars represent the 1-$\sigma$ confidence intervals. The middle column shows the posterior spectra for \stark, while the column on the right-hand side displays the corresponding results for \hans. Solid lines refer to the median spectra from the posterior sample, while the shaded areas correspond to the 1-$\sigma$ intervals. We note that the retrievals for the \stark reductions were made for relative occultation depths, i.e.\ the mean occultation depths in the middle column are close to zero.}
    \label{fig:retrieval_posterior_spectra}
\end{figure*}

\begin{figure*}
    \centering
    \includegraphics[width=0.7\textwidth]{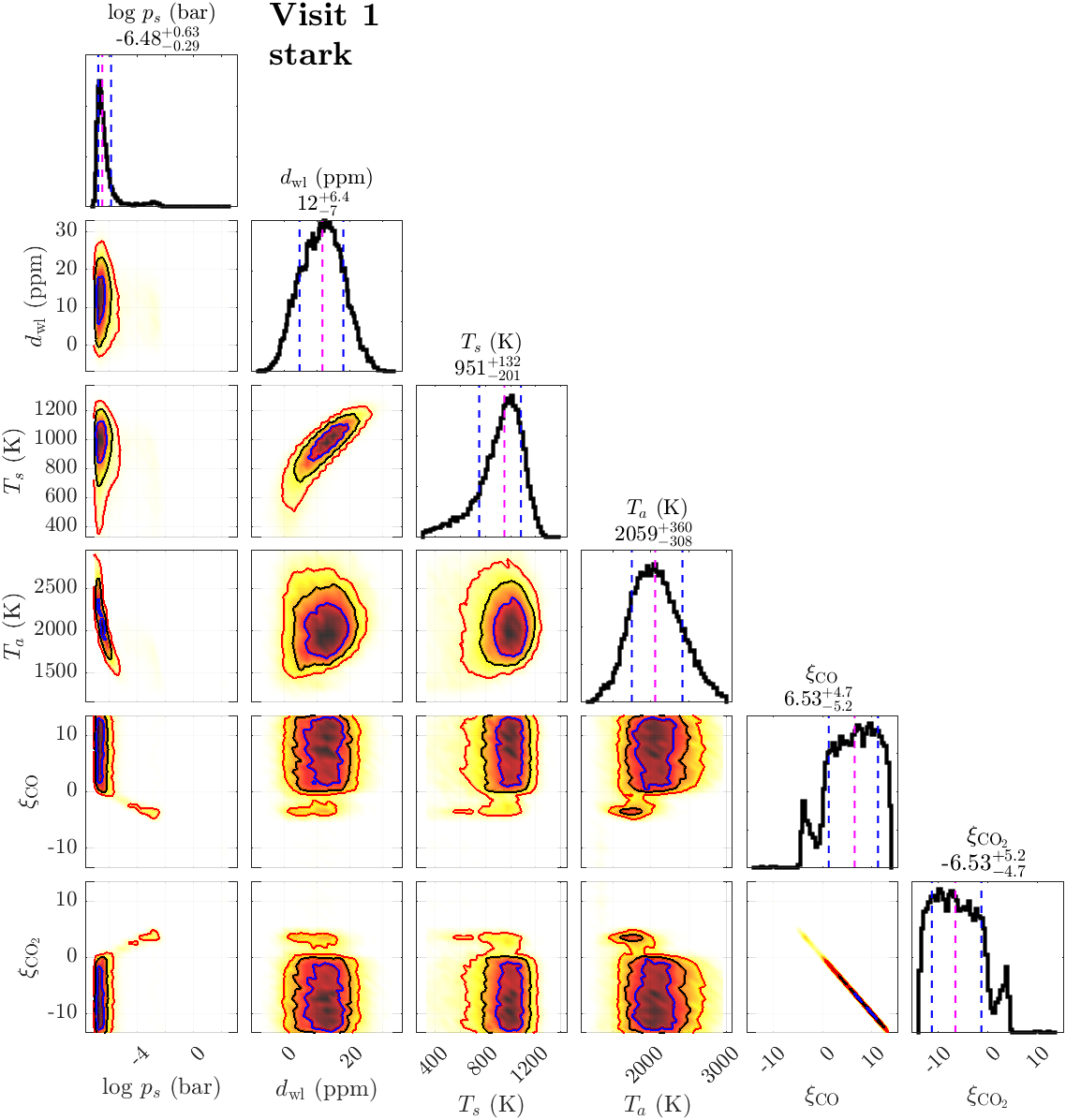}
    \caption{Posterior distributions of the free parameters for the first visit, representing the CO/\ch{CO2}-atmosphere scenario. Results are shown for the \stark reduction. We note that $\xi_\mathrm{CO_2}$  is not a free parameter in the retrieval but was calculated during a postprocess procedure following the requirement that in each posterior sample the sum of all $\xi$ values must be zero.}
    \label{fig:retrieval_posterior_visit1_red1}
\end{figure*}

\begin{figure*}
    \centering
    \includegraphics[width=0.85\textwidth]{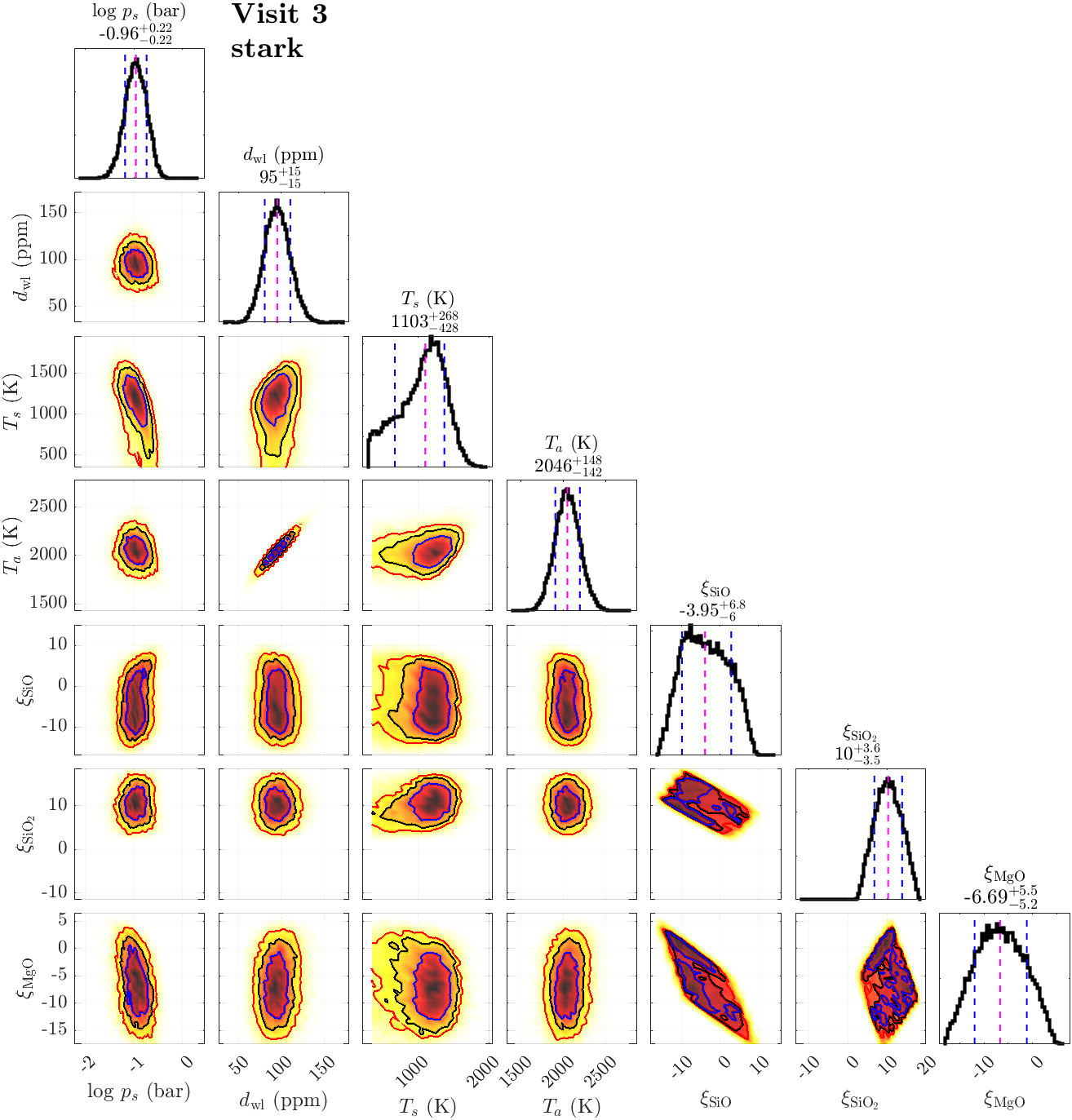}
    \caption{Posterior distributions of the free parameters for the third visit, representing the SiO/\ch{SiO2}/MgO-atmosphere scenario. Results are shown for the \stark reduction. We note that $\xi_\mathrm{MgO}$ is not a free parameter in the retrieval but was calculated during a postprocess procedure following the requirement that in each posterior sample, the sum of all $\xi$ values must be zero.}
    \label{fig:retrieval_posterior_visit3_red1}
\end{figure*}

\subsubsection{Summary of the retrieval results}

The retrieval results for the two different reductions across all five visits and for the four different model scenarios described in Sect. \ref{subsec:retrieval_setup} are summarised in Table~\ref{table:retrieval_results}. The table shows the resulting Bayesian evidence values $\ln \mathcal Z$ and the Bayes factors $B$ with respect to the models with the highest likelihood value. The former are marked in bold for every visit. Fig.~\ref{fig:retrieval_posterior_spectra} additionally shows the posterior spectra for all models, visits, and reductions. The detailed posterior distributions for all atmospheric retrievals can be found in Fig.~\ref{fig:retrieval_posterior_visit1_red1} \& \ref{fig:retrieval_posterior_visit3_red1}, as well as in Appendix~\ref{sec:retrieval_posteriors_appendix}.

The results presented in Table~\ref{table:retrieval_results} suggest that for the \hans reduction, the planetary blackbody model is always the preferred model. This is likely caused by the relatively large errors of the reduction that results in the retrieval favouring a simpler model as can clearly be noticed in the spectra shown in the right column of Fig.~\ref{fig:retrieval_posterior_spectra}. 

However, for most visits, the preference for the simple blackbody model is not statistically significant. The more complex atmospheric scenarios usually have a Bayes factor of less than three, which suggests that they are essentially equally likely. For the first three visits, a flat-line fit to the measured spectrum is effectively ruled out by the Bayesian evidence. The last two visits, on the other hand, can be fit with any of the four models. There seems to be little statistical preference for any of the different modelling scenarios.

The results for the \stark reduction show a much broader range of different models that are statistically preferred. As suggested by Table~\ref{table:retrieval_results}, the first visit strongly prefers a CO/\ch{CO2} atmosphere, the second visit can be explained by either a CO/\ch{CO2} atmosphere, a planetary blackbody, or a siliciate vapour atmosphere, while the third model overwhelmingly prefers the SiO/\ch{SiO2} scenario. The fourth visit is consistent with a planetary blackbody spectrum, as well as an atmosphere with CO \& \ch{CO2}, or SiO, \ch{SiO2} \& MgO. Finally, the last visit strongly prefers a flat-line model.

\subsubsection{Detailed posterior distributions}

Detailed posterior distributions for the preferred model \boldchange{from the \stark reduction} of Visit 1 (CO \& \ch{CO2}) and Visit 3 (SiO, \ch{SiO2} \& MgO), where atmospheric models are favoured, are shown in Fig.~\ref{fig:retrieval_posterior_visit1_red1} \& \ref{fig:retrieval_posterior_visit3_red1}. The posterior distributions for the first visit reveal a bimodal distribution for the surface pressure $p_\mathrm{surf}$ and the abundances of CO and \ch{CO2}. As the two-dimensional correlation plots suggest, the surface pressure has a solution with a very low value of about $10^{-6.5}$\,bar that is dominated by CO in composition, as well as a higher-pressure mode at about $10^{-3}$\,bar that contains mostly \ch{CO2}. \boldchange{For comparison, if the outgassing flux were to be balanced by flux-limited atmospheric escape then the implied surface pressure is $\sim 10^{-7}$ bar \citep{Heng2023}}.  At about 2000\,K, the atmosphere temperature is much warmer than the retrieved temperature for the surface. It is also important to note, that the posterior distribution for the white-light occultation depths $d_\mathrm{wl}$ is shifted from its prior value of $7\pm 9$\,ppm, though they are both still within their 1-$\sigma$ intervals.

The posterior distribution for the SiO/\ch{SiO2}/MgO model shown in Fig.~\ref{fig:retrieval_posterior_visit3_red1} for the third visit, on the other hand, exhibits a unimodal pressure distribution with a median value of about 0.1\,bar. Here, the atmosphere is clearly dominated by \ch{SiO2} with only an upper limit for SiO and essentially no constraints on MgO. The posterior spectra shown in Fig.~\ref{fig:retrieval_posterior_spectra} clearly show the drop-off in the occulation depth near a wavelength of 4.8\,$\mu$m caused by \ch{SiO2}. Just like in the previous CO/\ch{CO2} scenario for Visit 1, the retrieved atmosphere temperature is again much higher than the one of the surface.

\subsubsection{Blackbody temperatures}

\begin{figure}
    \centering
    \includegraphics[width=0.9\columnwidth]{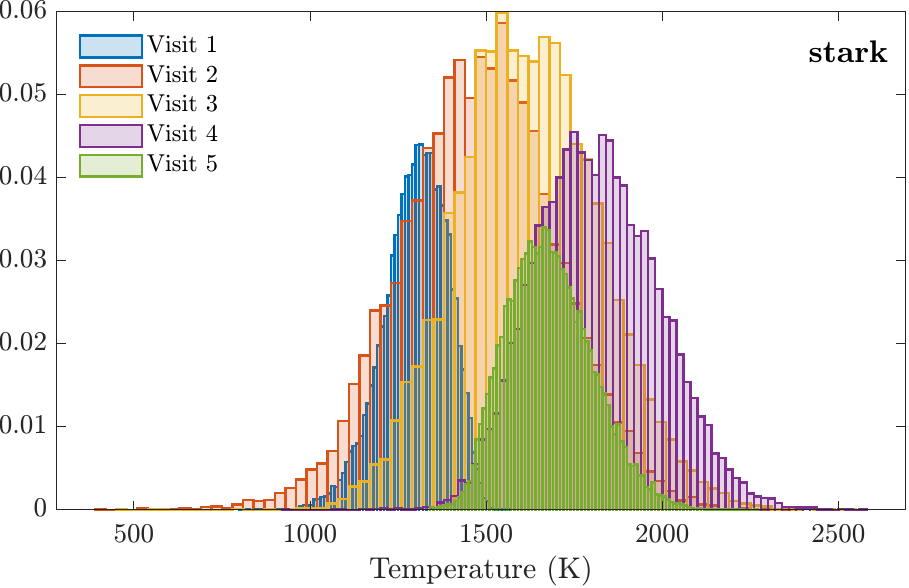}
    \includegraphics[width=0.9\columnwidth]{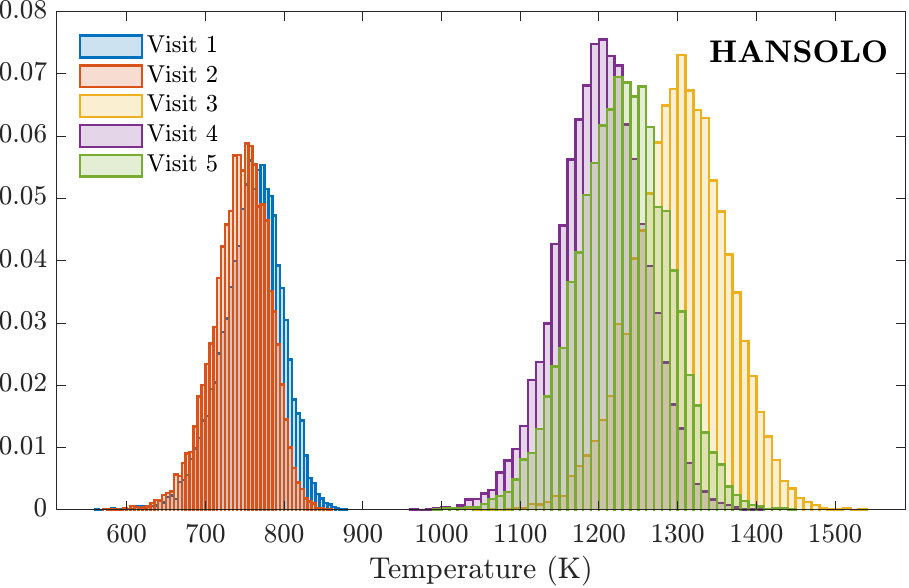}
    \caption{Retrieved temperatures for all five visits using the blackbody model. Top panel: results for the \stark reduction, bottom panel: \hans.}
    \label{fig:posterior_bb}
\end{figure}

The resulting posterior distributions of the blackbody temperature models are shown in Fig.~\ref{fig:posterior_bb} for all visits and the two different reductions. In the case of the \hans reduction, the blackbody is always the preferred model according to the Bayesian evidence, though, as previously mentioned, this preference is statistically not very significant.

As the distributions depicted in the figure suggest, the temperatures retrieved from the \hans observational data are found in two different clusters. A low-temperature mode near 750\,K is found for Visits 1 and 2 and a second one at about 1200\,K to 1300\,K for the other three visits. The temperatures are quite well constrained with 1-$\sigma$ intervals usually in the range of about $\pm 100$\,K, despite the rather large errors on the observational data points (see Fig.~\ref{fig:retrieval_posterior_spectra}).

For \stark, the temperatures are clustered much closer together around a mean temperature of 1500\,K. In comparison to the \hans reduction, however, these temperatures are less well constrained with 1-$\sigma$ intervals typically covering a range of several 100\,K. This is likely caused by the white-light occultation depths that are directly correlated with these temperatures. Following Table~\ref{tab:obs_log_res}, they have in general quite large associated errors that translate into less well-constrained temperatures.

\subsubsection{Surface pressures}

\begin{figure}
    \centering
    \includegraphics[width=0.9\columnwidth]{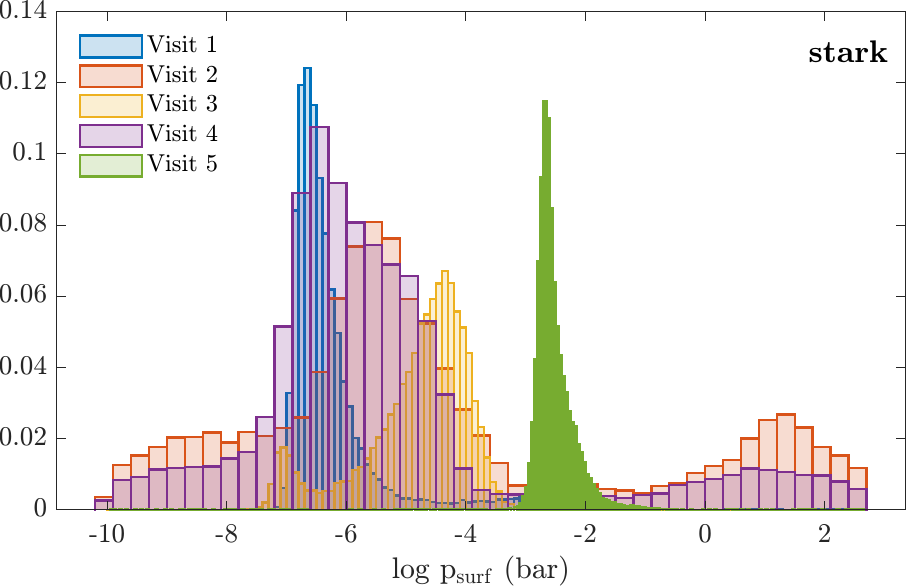}
    \includegraphics[width=0.9\columnwidth]{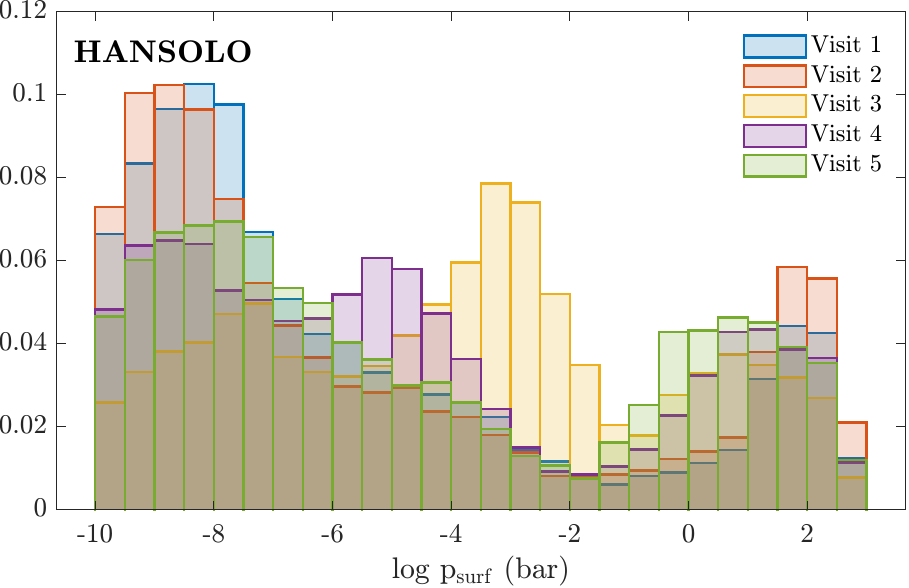}
    \caption{Surface pressure posterior distribution from the CO/\ch{CO2} model for all five visits. Top panel: results for the \stark reduction, bottom panel: \hans.}
    \label{fig:posterior_pressure_co}
\end{figure}

\begin{figure}
    \centering
    \includegraphics[width=0.9\columnwidth]{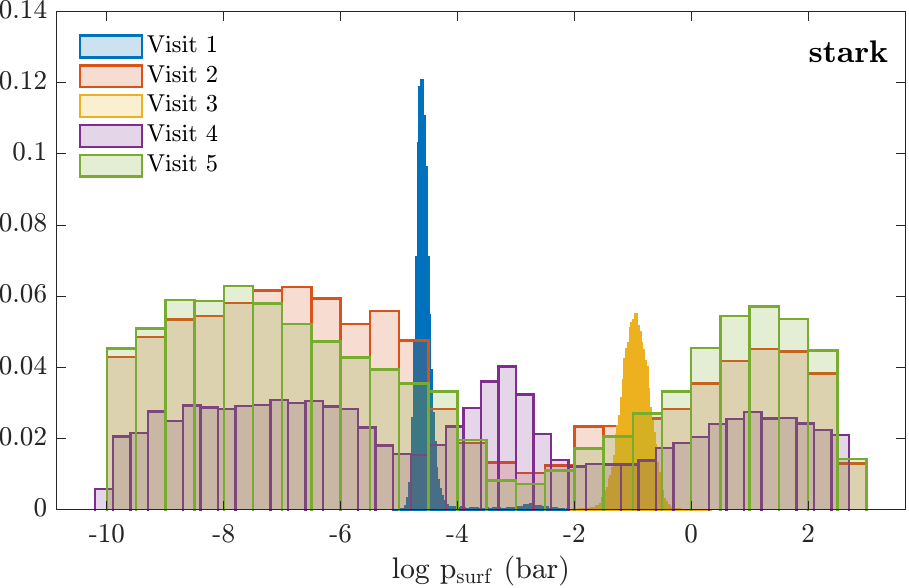}
    \includegraphics[width=0.9\columnwidth]{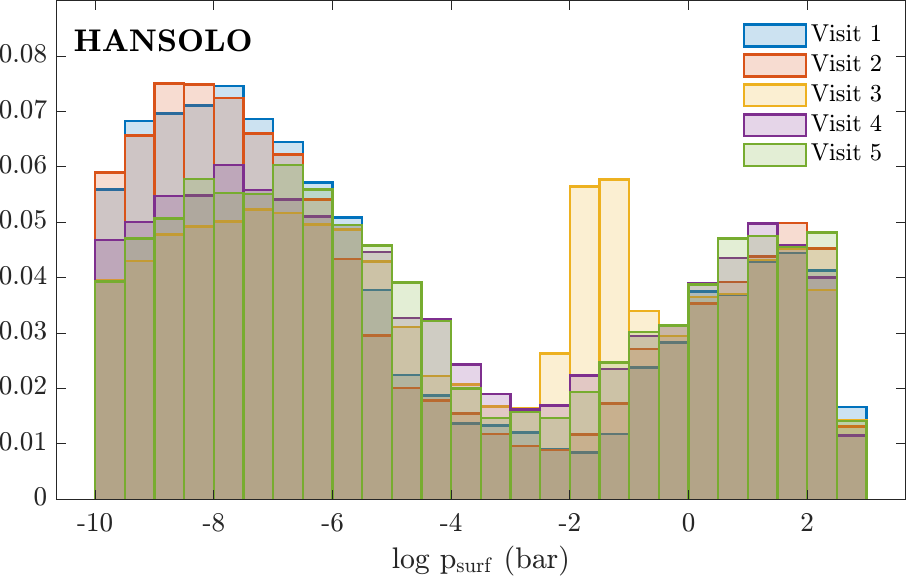}
    \caption{Surface pressure posterior distribution from the SiO/\ch{SiO2}/MgO model for all five visits. Top panel: results for the \stark reduction, bottom panel: \hans.}
    \label{fig:posterior_pressure_sio}
\end{figure}

For the two model scenarios that involve atmospheres, we also retrieve the surface pressure. For the CO/\ch{CO2} model, the corresponding posterior distributions are shown in Fig.~\ref{fig:posterior_pressure_co}, while those for the SiO/\ch{SiO2}/MgO scenario are shown in Fig.~\ref{fig:posterior_pressure_sio}.

In general, the \hans reduction only weakly constrains the surface pressure with posteriors that usually cover the entire prior range of the pressure from $10^{-10}$\,bar to 500\,bar. The posterior distributions seem to be essentially bimodal for almost every visit, with a very low-pressure mode and a high-pressure one. These more or less unconstrained pressures are the result of the rather large errors of the observational data from the \hans reduction. Those make it difficult to provide good constraints for actual atmospheric models.

For the \stark reduction, the results are more diverse. Some visits seem to result in very well-constrained surface pressures. This includes Visits 1 and 5 for the CO/\ch{CO2} model (see upper panel of Fig.~\ref{fig:posterior_pressure_co}) and Visits 1 and \boldchange{3} for the SiO/\ch{SiO2}/MgO case (see upper panel of Fig.~\ref{fig:posterior_pressure_sio}).

Other visits show the same behaviour as for the \hans reduction: rather unconstrained surface pressures with usually a bimodal posterior distribution. Even though not very visible in Fig.~\ref{fig:posterior_pressure_co}, the posterior distribution for Visit 1 is also bimodal in shape, with a smaller, high-pressure mode of an atmosphere dominated by \ch{CO2}, as discussed above.

We note that our retrieved surface pressures differ from the one reported by \citet{2024arXiv240504744H}, which corresponds to our Visit 4 and is based on the JWST program by \citet{2021jwst.prop.1952H}. However, given that even the two reductions of the same data in our study produce different results regarding the atmospheric properties, this is not too surprising. 
Furthermore, \citet{2024arXiv240504744H} employed a different retrieval approach. This includes not using the white-light eclipse depths of the NIRCam data, imposing a lower limit on the surface temperature and allowing for a non-radiatively interacting background gas. Especially the latter assumption will affect the posterior distributions of the surface pressure.

\subsubsection{Surface and atmosphere temperatures}

\begin{figure}
    \centering
    \includegraphics[width=0.9\columnwidth]{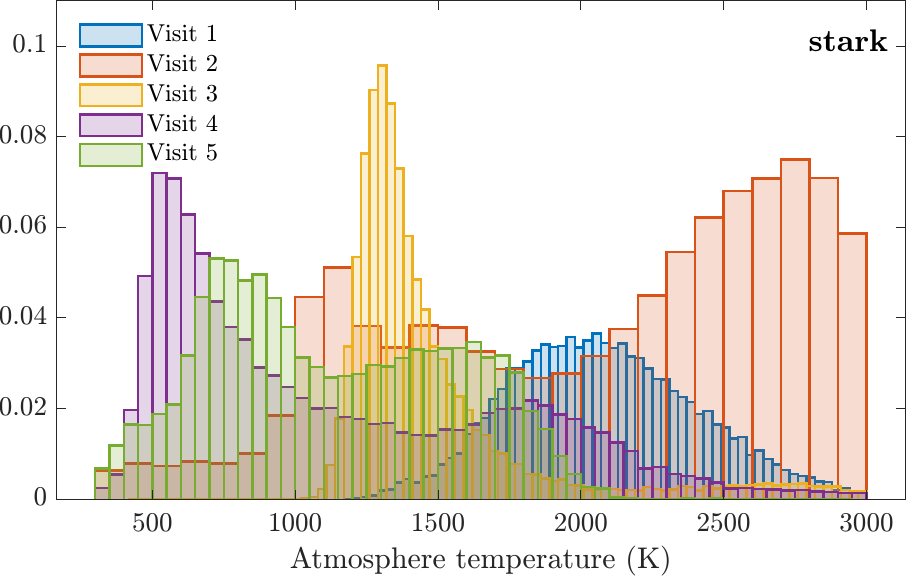}
    \includegraphics[width=0.9\columnwidth]{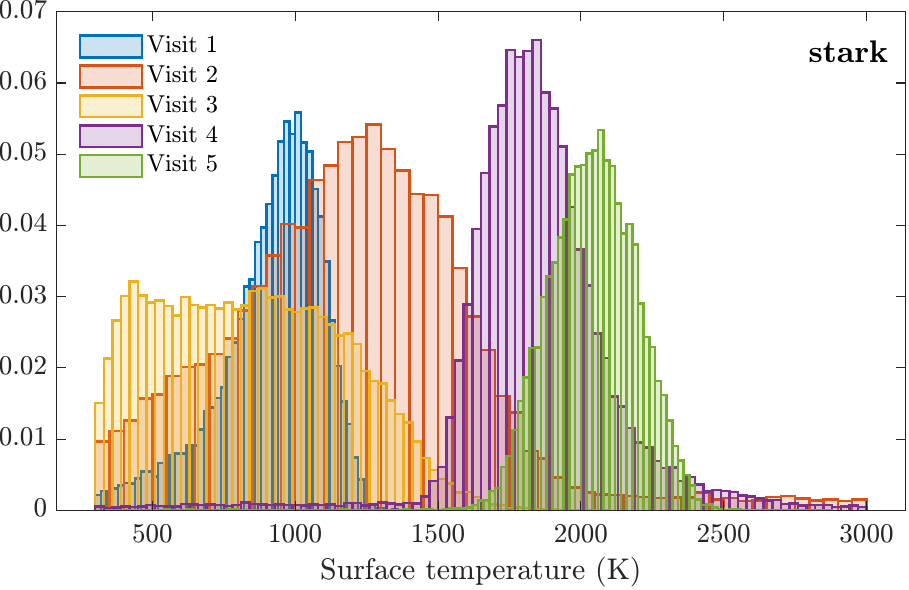}
    \caption{Posterior distributions for the atmosphere (top) and surface temperatures (bottom). The distributions are shown for the CO/\ch{CO2} model and the \stark reduction.}
    \label{fig:posterior_temp_co}
\end{figure}

For the CO/\ch{CO2} model, we present the posteriors for the surface and atmosphere temperatures in Fig.~\ref{fig:posterior_temp_co}. As discussed in Sect. \ref{subsec:retrieval_setup}, we allow these two temperatures to have distinct values. We only present the posteriors for the \stark reduction since, as shown above, the \hans one does not provide good constraints on the atmospheric properties.

Just like the surface pressure, the temperatures are rather well-constrained for some visits, such as the surface temperatures for Visits 4 and 5. Observational data from other visits yield much broader distributions, such as Visit 2, some of which also seem to possess a bimodal shape or only provide upper limits. 

Visit 1 is the only case where the atmosphere seems to have a distinctly higher temperature than the surface. For other visits, this trend is less clear. For example, Visit 5 yields a very high surface temperature but the atmospheric one is less well-constrained and only seems to provide an upper limit that is roughly equal to the surface temperature. In the case of Visit 3, this situation is reversed. Here, the atmosphere temperature is constrained with a median value of roughly 1400\,K, while the surface temperature only has an upper limit of about the same value.

\section{Interpretation of observations}\label{sec:interpretations}

As mentioned in Section \ref{subsec:3:2Theory}, if the variability in the emission from the planet is caused by the planet showing different faces during consecutive occultations, we would expect the occultation depth to be correlated with the orbit number. However, Fig.~\ref{fig:sw_lw_ecl_dep_comp}, which plots the occultation depths as a function of orbit number, shows that this is not the case. This means that the observations give no support for a 3:2 spin-orbit resonance being the root cause for the variability. It is still possible that the planet is trapped in some higher-order spin-orbit resonance, but to show this by establishing a pattern would require many more occultation observations than we currently have.

There are several hypotheses that could potentially explain the full or part of the observations. We outline two such models in the subsections below: a transient outgassing atmosphere model and a circumstellar material supported by the volcanism model. Moreover, the NIRCam data also constrain the presence of spectral features from a mineral atmosphere resulting from a purported lava ocean, as described in Section.~\ref{sec:mineral_atmo} below.

\subsection{Constraints on silicate atmosphere on 55\,Cnc\,e\label{sec:mineral_atmo}}

\begin{figure}
    \centering
    \includegraphics[width=\columnwidth]{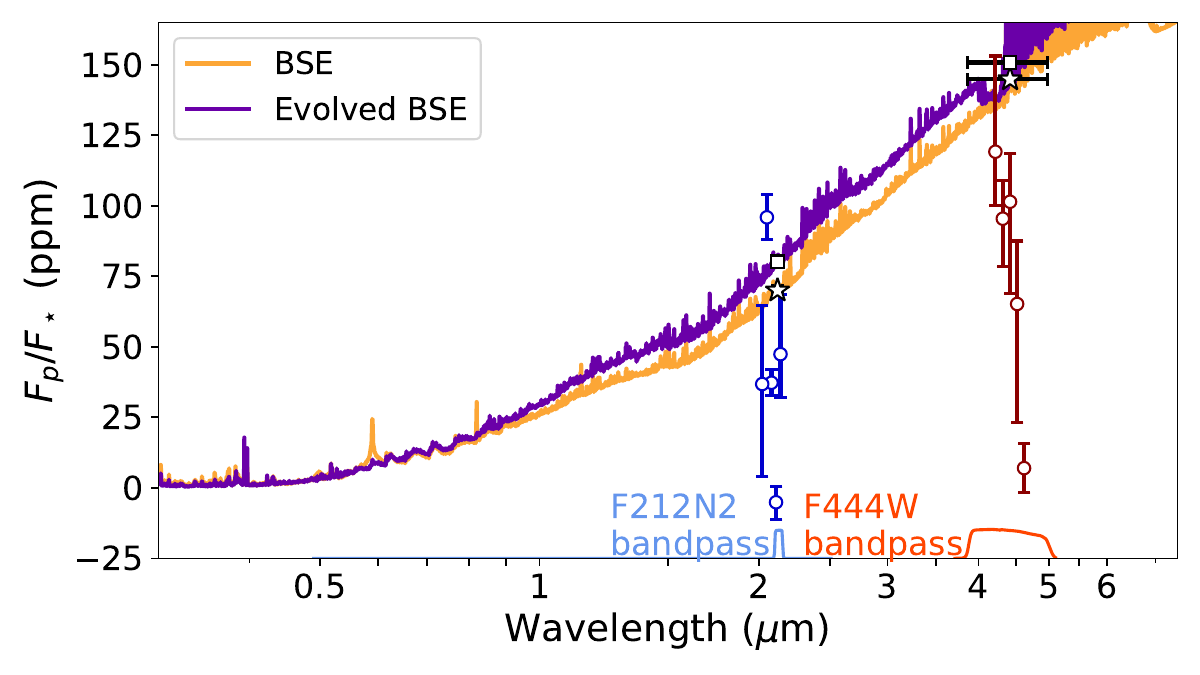}
    \caption{Theoretical models of evaporating lava atmospheres for 55\,Cnc\,e from \citet{2022A&A...661A.126Z}. Two models are for bulk-silicate composition (in yellow) and for evolved bulk-silicate composition (in purple). Also overplotted are photometric occultation depths from 2.1\,$\mu$m channel (in blue) and white-light occultation depths for 4.5\,$\mu$m channel (in maroon). \boldchange{The blue and maroon points are slightly spread in wavelength near their bandpasses to avoid overlap.} We show the two bandpasses corresponding to both of these channels. The black points show the predicted occultation depths for both NIRCam bandpasses with different shapes representing bulk-silicate (stars) and evolved bulk-silicate (squares) compositions.}
    \label{fig:zilinskas22}
\end{figure}

Being in proximity to its host, the substellar temperature on 55\,Cnc\,e can reach >\,2000\,K. The surface of the planet at such a high temperature is expected to be molten if there is no atmosphere on the planet. A molten surface on the planet could then produce a thin rock vapour atmosphere on the planet. \citet{2022A&A...661A.126Z} recently calculated self-consistent models of outgassed atmospheres for all USPs at the time. They solved the radiative transfer equations along with equilibrium chemistry models for the outgassed atmosphere to compute temperature-pressure profile and emission spectra. They showed that gases such as SiO, SiO$_2$, Na, MgO etc. are some of the main constituents of these outgassed atmospheres. Their models for 55\,Cnc\,e\footnote{All models are publicly available at \url{https://github.com/zmantas/LavaPlanets}} are shown in Fig.~\ref{fig:zilinskas22} overplotted with our observations. The models assume bulk silicate (oxidised) Earth (BSE) composition for the planet \boldchange{with unevolved} and evolved surface with 80\% outgassed efficiency (evolved BSE composition).

All of their models with different outgassing efficiencies predict occultation depths of 70--80 and 145--150\,ppm for NIRCam 2.1 and 4.5\,$\mu$m channels. As depicted in Fig.~\ref{fig:zilinskas22} these values are larger compared to our observations. Some occultation depths are, however, consistent with models at 1--3$\sigma$.
One occultation depth at 2.1\,$\mu$m in Visit 5 produces a larger depth compared to the models. 
This hints towards a lack of short-wave absorbers such as SiO and/or SiO$_2$ from the atmosphere that are responsible for thermal inversion and, in turn, larger occultation depths in NIRCam bandpasses.
Indeed, only one visit (Visit 3) favoured the SiO/\ch{SiO2}/\ch{MgO} model in the retrieval analysis. The band-averaged occultation depth for this visit at 4.5\,$\mu$m agrees with the model prediction (145\,ppm for BSE case) at 2.4$\sigma$. However, the short-wave occultation depth in this visit is inconsistent with the model prediction at 7$\sigma$. We here note that \citet{2024arXiv240504744H} found that the occultation depths in the MIRI bandpass are significantly lower than what is predicted by \citet{2022A&A...661A.126Z} models and thus do not support the presence of the silicate-rich atmosphere.

At the same time, lower occultation depths in the NIRCam bandpasses could imply the presence of a gaseous species that have opacity sources in our NIRCam bandpasses. Alternatively, the lower occultation depths, translated into lower brightness temperatures, suggest a thick atmosphere with a strong heat re-distribution \citep[e.g.,][]{2017ApJ...849..152H}. The estimated dayside brightness temperatures \boldchange{(see, Table \ref{tab:obs_log_res})} at 4.5\,$\mu$m (Table~\ref{tab:obs_log_res}) in all visits are smaller than the expected dayside temperature\footnote{Computed using $T_{\textnormal{day}} = T_\star \sqrt{\frac{R_\star}{a}} (1-A_B)^{1/4} f^{1/4}$, while using zero bond albedo and the heat re-distribution factor $f=2/3$, for a bare rock with no heat re-distribution \citep{2014PNAS..11112601B, 2019ApJ...886..140K}.}
of 2537\,K indicating the presence of heat transfer. In either case, our observations seem to indicate the existence of volatiles in the atmosphere of 55\,Cnc\,e. However, it is still challenging to explain the very large occultation depth (and, thus, hot brightness temperature \boldchange{--- 3138\,K; see, Table~\ref{tab:obs_log_res}}) observed at 2.1\,$\mu$m in Visit 5.

\subsection{Constraints on an outgassed secondary atmosphere}\label{subsec:heng23}


\cite{Heng2023} previously suggested that a transient, outgassed secondary atmosphere is capable of simultaneously explaining the observed variability of 55\,Cnc\,e in both the optical/visible and infrared range of wavelengths. Specifically, atmospheres of several \boldchange{tens of} bars of pure carbon monoxide (CO) are capable of producing occultation depths of about 21\,ppm in the CHEOPS and TESS bandpasses, which are consistent with most of the occultation depths measured by CHEOPS \citep{2023A&A...677A.112M} \boldchange{and TESS \citep{2022A&A...663A..95M}}.
However, a change in atmospheric surface pressure of several \boldchange{tens of} bars through loss processes or outgassing over the observed variability time scale in the CHEOPS data is difficult to explain.
Such outgassed atmospheres are incapable of producing occultation depths as high as $\approx 40$--50\,ppm, which were measured thrice in Fig.~3 of \cite{2023A&A...677A.112M}. Similarly, they cannot produce phase variations as high as 110\,ppm as measured by MOST \citep{2019A&A...631A.129S}. It cannot be ruled out that these anomalously high occultation depths are associated with stellar activity. 


For the first data reduction (\texttt{stark}), the outgassed atmosphere with CO and \ch{CO2} is associated with the highest Bayesian evidence in Visits 1 \& 2. Bayesian model comparison does not disfavour this interpretation of Visit 4 as well.  Fig.~\ref{fig:retrieval_posterior_visit1_red1} shows the interpretation of the spectrum from Visit 1 using a CO+CO$_2$ atmosphere.  For Visit 3, a silicate-vapour atmosphere is strongly preferred over an outgassed atmosphere (with the logarithm of the Bayes factor being 9.8; Fig.~\ref{fig:retrieval_posterior_visit3_red1}).  For the more conservative second data reduction (\texttt{HANSOLO}), the retrieval associated with the highest Bayesian evidence is a blackbody curve over all 5 visits.


The simplest interpretation of the spectra is using a blackbody curve, which is consistent with the data in Visits 2 and 4 of the \texttt{stark} reduction and all 5 visits of the \texttt{HANSOLO} reduction.  Fig.~\ref{fig:posterior_bb} shows the posterior distributions of the blackbody temperature.  For Visits 2 and 4 of the \texttt{stark} reduction, the blackbody temperature is broadly between 1500\,K and 2000\,K.  Note that a blackbody curve does not automatically imply that one is probing a bare rocky surface, since an optically thick, isothermal atmosphere may also produce a blackbody curve \citep{Heng2023}.  For the \texttt{HANSOLO} reductions, the blackbody temperature is about 750\,K for Visits 1 and 2 and increases to about 1250\,K for Visits 3, 4 and 5 over a period of about 2.2 days (between Visits 2 and 3).  Such a duration is not inconsistent with the radiative timescale, which is under an Earth day for $\sim 1$\,bar atmospheres \citep{Heng2023}.  If 55\,Cnc\,e has a bare rocky surface and negligible albedo, then its temperature would be the equilibrium temperature of about 2000\,K.  If we take these blackbody temperatures (750\,K and 1250\,K) seriously, then it implies that the observations are not probing a bare rocky surface that has reached a steady state with the stellar instellation, unless one assumes implausibly high surface albedos.

If we focus on the interpretation of the spectra using CO-CO$_2$ atmospheres, then Figs.~\ref{fig:posterior_pressure_co} and \ref{fig:posterior_temp_co} show the posterior distributions of surface pressures, atmospheric temperatures and surface pressures.  For the \texttt{HANSOLO} data reductions, the surface pressure is unconstrained.  For Visits 1, 2 and 4 of the \texttt{stark} reduction, the inferred surface pressure is $\sim 1$ $\mu$bar.  The surface temperature is $\sim 1000$\,K, which is only possible if the surface has not come to radiative equilibrium with the stellar instellation because of the presence of an atmosphere.  The atmospheric temperature jumps from $\sim 2000$\,K to $\sim 2500$\,K to $\sim 1500$\,K from Visits 1 to 2 to 3.  While this is not implausible because of the short radiative timescales, we do not have a mechanism to explain how and why this happens.

\subsection{Can a circumstellar inhomogeneous dusty torus explain variability?}\label{subsec:torus}



Two of our observations, Visit 1 at 4.5\,$\mu$m and Visit 2 at 2.1\,$\mu$m, show occultation depths that are consistent with zero at 1-$\sigma$. These non-detections are challenging to explain with any kind of atmospheric phenomena. Moreover, the occultation depths observed at 2.1\,$\mu$m and 4.5\,$\mu$m are not correlated with each other (Fig.~\ref{fig:sw_lw_ecl_dep_comp}), which potentially hints towards different origins of variability in different wavelength channels.

A grey absorber could explain the optical and 2.1\,$\mu$m channel variability. A natural candidate for this grey absorber is a circumstellar dust torus \citep{2019A&A...631A.129S, 2023A&A...677A.112M}. The progenitor of the dusty torus could be the volcanism on 55\,Cnc\,e developed by the extreme tidal heating akin to Io \citep[e.g.,][]{2019ApJ...885..168O, 2020MNRAS.497.5271G}. 
The most common gases from volcanism seen on the Earth, Io, and Venus - e.g., SO$_2$, CO$_2$, generate a tenuous atmosphere on the planet. Volcanism, supported by significant tidal heating, is expected to expel a prodigious quantity of dust grains into the upper atmosphere, which ultimately escape the planet's gravitational sphere of influence due to impinging stellar ions. 
Upon escape, such a mechanism may eventually generate a patchy, circumstellar dust torus, which has been shown to be sufficiently opaque in visible light to produce optical variability \citep{2023A&A...677A.112M}. Volcanic gases are additional non-trivial sources of opacity in our NIRCam 4.5\,$\mu$m channel. Analytical models showed that an optically thin \citep[e.g.,][]{2020MNRAS.497.5271G} SO$_2$ \boldchange{atmosphere} with a range of pressures can produce the IR variability observed with Spitzer. Since the Spitzer/IRAC bandpass at 4.5\,$\mu$m and our NIRCam/F444W bandpass have a large overlap in wavelength, it remains a possibility that a similar thin SO$_2$ (\boldchange{or, any other volcanic gases such as \ch{CO2} which also absorbs at 4.5\,$\mu$m}) atmosphere with several tens of $\mu$bar could explain the observed variability in our NIRCam dataset. \boldchange{To evaluate this idea in detail is however beyond the scope of the present work and instead planned for an upcoming publication (Oza et al., \textit{in prep.}).}



The variability at 2.1\,$\mu$m is difficult to explain with a thin
\boldchange{atmosphere consisting volcanic gases such as \ch{SO2} or \ch{CO2} since they do}
not have significant opacity in the 2.1\,$\mu$m bandpass. Instead, the dust grains present in the torus could be a cause of this variability, which was also hypothesised by \citet{2023A&A...677A.112M}.
If the grain size is larger than 0.3\,$\mu$m from the size range of 0.1--0.7\,$\mu$m discussed in \citet{2021A&A...653A.173M} and \citet{2023A&A...677A.112M}, the particles will be opaque in the 2.1\,$\mu$m channel, but transparent in the 4.5\,$\mu$m channel. Although many Earth-like dust species do not survive long enough in the circumstellar environment, dust made of quartz, silicon carbide and graphite can survive a significant fraction of an orbit to generate a patchy torus \citep{2023A&A...677A.112M}. 
Following the same formalism from \citet{2023A&A...677A.112M},
the mass loss needed to account for the maximum change in occultation depth (95.9\,ppm, in visit 5) 2.5--5.7 $\times$ 10$^6$ kg\,s$^{-1}$  is within a factor of two of the maximum escape rate derived by CHEOPS, reported to be as large as $\sim$ 2.9 $\times$ 10$^6$ kg\,s$^{-1}$ \citep{2023A&A...677A.112M}. \boldchange{If the particle size is larger than 0.7\,$\mu$m, they can, in principle, even explain the variability at 4.5\,$\mu$m channel. However, the non-correlation of occultation depths at 2.1\,$\mu$m and 4.5\,$\mu$m channels suggests that although the two sources may be linked, they are indeed distinct absorbers, e.g, grains and gas at 2.1 and 4.5 $\mu$m, respectively, as mentioned above.} 
However, the effect of the dust torus on the transit observations is yet to be found observationally. \boldchange{In particular, if the dust escape happens during a transit event, dust could float in the Hill sphere of the planet or form a comet-like tail \citep[e.g.,][]{2012A&A...545L...5B}. Both processes should affect the transit light curve in the form of a significantly large transit depth and an asymmetric transit shape, respectively, unless dust very quickly leaves the vicinity of the planet.}

It is unknown what escape mechanism is currently operating at 55\,Cnc\,e, and therefore more phase curve observations, especially at shorter wavelengths \boldchange{where Si in the dust have emission lines}, are needed to monitor the variability. \boldchange{Multiple phase curves would scan the whole circumstellar region over time to determine the location of the dusty torus and how it evolves, helping in a better understanding of the escape mechanism and thus variability.} However, based on its close proximity several mechanisms including canonical photoevaporation and boil-off \citep{affolter2023} are able to reproduce the \boldchange{estimated} escape rate. For close-in rocky bodies like 55\,Cnc\,e, more energetic \boldchange{plasma} escape mechanisms including \boldchange{ion-neutral interactions such as} atmospheric sputtering \citep{2019ApJ...885..168O, 2024JGRE..12907935M}, which, similar to Io, drive a feedback process sourced by the melting and degassing of the rocky body itself via induction-heating \citep{Lanza2021} and two body tidal-heating \citep{o2019_55, Quick2020, Charnoz2021}. 

The aforementioned escape mechanisms are source-limited by geological activity and expected to vary on orbital timescales in phase-curve observations \citep{2024JGRE..12907935M}. \boldchange{Source-limited implies that the escape rate is ultimately limited by the outgassing rate below the escape layer, such that if the supply rate were zero, escape would not occur.}
Effectively, the discussed energetic escape mechanisms naturally generate extended neutral and grain clouds that provide a toroidal opacity source in the circumstellar environment.


\subsection{Can stellar activity cause the occultation depth variability?}

\boldchange{Stellar activity can, in principle, cause the occultation depth variability of 55\,Cnc\,e. \citet{2023A&A...669A..64D} checked whether stellar granulation could explain the optical occultation depth variability found with CHEOPS. They, however, rejected stellar activity as a source of variability due to very low occultation depths in some visits and their detection of a sinusoidal temporal trend of the variability. Furthermore, the photometric monitoring of the star for about 11 years in the optical from the ground revealed a photometric variability of 0.006\,mag which is too small to explain the $\sim$\,50\,ppm occultation depth variability observed with CHEOPS \citep{2008ApJ...675..790F, 2023A&A...669A..64D}. The stellar activity signal is expected to decrease at longer wavelengths. This means that it is challenging to explain IR variability with the photometric variation of mmag level observed by \citet{2008ApJ...675..790F} in the optical. Moreover, the activity has to happen every instance during the short time window around the occultation, which is improbable. In any case, the inflation of uncertainties with the injection-retrieval method accounts for any noise, including the correlated noise. The fact that the maximum difference in the occultation depths is significant even with inflated uncertainties suggests that the origin of the occultation depth variability is not related to the star.}

\section{Conclusions}\label{sec:conclusions}


We obtained time on JWST/NIRCam to study the dayside emission variability of 55\,Cnc\,e (GO\,2084: PI Brandeker and GO\,1952: PI Hu). In particular, we test the hypothesis that 55\,Cnc\,e is in a 3:2 spin-orbit resonance, thus showing different faces at every occultation and thereby explaining the observed dayside variability and also the hot-spot displacement from the sub-stellar location. The prediction was that this would result in occultation depths highly correlated with their orbital number parity, at least over short time scales.

We observed five occultations of 55\,Cnc\,e in two wavelength bands, or channels, a spectroscopic band at 4.5\,$\mu$m and a single photometric band at 2.1\,$\mu$m. Four of them are observed within a week, i.e., in the duration of eight planetary orbits, while the last was observed after five months. We analysed the data using six different pipelines. 
Our main finding is that the occultation depths change strongly, \boldchange{from a non-detection to 100\,ppm}, and rapidly (within a week). The variability is however not observed to correlate with the occultation number parity, \boldchange{implying} that a planet 3:2 spin-orbit resonance is not the reason for its variability. The variability is observed in both 2.1 and 4.5\,$\mu$m channels, but is curiously not correlated between channels. The estimated brightness temperature at 4.5\,$\mu$m varies between 873\,K -- 2256\,K. These values are less than the predicted dayside temperature in case of zero heat re-distribution and zero albedo, 2537\,K, which hints at the presence of a planetary atmosphere enabling the heat re-distribution.

The spectroscopic data at 4.5\,$\mu$m is affected by correlated noise of unknown origin. Although the results from different reductions overall agree well with each other, there are several differences in white-light occultation depths and emission spectra that can be attributed to different treatments of correlated noise. We select two representative reductions, \texttt{stark} and \texttt{HANSOLO}, to perform atmospheric retrieval. Our atmospheric retrieval was performed using two simple atmospheric models containing an isothermal atmosphere made up of either CO/\ch{CO2} or \ch{SiO}/\ch{SiO2}/\ch{MgO}. Additionally, we also tested a blackbody model and a flat line model with no atmospheric features. Retrievals performed with \hans results mainly favour a blackbody model owing to larger errorbars on the occultation depths. However, other models with CO/\ch{CO2} or \ch{SiO}/\ch{SiO2}/\ch{MgO} were not discarded either, statistically. The retrievals with \stark prefer CO/\ch{CO2} atmospheres in at least two visits, SiO/\ch{SiO2}/\ch{MgO} atmosphere in one visit and blackbody and flat line models in the remaining two visits. The CO/\ch{CO2} atmosphere could be generated from outgassing of the surface \citep[e.g.,][]{Heng2023}. The outgassing could be stochastic and thus can potentially explain the variability. \boldchange{As already advocated by \citet{Heng2023}, simultaneous observations in the optical and infrared are needed to corroborate (or refute) the presence of a transient outgassed CO/\ch{CO2} atmosphere.}

The occultation depth variability in the 2.1\,$\mu$m channel, especially its uncorrelated behaviour with its 4.5\,$\mu$m channel counterpart, is challenging to explain with a simple atmospheric model. It is possible that the variability seen at 2.1\,$\mu$m and that at 4.5\,$\mu$m have different origins. A circumstellar inhomogeneous cloud of dust could potentially describe the variability at 2.1\,$\mu$m. Volcanism induced by extreme tidal heating of 55\,Cnc\,e could be a natural source of dust in the atmosphere of the planet which would eventually escape the planet and generate a patchy dusty torus in the circumstellar environment. The presence of dust in the circumstellar environment could also be helpful in the interpretation of several non-detection of occultation depths found in our observations as it could hide our view of the planet. More observations at shorter wavelengths, e.g., in ultraviolet, would help to more strongly constrain the presence of a circumstellar patchy dust torus. \boldchange{Simultaneous observations in near and mid-IR around 4 and 8\,$\mu$m where volcanic gases \ch{CO2}/\ch{SO2} have opacity would be helpful in constraining their presence. Such multiple observations in the optical and IR would not only constrain the presence of a circumstellar dust torus and atmosphere on the planet but also probe how these components evolve with time, essentially distinguishing both scenarios discussed in this work.}

While we do find a hint of an atmosphere on the planet in at least some visits, corroborating \citet{2024arXiv240504744H}, the simple picture of a static atmosphere cannot explain all observational features. A more complex model, including an outgassed atmosphere, circumstellar material, and perhaps dynamical processes in the atmosphere, would probably be needed to explain the entire range of observations. Moreover, given the strong variability of the system, simultaneous multi-wavelength observations would go a long way to distinguish between possible explanations and help probe the true nature of 55\,Cnc\,e.

\begin{acknowledgements}
We would like to thank an anonymous referee for their detailed referee report and suggestions which significantly improved the manuscript.
JAP acknowledges N\'{e}stor Espinoza for discussing the peculiarities of JWST data analysis. JAP would like to thank Ludmila Carone for an insightful dialogue about theoretical models of the planet. JAP and ABr were supported by the Swedish National Space Agency (SNSA).
The contributions of DP and ML have been carried out within the framework of the NCCR PlanetS supported by the Swiss National Science Foundation under grants 51NF40$\_$182901 and 51NF40$\_$205606. DP and ML also acknowledge support of the Swiss National Science Foundation under grant number PCEFP2\_194576. 
EMV acknowledges support from the Centre for Space and Habitability (CSH). This work has been carried out within the framework of the National Centre of Competence in Research PlanetS supported by the Swiss National Science Foundation under grant 51NF40\_205606. EMV acknowledges the financial support of the SNSF.
This project has received funding from the European Research Council (ERC) under the European Union's Horizon 2020 research and innovation programme (project {\sc Spice Dune}, grant agreement No 947634, and {\sc Four Aces}; grant agreement No 724427).
ADe and DEh have received funding from the Swiss National Science Foundation for project 200021\_200726. This work has also been carried out within the framework of the National Centre of Competence in Research PlanetS supported by the Swiss National Science Foundation under grant 51NF40\_205606.
This research has made use of the Spanish Virtual Observatory (\url{https://svo.cab.inta-csic.es}) project funded by MCIN/AEI/10.13039/501100011033/ through grant PID2020-112949GB-I00.
CMP\ and MF\ gratefully acknowledge the support of the SNSA (DNR 65/19, 177/19). 
BOD acknowledges support from the Swiss State Secretariat for Education, Research and Innovation (SERI) under contract number MB22.00046.
Part of this research was carried out at the Jet Propulsion Laboratory, California Institute of Technology, under a contract with the National Aeronautics and Space Administration (80NM0018D0004). Part of the High Performance Computing resources used in this investigation were provided by funding from the JPL Information and Technology Solutions Directorate.
Finally, we thank ERASMUS student Charlotte Zimmermann for her contributions to the initial studies of this work.
\end{acknowledgements}

%
\bibliographystyle{aa} 
\bibliography{references} 
%

\begin{appendix}

\section{Data analysis methods}\label{app:a}

\begin{figure*}
    \centering
    \includegraphics[width=14cm]{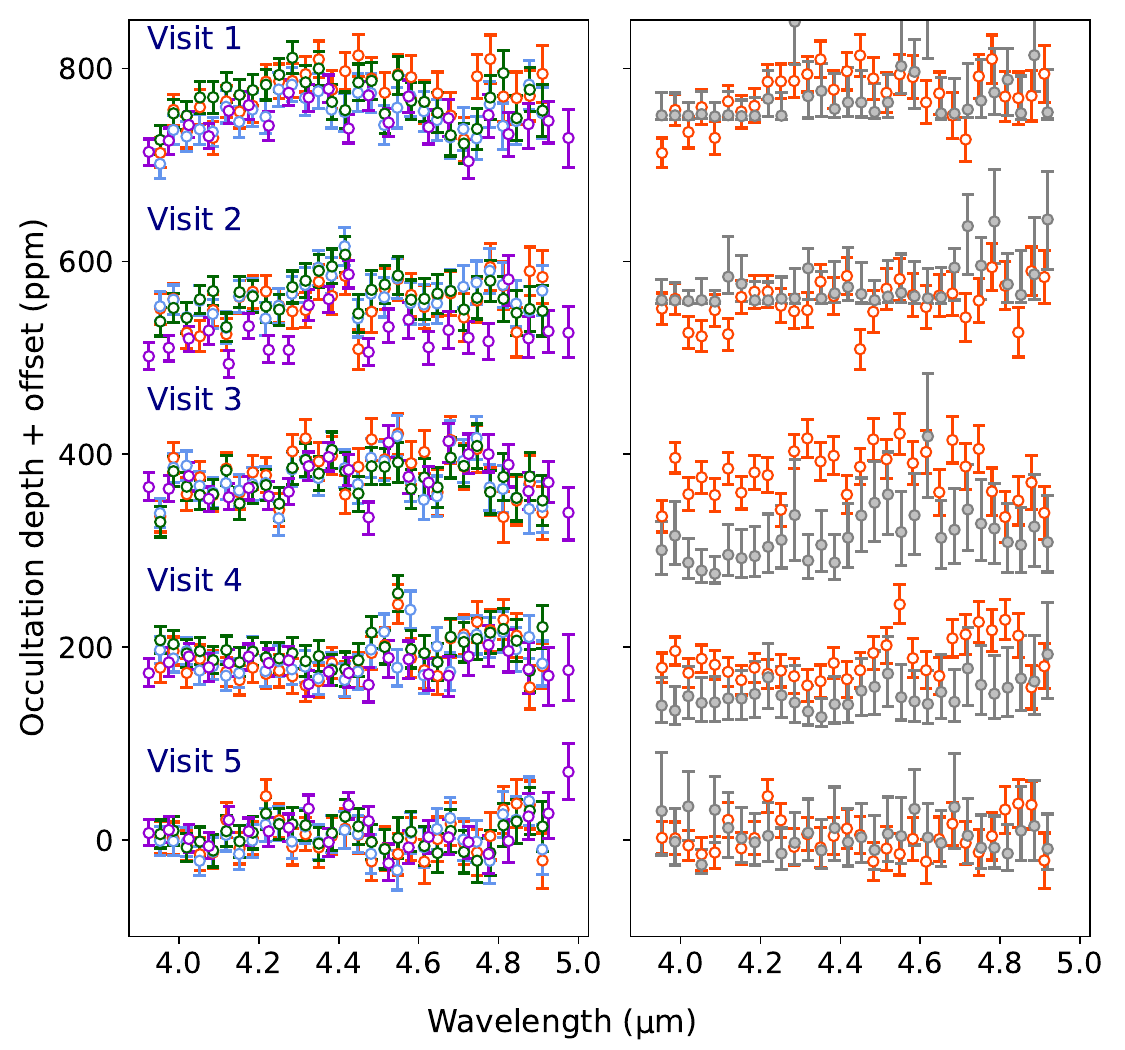}
    \caption{Comparison of occultation depth spectra for all observations from different methods: (\textit{Left}) Relative occultation depth spectra from \texttt{stark} (baseline spectra, in orange), \texttt{Eureka!} (in blue) and \texttt{transitspectroscopy} (in green), and absolute occultation depth spectra minus white-light depth for \texttt{SPARTA} (in purple). (\textit{Right}) \texttt{stark} relative occultation depth spectra (in orange) and \texttt{HANSOLO} absolute occultation depth spectra minus white-light depth (in grey).}
    \label{fig:spec_comp}
\end{figure*}

This section details six independent methods of analysing the JWST/NIRCam data. In Table~\ref{tab:ecl_dep_comp}, we summarise the white-light occultation depths between about 4 and 5\,$\mu$m (see, below for exact wavelength range for different methods) and photometric occultation depths at 2.1\,$\mu$m. Figure \ref{fig:spec_comp} compares the relative occultation depth spectra for all visits from different methods. It can be seen from Figure \ref{fig:spec_comp} and Table~\ref{tab:ecl_dep_comp} that the results obtained with various independent analysis methods overall agree with each other, however, there are some differences which could be attributed to the different handling of correlated noise in the data. For example, \texttt{HANSOLO} reduction uses Gaussian processes (GP) to model the correlated noise and thus produces results, white-light and spectroscopic occultation depths, that are the most distinct from the rest of the methods. On the other hand, reduction methods from, e.g., \stark, inflate errorbars on occultation depths to account for correlated noise.
We use results from \hans and \stark as two representative methods in our atmospheric retrieval analysis and interpretation. We describe each analysis method below.

\begin{table*}
    \centering
    \caption{Comparison of white-light and photometric occultation depths from different methods}
    \begin{tabular}{lcccccc}
    \hline
    \hline
    \noalign{\smallskip}
    Visit & \texttt{stark} & \texttt{Eureka!} & \texttt{Eureka!} & \texttt{HANSOLO} & \texttt{transitspectroscopy} & \texttt{SPARTA}  \\
     & (ppm) & R1 (ppm) & R2 (ppm) & (ppm) & (ppm) & (ppm) \\
    \noalign{\smallskip}
    \hline
    \hline
    \noalign{\smallskip}
    White-light occultation depths & & & & & \\ \hline
    \noalign{\smallskip}
    Visit 1 (Nov 18, 2022) & {\small $7.0 ^{+8.8} _{-8.8}$} & $-$ & {\small $49.2 ^{+12.4} _{-12.3}$} & {\small $2.6 ^{+14.1} _{-2.6}$} & {\small $15.9 ^{+11.6} _{-11.4}$} & {\small $52.1 ^{+11.1} _{-10.3}$} \\
    \noalign{\smallskip}
    Visit 2 (Nov 20, 2022) & {\small $65.2 ^{+22.3} _{-42.2}$} & $-$ & {\small $85.1 ^{+9.6} _{-9.8}$} & {\small $6.4 ^{+31.0} _{-6.1}$} & {\small $52.2 ^{+11.3} _{-11.5}$} & {\small $79.0 ^{+10.0} _{-9.5}$} \\
    \noalign{\smallskip}
    Visit 3 (Nov 23, 2022) & {\small $101.4 ^{+17.1} _{-32.4}$} & $-$ & {\small $130.9 ^{+10.3} _{-11.3}$} & {\small $112.1 ^{+28.4} _{-31.9}$} & {\small $141.9 ^{+11.5} _{-12.0}$} & {\small $119.1 ^{+10.8} _{-10.3}$} \\
    \noalign{\smallskip}
    Visit 4 (Nov 24, 2022) & {\small $119.2 ^{+34.0} _{-19.0}$} & $-$ & {\small $134.1 ^{+9.6} _{-9.5}$} & {\small $37.8 ^{+28.8} _{-24.1} $} & {\small $115.5 ^{+8.9} _{-8.9}$} & {\small $82.9 ^{+18.0} _{-18.3}$} \\
    \noalign{\smallskip}
    Visit 5 (Apr 24, 2023) & {\small $95.4 ^{+13.5} _{-16.8}$} & $-$ & {\small $106.7 ^{+9.2} _{-11.7}$} & {\small $73.5 ^{+21.3} _{-21.4}$} & {\small $98.6 ^{+11.0} _{-10.8}$} & {\small $95.9 ^{+11.3} _{-10.1}$} \\
    \noalign{\smallskip}
    \hline
    \noalign{\smallskip}
    Photometric occultation depths & & & & & \\ \hline
    \noalign{\smallskip}
    Visit 1 (Nov 18, 2022) & {\small $47.4 ^{+21.0} _{-15.5}$} & {\small $42.8 ^ {+4.9} _{-4.7}$} & $-$ & $-$ & $-$ & $-$ \\
    \noalign{\smallskip}
    Visit 2 (Nov 20, 2022) & {\small $-5.1 ^{+5.5} _{-6.0}$} & {\small $-9.8 ^ {+5.6} _{-6.0}$} & $-$ & $-$ & $-$ & $-$ \\
    \noalign{\smallskip}
    Visit 3 (Nov 23, 2022) & {\small $37.3 ^{+4.7} _{-4.6}$} & {\small $28.2 ^ {+5.5} _{-5.6}$} & $-$ & $-$ & $-$ & $-$ \\
    \noalign{\smallskip}
    Visit 4 (Nov 24, 2022) & {\small $36.8 ^{+27.7} _{-32.9}$} & {\small $39.5 ^{+6.0} _{-5.6}$} & $-$ & $-$ & $-$ & $-$ \\
    \noalign{\smallskip}
    Visit 5 (Apr 24, 2023) & {\small $95.9 ^{+8.1} _{-7.9}$} & {\small $92.4 ^{+5.9} _{-5.5}$} & $-$ & $-$ & $-$ & $-$ \\
    \noalign{\smallskip}
  \hline
    \end{tabular}
    \tablefoot{The uncertainties are 68 percentile of the corresponding posterior distribution. Visit 4 is the archival observation from \citet{2024arXiv240504744H}.}
    \label{tab:ecl_dep_comp}
\end{table*}

\subsection{\texttt{stark}}\label{app:stark}

As described in Section \ref{subsec:obs}, the observations were carried out using NIRCam grism timeseries observing mode, which has two channels, a long-wave (LW, 4.5\,$\mu$m channel) spectroscopic channel and a short-wave (SW, 2.1\,$\mu$m channel) photometric channel. We analysed both datasets with our pipeline.

\begin{figure*}[h!]
    \centering
    \includegraphics[width=1.01\textwidth]{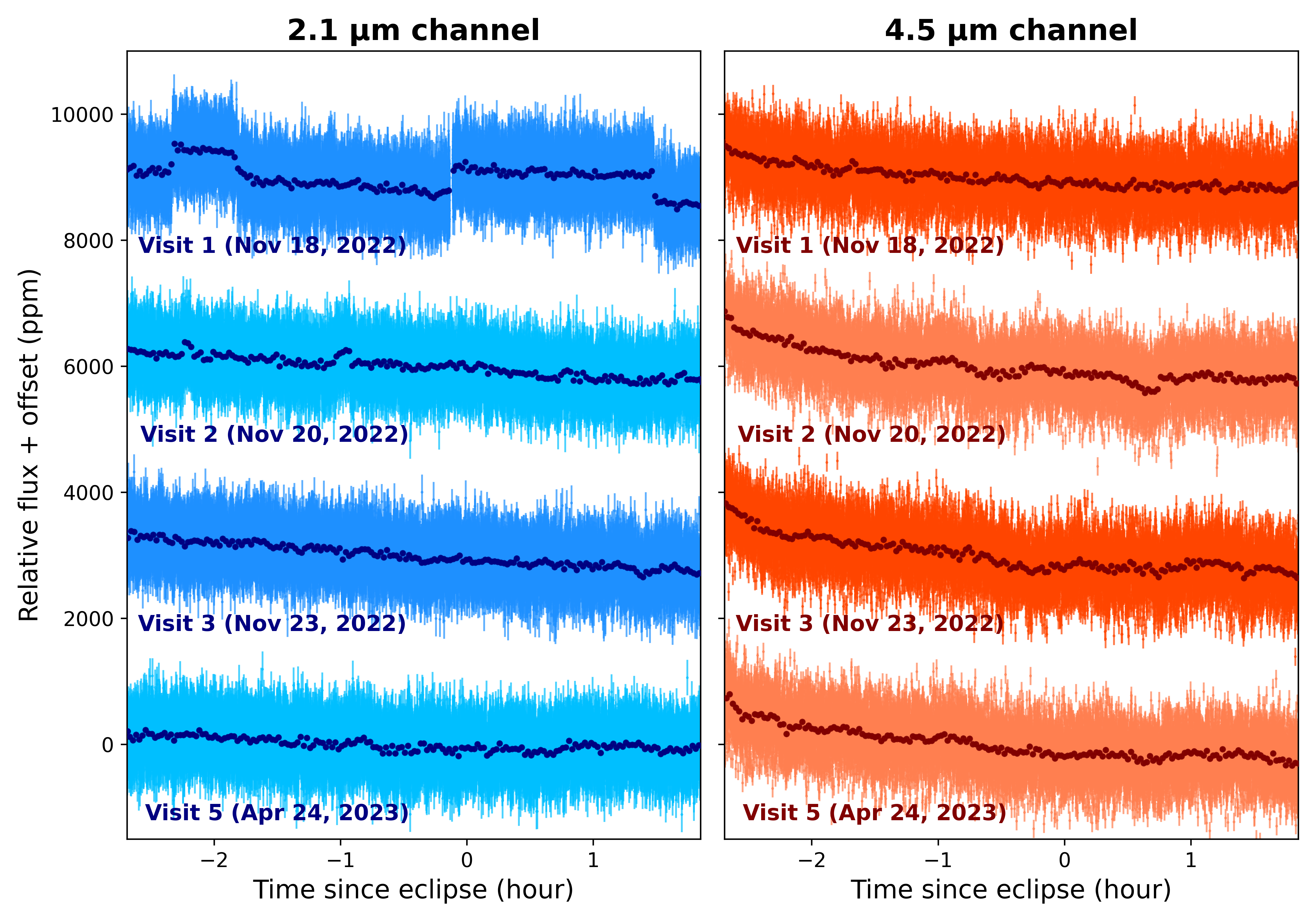}
    \includegraphics[width=1.01\textwidth]{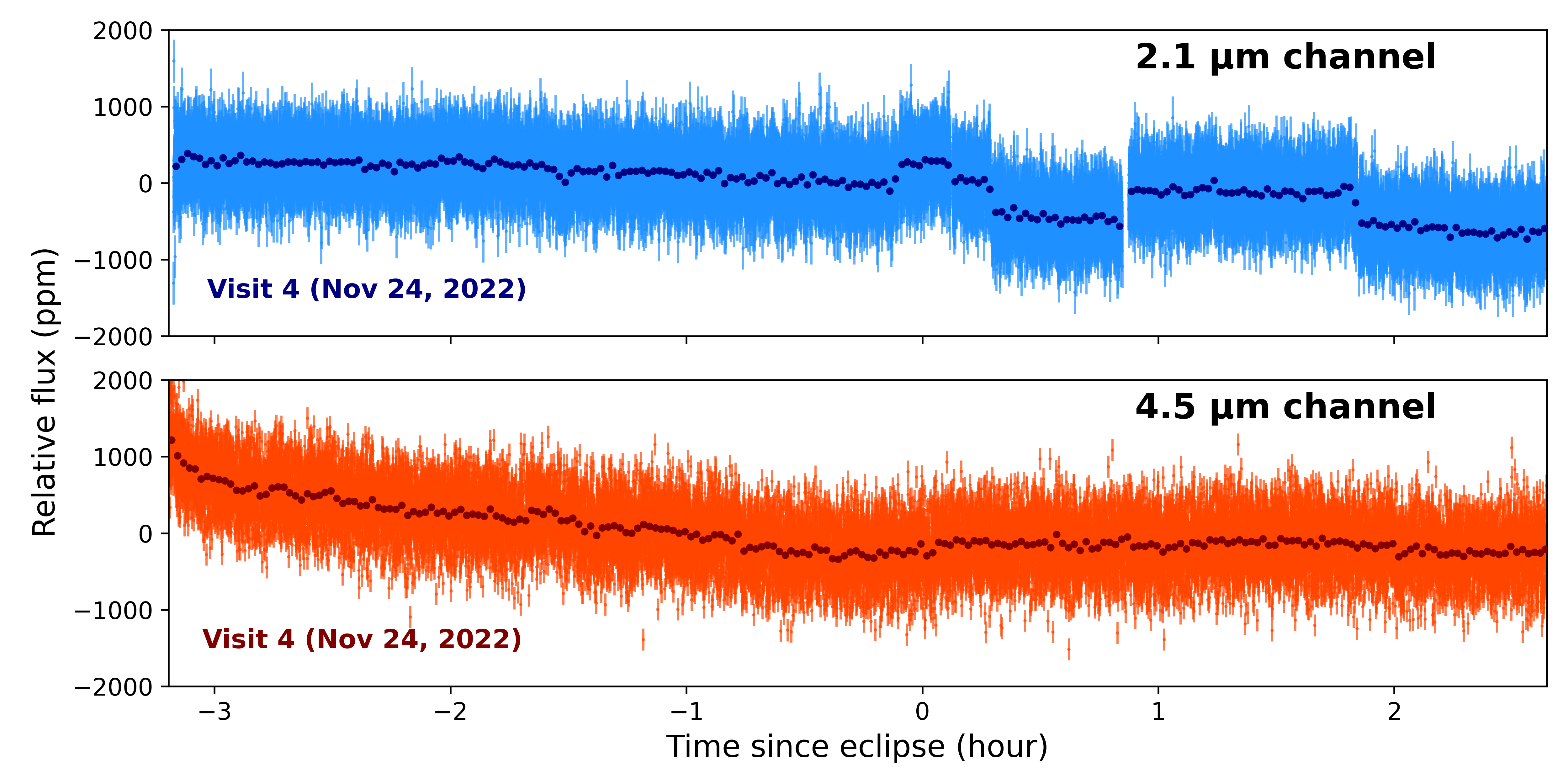}
    \caption{Raw photometric light curves from the short-wave channel at 2.1\,$\mu$m (in blue) and raw white-light light curves from the long-wave channel at 4.5\,$\mu$m (in orange) for Visit 1 to 3 and 5 (GO\,2084, in the top panel) and for Visit 4 (GO\,1952, bottom panel). A darker and lighter shade of colours depicts the even and odd parity of the observations. The darker points on the top of the main data show the binned data points.}
    \label{fig:raw_lc}
\end{figure*}

\begin{figure*}
    \centering
    \includegraphics[width=\columnwidth]{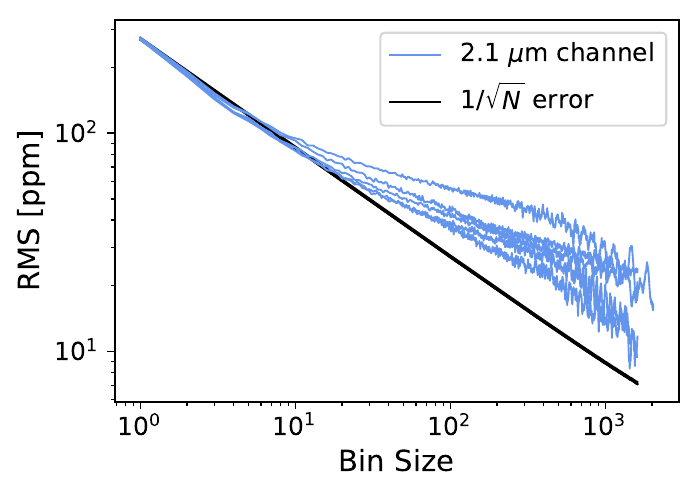}
    \includegraphics[width=\columnwidth]{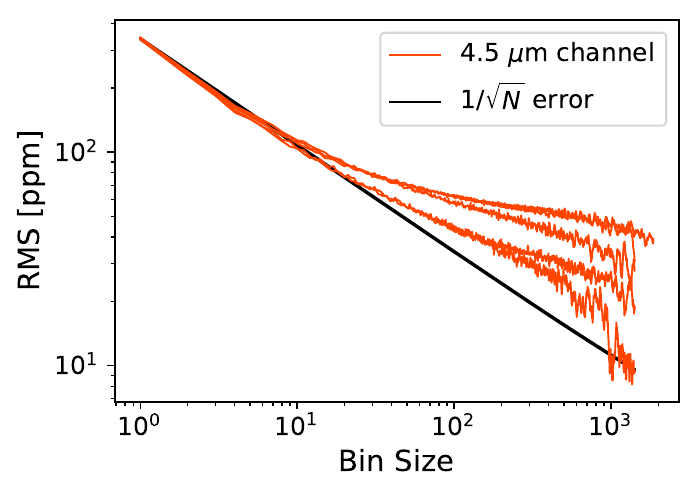}
    \caption{Allan deviation plots of residuals from photometric light curve analysis from 2.1\,$\mu$m (SW) channel (left panel, in blue) and 4.5\,$\mu$m (LW) channel white-light light curve analysis (right panel, in orange).}
    \label{fig:allan_deviation}
\end{figure*}

\begin{figure*}
    \centering
    \includegraphics[width=\textwidth]{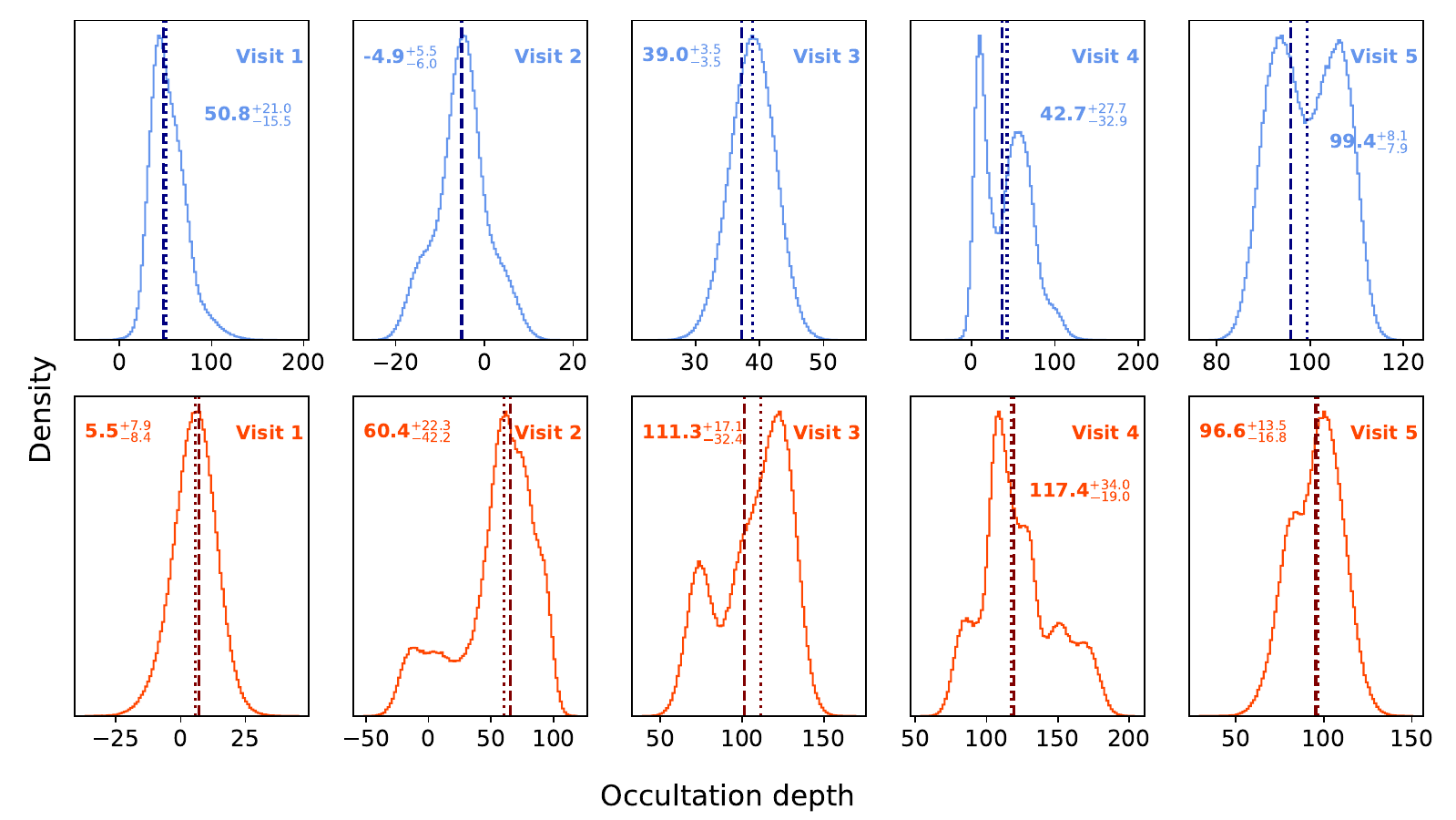}
    \caption{Posteriors of occultation depths from injection-retrieval exercise (see, text) for 2.1\,$\mu$m (SW) channel (the top row, in blue) and 4.5\,$\mu$m (LW) channel (the bottom row, in orange). The dashed and dotted vertical lines are injected and retrieved -- a median of the posteriors -- values of occultation depths, respectively. The median and 68-percentile confidence intervals of the posterior are written on the top of the plots.}
    \label{fig:post-inj-ret}
\end{figure*}

\subsubsection{Long-wave data analysis}\label{app:stark_long}

We downloaded uncalibrated data files (\texttt{uncal} files) from the MAST archive and used the official \texttt{jwst} pipeline to produce calibrated from them. We ran Stage 1 of the \texttt{jwst} pipeline with some modifications to the \texttt{uncal} files. The main change in Stage 1 is that we skipped the \texttt{dark current} step and \texttt{jump} step. This is justified because the dark current level in NIRCam detectors is low. Furthermore, since our observations were carried out using only two groups per integration, the \texttt{jump} step would become obsolete. Once we have \texttt{rateints} data from Stage 1 processing, we replace all \texttt{NaN} values in data and error arrays with average values of their neighbouring pixels. We add these pixels to the default bad-pixel map generated by the \texttt{jwst} pipeline. We performed a column-by-column and row-by-row background subtraction to reduce $1/f$ noise from the data. In this process, we subtracted a median of background pixels from each row while we fitted a line to the column background pixels and subtracted the estimated background from each column pixel. We then searched for cosmic ray events in the data file by comparing each frame with a median frame. We replaced all detected events with the mean of neighbouring pixels. However, we added these events to the bad-pixel map in the end. We did not run Stage 2 of the \texttt{jwst} pipeline because it does not change the science images.

Once we have corrected timeseries data, we used an open-source package \texttt{stark}\footnote{\url{https://stark-package.readthedocs.io/en/latest/}} to extract spectra. \texttt{stark} fits one and two-dimensional splines to the spectral data to find a robust estimate of PSF which can later be used to extract the spectrum. Before spectral extraction, we computed the location of the spectral trace using the centre-of-flux method. We found that the location of the trace on the detector remains extremely stable and varies only within 0.03 pixels. To estimate the stellar spectrum, we first need to compute the stellar PSF, which we did by fitting splines to the data. As a first approximation, we assume that the PSF does not change with wavelength and with time, so we fitted a 1D spline to the data as a function of distance from trace (known as pixel coordinates). This is a poor assumption because while the PSF stays constant in time, it varies significantly with wavelength. We improved our PSF estimate by fitting a 2D spline to the data as a function of pixel coordinates and wavelength. This robust PSF is then used to find stellar timeseries spectra. We used aperture half-widths of 9 and 2 pixels to fit PSF and extract spectra, respectively. We ran this procedure iteratively. At the end of each iteration, we subtracted the median static residual noise from the raw data. The median static noise is defined as a median difference between data and synthetic images constructed using stellar PSF and spectra. Only two iterations were sufficient to find robust stellar spectra. We compute the white-light light curve by taking a weighted average of light curves in all spectroscopic channels between 3.8612 and 4.9771\,$\mu$m. The raw white-light light curves for all visits are shown in Figure \ref{fig:raw_lc}.

Now that we have generated light curves we can fit an occultation model to the data. The light curves show a strong ramp in the beginning of each visit (see, Figure \ref{fig:raw_lc}), so we discarded the first 35\,min of the data before the analysis. In the light curve analysis, we fixed all planetary parameters except occultation depth to their values from the literature \citep{2018A&A...619A...1B, 2022A&A...663A..95M}. We used a wide uniform prior between -500 to 500\,ppm to the occultation depth parameter. We analysed white-light light curves from all five visits together. We used \texttt{juliet} \citep{2019MNRAS.490.2262E} to fit an occultation model to the data, which uses an occultation model from \texttt{batman} \citep{2015PASP..127.1161K} and samples posteriors using \texttt{dynesty} \citep{2020MNRAS.493.3132S}. In addition to the planetary model, we added linear and quadratic polynomials in time to correct for long-term trends seen in the light curve. The best-fitted values of white-light occultation depths are tabulated in Table~\ref{tab:ecl_dep_comp}. We could not, however, model hour-long correlated noise (see, e.g., Fig.~\ref{fig:detrended-lcs}), with this simple polynomial model. This is also evident from the Allan deviation plots, shown in Fig.~\ref{fig:allan_deviation}, of residuals that show additional noise at larger bin sizes. The presence of uncorrected correlated noise means that the uncertainties found on the occultation depths are underestimated. We could not determine the origin of this noise: we searched engineering data but could not find any parameter that correlates with the noise, pointing towards a possible astrophysical origin. However, recent transit observations of a bright star \citep[GJ\,341, K = 5.6\,mag,][]{2024AJ....167...90K} with the same observing mode also show a similar noise as our dataset (see, their Fig.~2). So, the correlated noise could be a previously unknown systematics of the instrument. We looked at the 2D spectral data at the group level to further test this possibility. Generally, the data from the first and last groups are discarded as they could be unreliable. We cannot do this since our dataset has only two groups. We took the 2D spectral data for both groups independently and extracted spectral timeseries from them in exactly the same manner described earlier. We finally computed and analysed white-light light curves from both groups. We found that the correlated noise similar to the integration level light curve is also present at ``group level'' white-light light curves. This suggests that the correlated noise does not originate from unreliable first and last groups (see also our companion paper for more details, Patel \& Brandeker, \textit{in prep}).

We perform injection-retrieval tests on the white-light light curves to estimate proper uncertainties on the occultation depths in the presence of correlated noise. We first subtract the normalised planetary signal from the raw white-light light curve keeping the long-term trend and the correlated noise as it is in the data. We next produced 1000 realisations of light curves by injecting an occultation signal at random times in the data. The depth of the signal is equal to the median value from the full light curve analysis presented earlier. In this process, we made sure that the full signal remained inside the data. We fit a full model, consisting of an occultation model and polynomial -- linear and quadratic -- trend, using \texttt{juliet} to each of the realisations. We build a posterior of occultation depth using randomly selected samples from the posteriors of occultation depth in each realisation. These posteriors, shown in Fig.~\ref{fig:post-inj-ret} for all visits, are clearly not Gaussian for most of the cases illustrating the effect of correlated noise. A 68-percentile confidence interval of this posterior should be more representative of uncertainties on white-light occultation depths. In the cases where the uncertainties obtained this way were smaller than the ``white'' uncertainties from the light curve analysis, we choose to report the larger value.

The correlated noise is also present in the spectroscopic light curves of each column. We first boosted the estimated errors of the spectroscopic light curves and the white-light light curve according to the scatter in the light curves. Then we divided spectroscopic light curves from each column with the white-light light curve to remove the correlated noise from the spectroscopic data. This mostly removed correlated noise from the spectroscopic light curves. Finally, we computed relative occultation depths as $1 - (F_{\mathrm{in}} / F_{\mathrm{out}})$, where $F_{\mathrm{in}}$ and $F_{\mathrm{out}}$ are the flux inside and outside of the occultation duration, respectively. Before computing this, we made sure that the baseline before and after the occultation signal was the same. Note that we compute relative occultation depths at the native resolution of the instrument before binning them to a lower resolution. This method minimizes the impact of any leftover 1/f noise in the data \citep[see, e.g.,][]{2023PASP..135a8002E}.

\subsubsection{Short-wave data analysis}

The Stage 1 processing of 2.1\,$\mu$m channel \texttt{uncal} files was mostly done in the same way as for the 4.5\,$\mu$m channel \texttt{uncal} files described above. The main difference is that here we only perform a row-by-row background subtraction. The short-wave PSF spreads to almost all pixel ranges along the column so that there are very few background pixels along the column making it impossible to perform background subtraction along columns. 

Once we got \texttt{rateints} data, we performed simple aperture photometry to 2.1\,$\mu$m channel data to obtain a photometric light curve. Before doing this, we computed the centroids of the PSF using the centre-of-flux method. We then computed a growth function -- flux inside an aperture as a function of increasing aperture radius -- to optimally select an aperture radius. We find that the growth function flattens out at around 45 pixel radius that we eventually used in our analysis. We adapted the \texttt{photutils}\footnote{\url{https://photutils.readthedocs.io/en/stable/index.html}} \citep{larry_bradley_2023_7946442} package to compute aperture photometry. \texttt{photutils} simply calculates the total flux inside the aperture. Since we already did a row-by-row background subtraction we did not perform another sky annulus subtraction. Uncorrected short-wave photometric light curves are plotted in Figures \ref{fig:raw_lc}.

We fitted an occultation model to thus-obtained SW light curves in almost the same manner as for the occultation model fitting of LW white-light light curves. The instrumental model used here was different from what was used in the LW case. Here we used a linear polynomial in time and PSF centroids as decorrelation vectors. Additionally, light curves from two of our visits (Visits 1 and 4) show abrupt flux jumps analogous to what was found in \citet{2023PASP..135a8001S} (see, Figure \ref{fig:raw_lc}). These flux jumps may or may not be caused by mirror tilting events as described in \citet{2023PASP..135a8001S} --- a thorough investigation of the origin of these jumps is ongoing (see also our companion work Patel \& Brandeker, in prep.). Here we model these flux jumps using multiple step functions; since the jumps are abrupt and affect only a few integrations, it is fairly easy to set the boundaries of step functions. For certainty, we masked all integrations near jumps, which is safe because the masked integrations consist of only a few per cent of the total number of data points and none of these are near the ingress or egress. Another source of noise in the SW light curves is the high-frequency periodic noise possibly caused by the thermal cycling of heaters in the Integrated Science Instrument Module on JWST \citep[see,][]{2023PASP..135a8002E}. This is clearly visible in the power spectrum of the light curve as a peak period near 3.8\,min in all visits. We performed a principal component analysis (PCA) of the PSF time series to see if we could capture this noise as a principal component (PC) or not. Indeed, one of the first PCs in all visits show a periodic pattern with a period of about 3.8\,min. While we are uncertain about the origin of this noise, we simply use this PC as a decorrelation vector in our light curve analysis. 

In summary, our total model fitted to the SW light curve includes an occultation model, linear models in time, PSF centroids and a PC. Step functions were also included as decorrelation vectors in Visits 1 and 4. We used \texttt{juliet} to fit the light curve data. The best-fitted occultation depths can be found in Table~\ref{tab:ecl_dep_comp}. These data are also affected by a correlated noise that we could not model using our simple model. This is also evident from the Allan deviation of the residuals shown in Figure \ref{fig:allan_deviation}. We performed injection-retrieval tests similar to the LW data analysis described in Appendix~\ref{app:stark_long} to properly estimate the uncertainties on the occultation depths.

\subsection{\textit{Eureka!} --- Reduction 1}

Here we provide an independent reduction of the short-wave (SW) observations of NIRCam. To reduce the \texttt{nrca1 uncal} files we used \texttt{Eureka!} \citep[version 0.11.dev276+g4e12d23d,][]{2022JOSS....7.4503B} pipeline. Stage 1 consists of running default \texttt{jwst} detector processing steps, but we skip the saturation step. On stage 2 we only correct for the flat field. On Stage 3, we crop the full array to a window between pixels 1400 and 2000 in the $x$-axis and between pixels 1 and 64 in the $y$-axis. We also mask pixels flagged as bad quality and reject outliers above 7$\sigma$ along time axis. We interpolate bad pixels with a linear function and perform row-by-row background subtraction and $1/f$ noise correction. Aperture photometry is performed using a circular 40 pixel radius aperture. We subtract the background region with an annulus with an inner edge of 45 pixels and an outer edge of 60 pixels. Finally, Stage 4 uses the calibrated files to produce the light-curve. Visit 1 and 4 exhibit strong discontinuities, dividing the light-curve into five and six clearly defined segments, respectively. To correct the discontinuities, first, we mask the occultation. To flatten the light-curve, we fit a linear function to each segment and then fit an occultation model with \texttt{exoplanet} in a Hamiltonian Monte Carlo algorithm with \texttt{PyMC3}. The rest of the visits did not exhibit such discontinuities and thus we fit only one linear function in time. The resulting occultation depths are shown in Table~\ref{tab:ecl_dep_comp}. Compared to the \texttt{stark} reduction and analysis, all occultation depths are consistent within 1$\sigma$.

\subsection{\texttt{Eureka!} --- Reduction 2}
We produced an independent reduction of the NIRCam spectra using the \texttt{jwst} \citep[version 1.12.5,][]{2023zndo...6984365B} and \texttt{Eureka!} \citep[version 0.9,][]{2022JOSS....7.4503B} pipelines, including purpose-built steps that we describe here. Starting from the uncalibrated raw data, we ran the default \texttt{jwst} detector processing steps up to (and including) the dark current step. Prior to the ramp fitting step, we subtracted from each row the median of the left-most 650 pixels in the corresponding row and group. By using these unilluminated pixels as a reference of the level of noise added during readout, this helps reduce 1/f noise. We then applied the remaining \texttt{jwst} calibration steps.

We ran the resulting calibrated files through \texttt{Eureka!}. We extracted columns 850 through 1945 and discarded the reference pixels. To straighten the trace, we vertically slid each detector column by an integer number of pixels. We performed background subtraction using the average value of each column, rejecting $7\sigma$ outliers and excluding a window with a half-width of 15 pixels centred on the trace. Constructing the spatial profile from the median frame, we performed optimal extraction on a region centred on the source and with a half-width of 5 pixels. We generated 30 spectroscopic light curves between 3.9365 and 4.9265\,$\mu$m, each spanning 0.033\,$\mu$m. In each light curve, we discarded values farther than 4$\sigma$ from the mean of a sliding window.

The flux in the light curves follows a downward trend with time, and they show significant time-correlated noise. After trimming the initial 20\,min of data, where the ramp is the steepest, we modelled the white light curve in each visit as the product of an exponential ramp, a linear polynomial and a \texttt{batman} occultation model, where the occultation depth acted as a free parameter. The fits included an estimated error multiplier to match the scatter in the residuals. We assumed a circular orbit, and fixed the orbital period and mid-transit time to the values in \citet{2021AJ....161..181Z}, and planet radius, orbital inclination and scaled semi-major axis to those reported by \citet{2018A&A...619A...1B}. For each visit, we also calculated the relative occultation depths following the methodology outlined in Appendix~\ref{app:stark_long}. 

\subsection{\texttt{HANSOLO}}\label{app:hansolo}
The \hans (atmospHeric trANsmission SpectrOscopy anaLysis cOde) pipeline was originally developed to analyse ground-based transmission spectra observed with 8m-class telescopes, but has been adapted to also enable its use on NIRCam data \citep{2016A&A...587A..67L,2017A&A...606A..18L,2023Natur.614..653A}. 
\hans takes calibrated \texttt{rateints} outputs of the \texttt{jwst} pipeline Stage 1 as input. 

We used the LACOSMIC algorithm \citep{2001PASP..113.1420V} to remove cosmic ray effects from the two-dimensional images and identified the spectral trace by using a Moffat function fit to each
column. The sky background was calculated on a column-by-column basis by calculating a linear trend in the column background, which was defined as at least 20 pixels away from the centre of the spectral trace. This linear trend was then subtracted from the whole column. We extracted the spectrum by summing over an aperture with a half-width of 4 pixels. 

Consistent with the other reductions, we generated a white light curve and 30 spectroscopic light curves from which we clipped the first 35\,min to remove the worst of the ramp that is present in all the data. For each light curve we applied a $5\sigma$ outlier rejection filter. We used the light curve and RV fitting code CONAN to fit the white light curves with an occultation model and a GP with a 3/2 Matern kernel to account for both the remaining ramp and the correlated red noise. We leave the occultation depth and the GP parameters (amplitude, lengthscale and a white noise factor) as free parameters and fix all orbital parameters to the literature values found by \citet{2018A&A...619A...1B}. The white light occultation depths are presented in Table~\ref{tab:ecl_dep_comp}. We then calculate the common mode for each visit by removing the fitted occultation from the white light curve and divide the common mode out of the spectroscopic light curves. Since the spectroscopic light curves still show some correlated noise even with the common mode removed, we then fit each spectroscopic light curve individually in the same way as the white light curves, with the orbital parameters held fixed and the occultation depth and GP parameters as free parameters. The resulting emission spectra are shown in Figure \ref{fig:spec_comp}. 

\subsection{\texttt{transitspectroscopy}}
We take the corrected timeseries data from \texttt{stark} long-wave analysis and use an open-source tool \texttt{transitspectroscopy} \citep{espinoza_nestor_2022_6960924}\footnote{\url{https://github.com/nespinoza/transitspectroscopy}}. We first use a centre of flux method to find the location of trace on the detector. We used the optimal extraction algorithm from \citet{1989PASP..101.1032M} to extract 1D stellar spectra from the timeseries data. In this procedure, we used an aperture half-width of 3 pixels. The optimal extraction naturally clips all outliers not identified by the pipeline. We masked all such 10$\sigma$ outliers. White-light light curves for each visit were computed by taking a weighted average of spectroscopic light curves between 3.8612 and 4.9771\,$\mu$m.

We used \texttt{juliet} to fit the occultation model to the white-light light curve data. In addition to the occultation model \citep[from \texttt{batman},][]{2015PASP..127.1161K}, our full model includes linear, quadratic and cubic polynomials to model a long-term decreasing trend. We also added white noise to the errors on the flux. We fixed all planetary parameters except occultation depth from the literature \citep{2018A&A...619A...1B, 2022A&A...663A..95M}. The median and 68-percentile confidence intervals for the best-fitted occultation depths are tabulated in Table~\ref{tab:ecl_dep_comp}. We also determined relative occultation depth spectra using the procedure described in Appendix~\ref{app:stark_long} and plotted in Fig.~\ref{fig:spec_comp}.

\subsection{\texttt{SPARTA}}
Our SPARTA reduction is very similar to that used in \citet{2024arXiv240504744H}, which analyzed the one occultation observed by GO~1952 (PI~Hu).  The steps that we used to go from the uncalibrated files to the spectroscopic light curves are identical.  In stage 1, we perform superbias subtraction, reference pixel subtraction, non-linearity correction, dark subtraction, and up-the-ramp fitting (which amounted to subtracting the two reads since we only have two).  In stage 2, we remove the background, which also removes some of the 1/f noise because we perform row-by-row subtraction in addition to column-by-column subtraction.  In stage 3, we perform sum extraction with a window half-width of 2 pixels, obtaining spectroscopic light curves.

Using \texttt{emcee}, we fit the white light curve with a model that has the occultation time and occultation depth as astrophysical free parameters, while the light curve normalization factor, exponential ramp amplitude and timescale, x and y linear correlation parameters, linear slope with time, and error inflation multiple are free systematics parameters.  We save the systematics model corresponding to the best fit to the white light curve.  To fit the spectroscopic light curves, we first divide each light curve by the aforementioned systematics model, and then fit the result with a model that includes every parameter in the white light curve fit except the occultation time (which we fix to the white light value).

\begin{table*}[h!] \label{Table: spectroscopic results}
\centering
 \caption{Spectroscopic  parameters for 55\,Cnc.}   
\begin{tabular}{llcccccc }
 \hline
     \noalign{\smallskip}
Method  & $T_\mathrm{eff}$  & $\log g_\star$ & [Fe/H]   & [Ca/H] &  [Mg/H]& [Na/H]&  $V \sin i$    \\  
& (K)  &(dex) &(cgs)  &(cgs)&(cgs)  &(cgs)   & (km~s$^{-1}$)   \\
    \noalign{\smallskip}
     \hline
\noalign{\smallskip} 
SME   &  $5234\pm 55$  & $4.33 \pm 0.05$ & $+0.31\pm0.05$   & $+0.33\pm0.05$ & $+0.44\pm0.12$ &$+0.60\pm0.11$ & $2.0 \pm 0.7$  \\
astroARIADNE\tablefoottext{a}   &  $5269 \pm 46$  & $4.34\pm 0.07$ & $+0.34\pm0.07$     &\ldots  &\ldots &\ldots &\ldots \\ 
\hline 
\end{tabular}  \label{Table: stellar spectroscopic parameters}
\tablefoot{
\tablefoottext{a}{Posteriors from the SED modelling.}
}
\end{table*}

\section{Properties of the star}

\subsection{Observed stellar spectrum}

\begin{figure}
    \centering
    \includegraphics[width=\columnwidth]{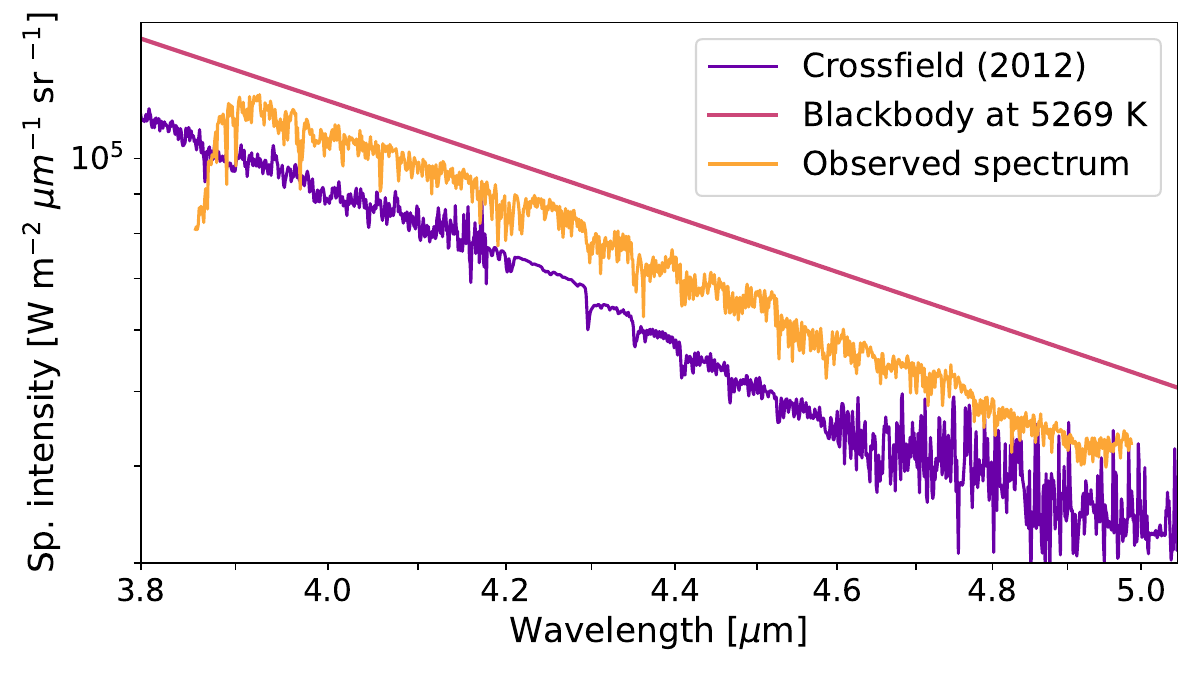}
    \caption{The observed stellar spectrum with NIRCam/JWST (in yellow) is shown with \citet{2012A&A...545A..97C} empirical spectrum and a blackbody at 5269\,K.}
    \label{fig:stelspec}
\end{figure}

We produced \texttt{rateints} files from uncalibrated data using the \texttt{jwst} pipeline using the same procedure as described in Appendix~\ref{app:stark_long}. We then ran Stage 2 of the \texttt{jwst} pipeline with some modifications, namely skipping the \texttt{flat fielding} and \texttt{extract1d} steps, to produce calibrated spectrum files. This was followed by correcting data and error files for \texttt{NaN} and cosmic rays as described in Appendix~\ref{app:stark_long}. Despite being classified as a point source by the \texttt{jwst} pipeline, the physical unit of calibrated 2D spectrum data is given as MJy/sr. We converted the units to Jy using the pixel area quoted in a header file of \texttt{calints} data products from Stage 2 of the \texttt{jwst} pipeline. We finally extracted the spectrum using \texttt{stark} as described in Appendix~\ref{app:stark_long}. We extracted a timeseries of spectra from part of the data from our most recent visit, Visit 5. A median spectrum of these timeseries spectra is plotted in Figure \ref{fig:stelspec} and compared with the \citet{2012A&A...545A..97C} empirical spectrum and black body spectrum. We found that similar to \citet{2024arXiv240504744H}, the NIRCam observed spectrum is discrepant with the \citet{2012A&A...545A..97C} empirical spectrum. We think that this may be because of improper photometric correction for bright stars provided by the \texttt{jwst} pipeline. Furthermore, \citet{2024arXiv240504744H} found that their MIRI observed spectrum agrees very well with \citet{2012A&A...545A..97C} spectrum. Here, we use the \citet{2012A&A...545A..97C} spectrum in our atmospheric retrieval analysis.

\FloatBarrier

\subsection{Stellar parameters from modelling}

We modelled 85 publically available spectra from the High 
Accuracy Radial velocity Planet Searcher \citep[HARPS;][]{2003Msngr.114...20M} spectrograph  with a resolution of 115\,000. The spectra were co-added and  modelled with Spectroscopy Made Easy\footnote{\url{http://www.stsci.edu/~valenti/sme.html}} \citep[SME;][]{vp96, pv2017} version 5.2.2 and the stellar atmosphere grid   Atlas12 \citep{Kurucz2013}. SME computes synthetic spectra and adjusts the chosen free parameters based on comparison with the observed spectrum. We modelled one parameter at a time, utilising spectral features sensitive to different photospheric parameters and iterating until all parameters converged. Throughout the modelling, we held the macro- and micro-turbulent velocities, $V_{\rm mac}$ and $V_{\rm mic}$,  fixed at 2.7\,km\,s$^{-1}$ \citep{Doyle2014} and 0.95\,km\,s$^{-1}$ \citep{bruntt08}. A description of the modeling procedure is detailed in    \citet{2018A&A...618A..33P}. The results are listed in Table~\ref{Table: spectroscopic results}.
 
The stellar radius was modelled with the SED fitting software 
astroARIADNE\footnote{\url{https://github.com/jvines/astroARIADNE}}
\citep[][]{2022arXiv220403769V} using priors from SME and   photometry from  the Johnson $B$ and $V$ magnitudes (APASS), 
$G G_{\rm BP} G_{\rm RP}$   (DR3),        
$JHK_S$ magnitudes ({\it 2MASS}), {\it WISE} W1-W2, and 
the \textit{Gaia} DR3 parallax. 
We utilized three different atmospheric model grids from {\tt {Phoenix~v2}} \citep{2013A&A...553A...6H}, \citet{Castelli2004},
and \citet{1993yCat.6039....0K}.  The final radius  was computed 
with Bayesian Model Averaging and was   
found to be $0.953\pm 0.011$~$R_{\odot}$. The luminosity is $0.63\pm 0.02$~$L_{\odot}$, and the visual extinction is consistent with zero 
($0.03\pm 0.03$). 
We derived a stellar mass  of $0.639^{+0.021}_{-0.020}$~$M_{\odot}$  interpolating   the MIST 
\citep{2016ApJ...823..102C} isochrones with astroARIADNE. 
Our results are very close to previous results;  \citet{2011ApJ...740...49V} derive a stellar radius of $0.943\pm 0.010$~$R_{\odot}$ 
based on interferometric measurements and the parallax from \citet{2007A&A...474..653V}. Updating this calculation with the \textit{Gaia}~DR3
parallax, this radius becomes $0.962\pm 0.010$~$R_{\odot}$ in good agreement with our results. 

\FloatBarrier

\section{Detailed retrieval posterior distributions}
\label{sec:retrieval_posteriors_appendix}

In this appendix we present all posterior distributions from our retrieval calculations for the CO/\ch{CO2} and SiO/\ch{SiO2}/MgO cases. The posterior distributions are ordered in chronological order and shown for the \stark and \hans reductions. Due to the fact that for the \hans reduction, the retrievals are performed on absolute occultation depths, the posterior distributions do not include the white-light occultation depths parameter $d_\mathrm{wl}$. 

It is also important to note that the depicted centre-log-ratio posterior $\xi_j$ for the last molecule is not a free parameter in the retrieval as mentioned in Sect. \ref{subsec:retrieval_setup}. Instead, we calculated the corresponding posterior distribution following the requirement that for each posterior sample, the sum of all $\xi$ values must be zero.

For Visits 1 and 3, the posterior distributions are already shown in Figs.~\ref{fig:retrieval_posterior_visit1_red1} \&  \ref{fig:retrieval_posterior_visit3_red1} in the main text and are not repeated here. The corresponding posterior spectra for the posteriors are shown in Fig.~\ref{fig:retrieval_posterior_spectra}.

\begin{figure*}
    \centering
    \includegraphics[width=0.6\textwidth]{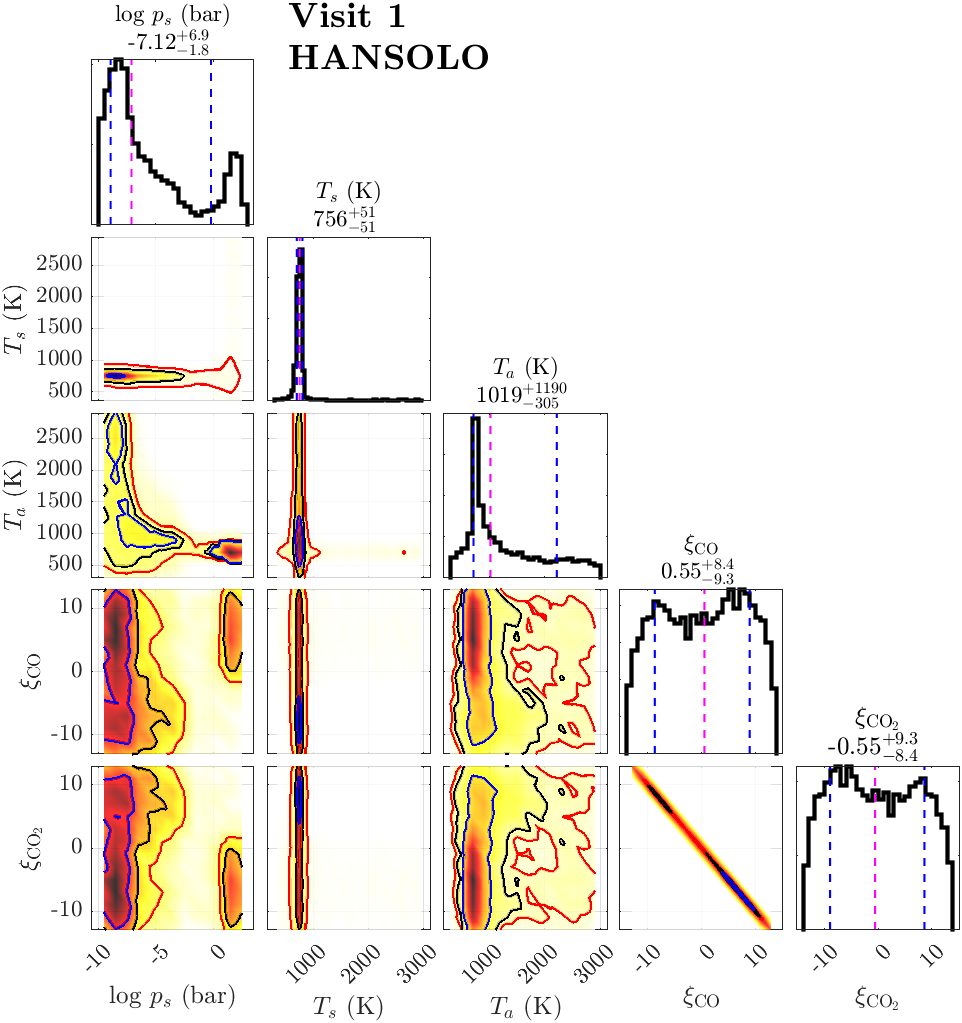}
    \caption{Posterior distributions of the free parameters for the first visit, representing the CO/\ch{CO2}-atmosphere scenario. Results are shown for the \hans reduction. The corresponding distributions for the \stark reduction are depicted in Fig.~\ref{fig:posterior_pressure_co}. We note that $\xi_\mathrm{CO_2}$ is not a free parameter in the retrieval but was calculated during a postprocess procedure following the requirement that in each posterior sample the sum of all $\xi$ values must be zero.}
    \label{fig:retrieval_posterior_co_visit1}
\end{figure*}

\begin{landscape}
\begin{figure}
    \begin{center}
    \includegraphics[width=0.6\textwidth]{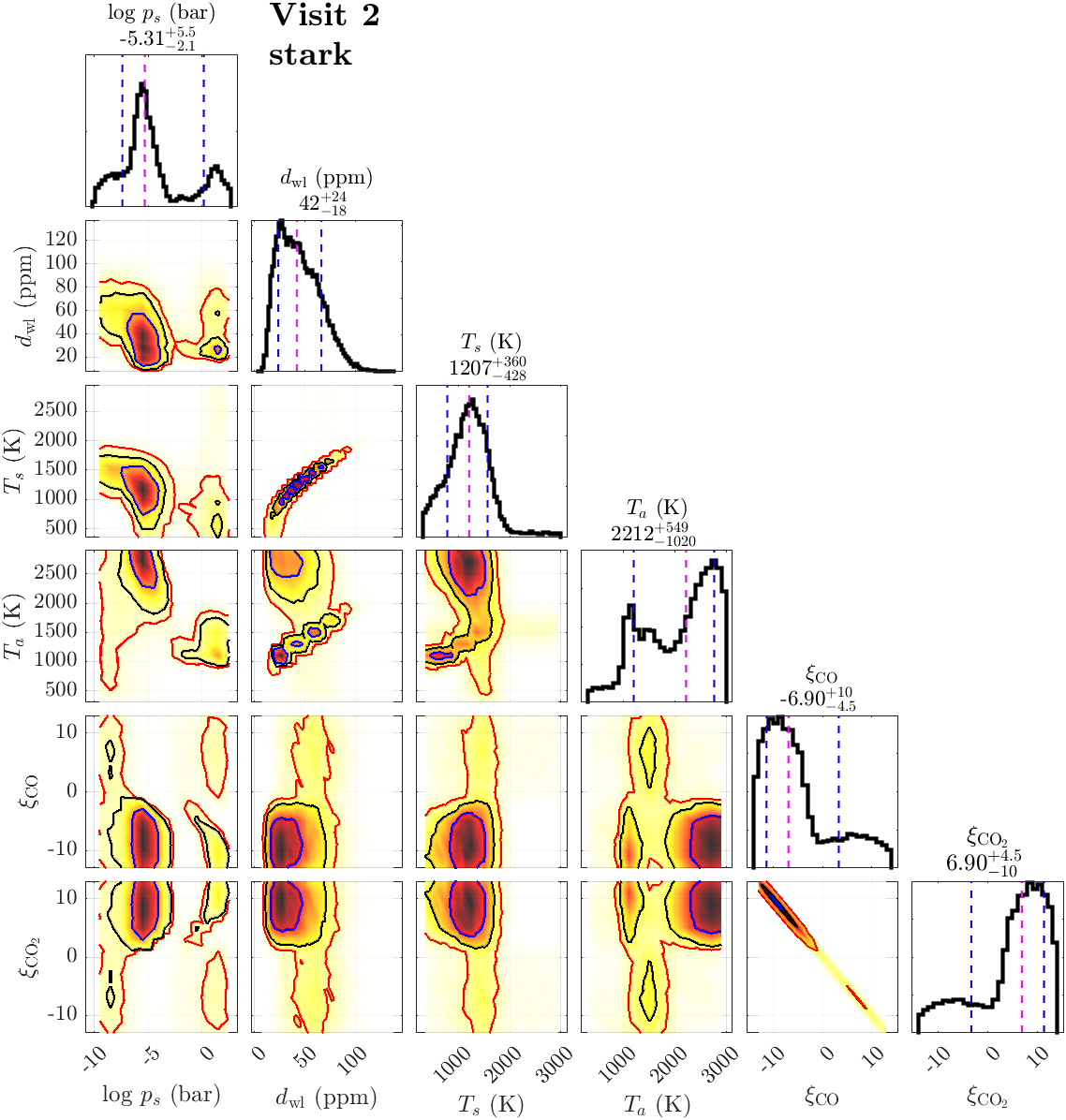} \includegraphics[width=0.5\textwidth]{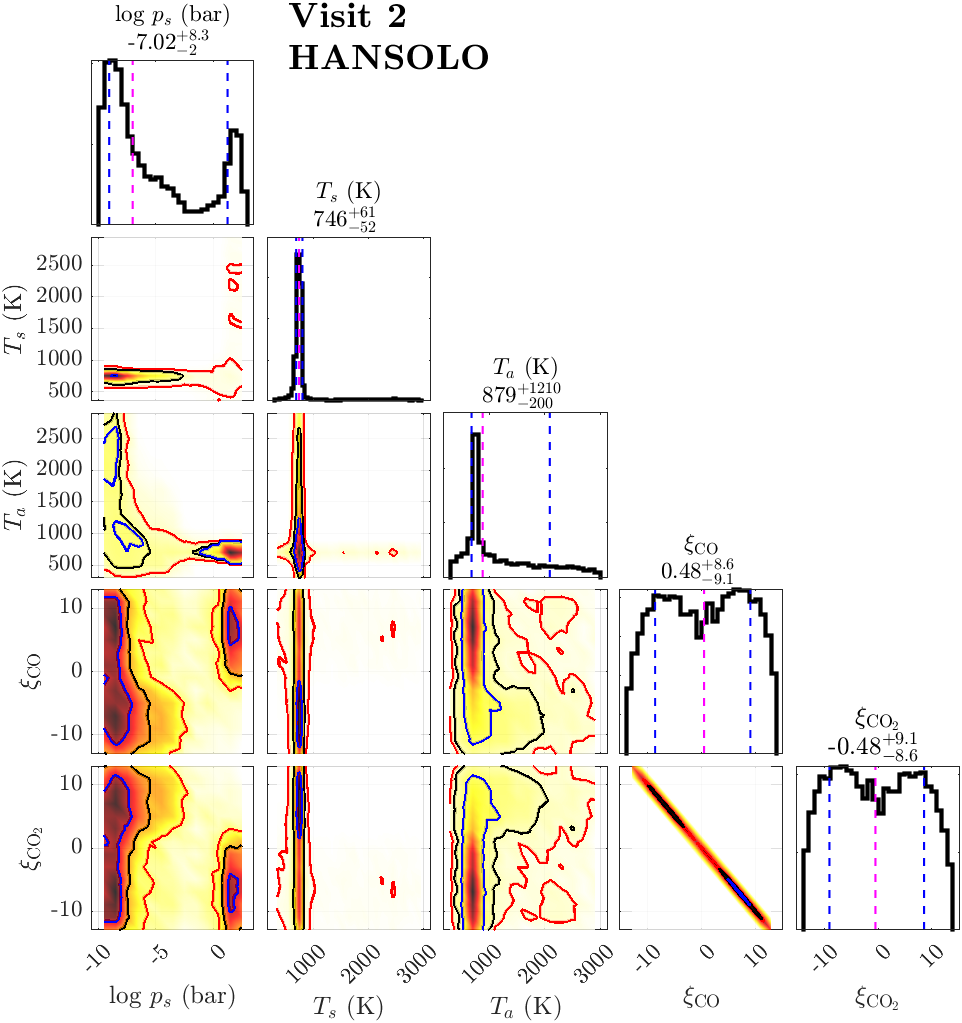}
    \caption{Posterior distributions of the free parameters for Visit 2, representing the CO/\ch{CO2}-atmosphere scenario. Results are shown for the \stark (left) and \hans (right) reductions.}
    \end{center}
    \label{fig:retrieval_posterior_co_visit2}
\end{figure}
\end{landscape}

\begin{landscape}
\begin{figure}
    \begin{center}
    \includegraphics[width=0.6\textwidth]{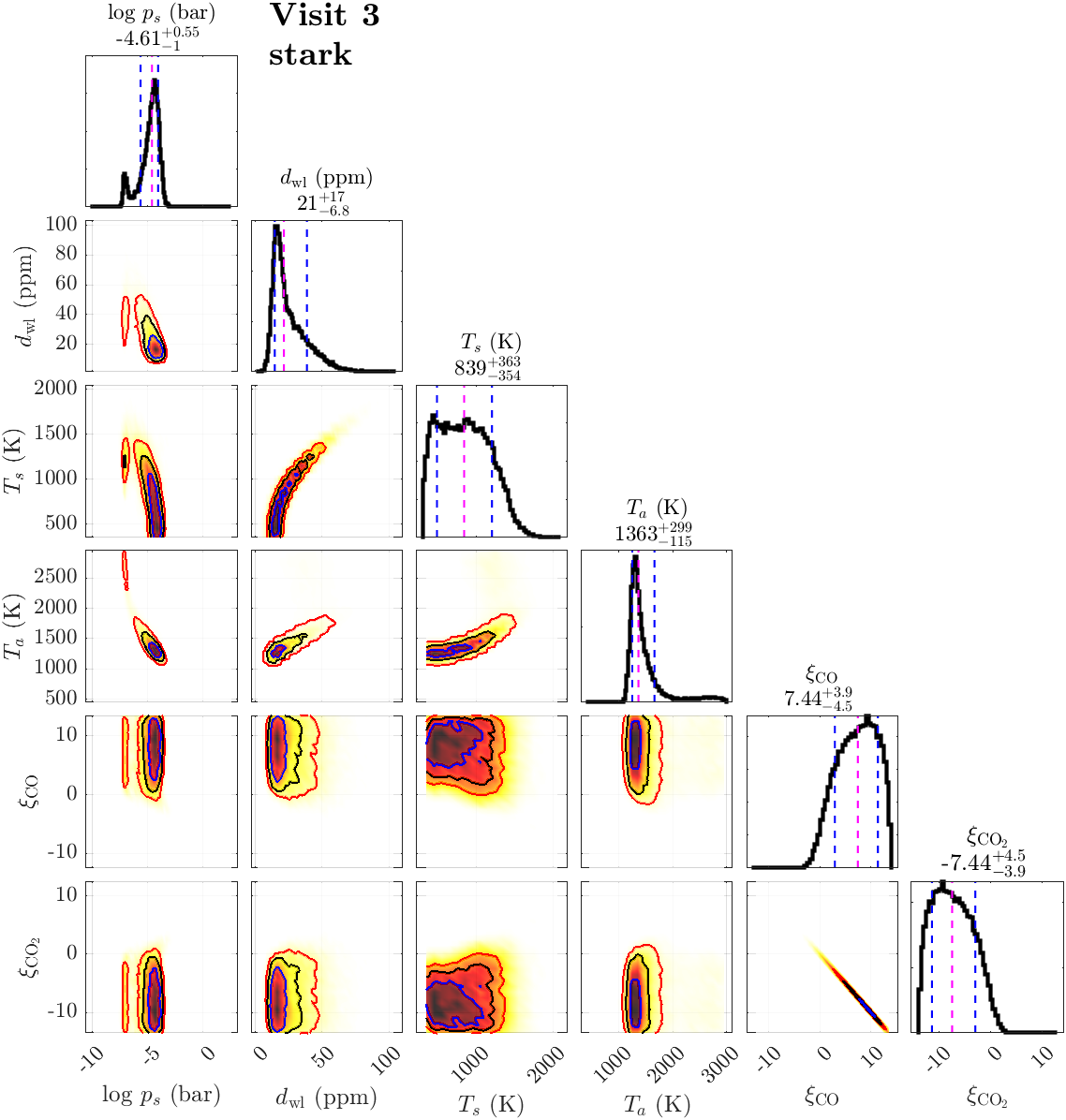} \includegraphics[width=0.5\textwidth]{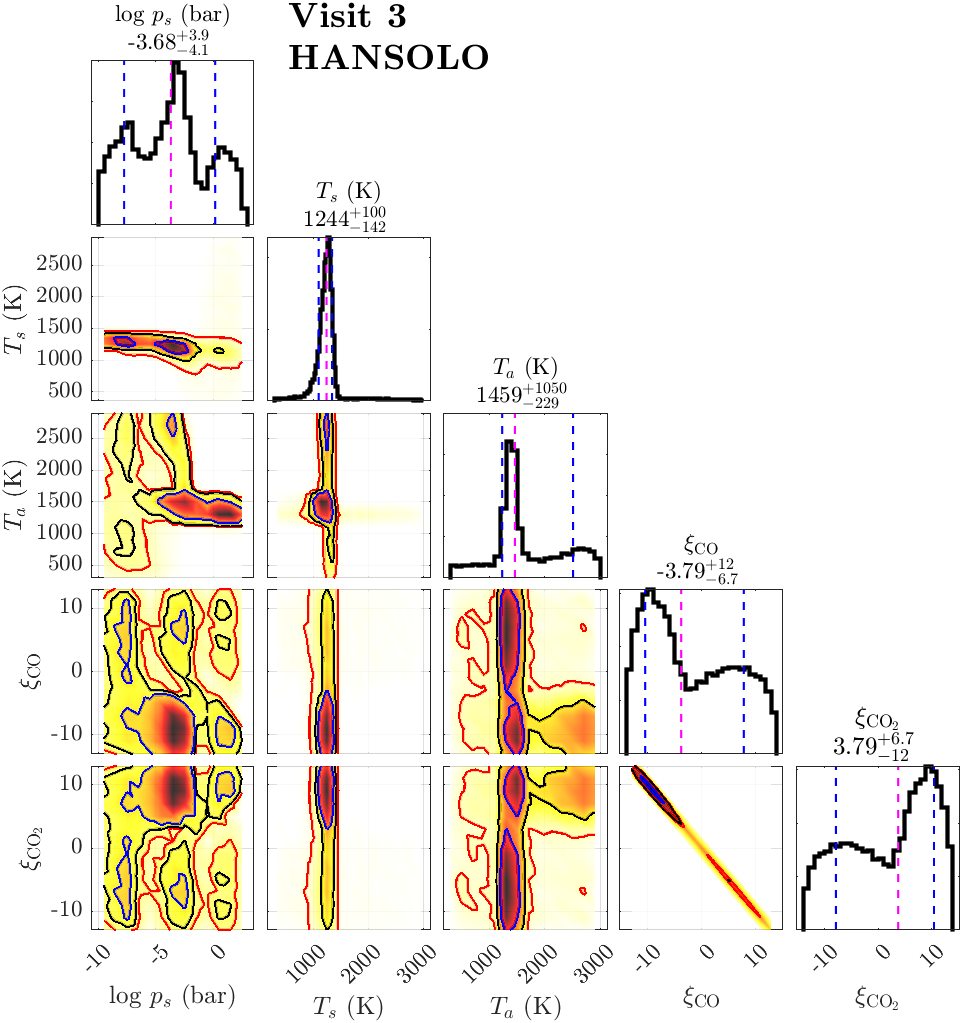}
    \caption{Posterior distributions of the free parameters for Visit 3, representing the CO/\ch{CO2}-atmosphere scenario. Results are shown for the \stark (left) and \hans (right) reductions.}
    \end{center}
    \label{fig:retrieval_posterior_co_visit3}
\end{figure}
\end{landscape}

\begin{landscape}
\begin{figure}
    \begin{center}
    \includegraphics[width=0.6\textwidth]{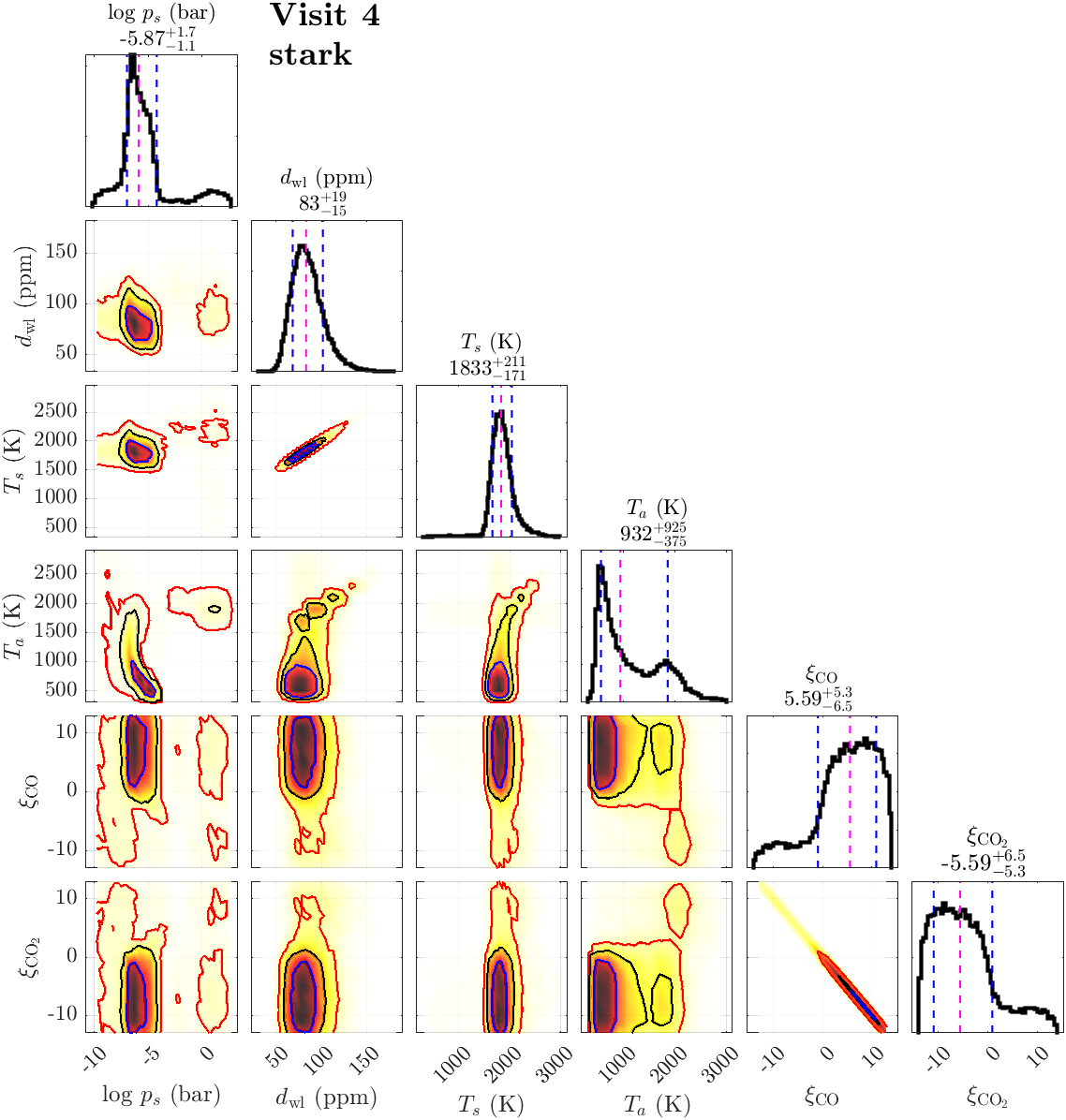} \includegraphics[width=0.5\textwidth]{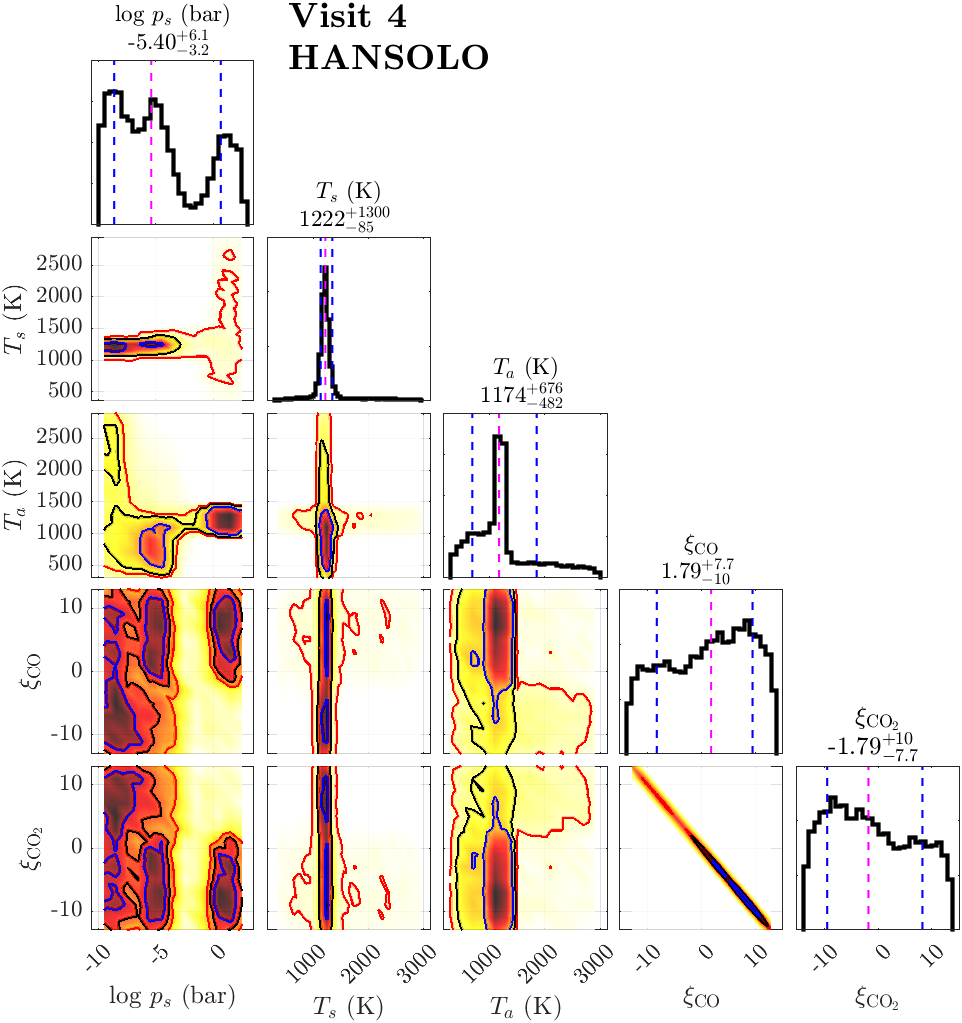}
    \caption{Posterior distributions of the free parameters for Visit 4, representing the CO/\ch{CO2}-atmosphere scenario. Results are shown for the \stark (left) and \hans (right) reductions.}
    \end{center}
    \label{fig:retrieval_posterior_co_visit4}
\end{figure}
\end{landscape}

\begin{landscape}
\begin{figure}
    \begin{center}
    \includegraphics[width=0.6\textwidth]{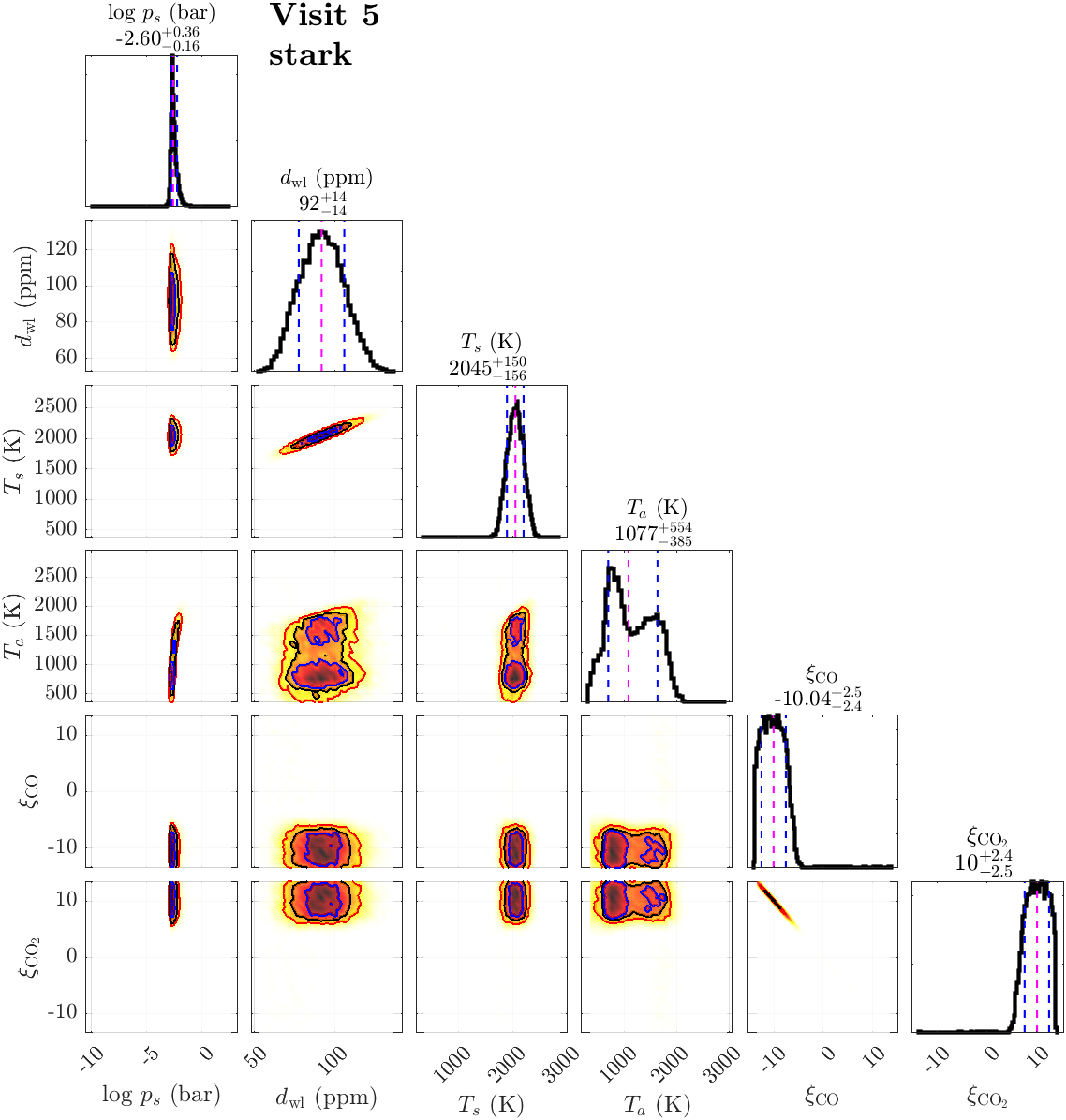} \includegraphics[width=0.5\textwidth]{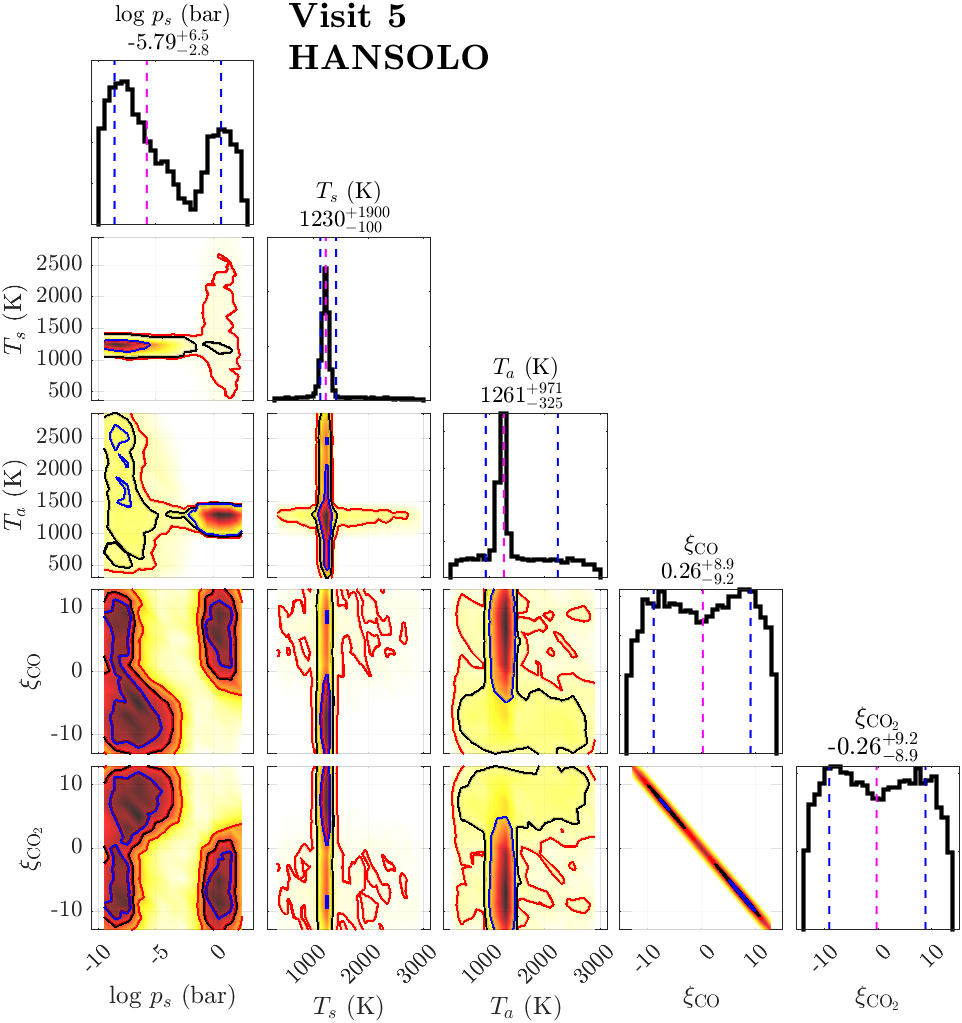}
    \caption{Posterior distributions of the free parameters for Visit 5, representing the CO/\ch{CO2}-atmosphere scenario. Results are shown for the \stark (left) and \hans (right) reductions.}
    \end{center}
    \label{fig:retrieval_posterior_co_visit5}
\end{figure}
\end{landscape}

\begin{landscape}
\begin{figure}
    \begin{center}
    \includegraphics[width=0.7\textwidth]{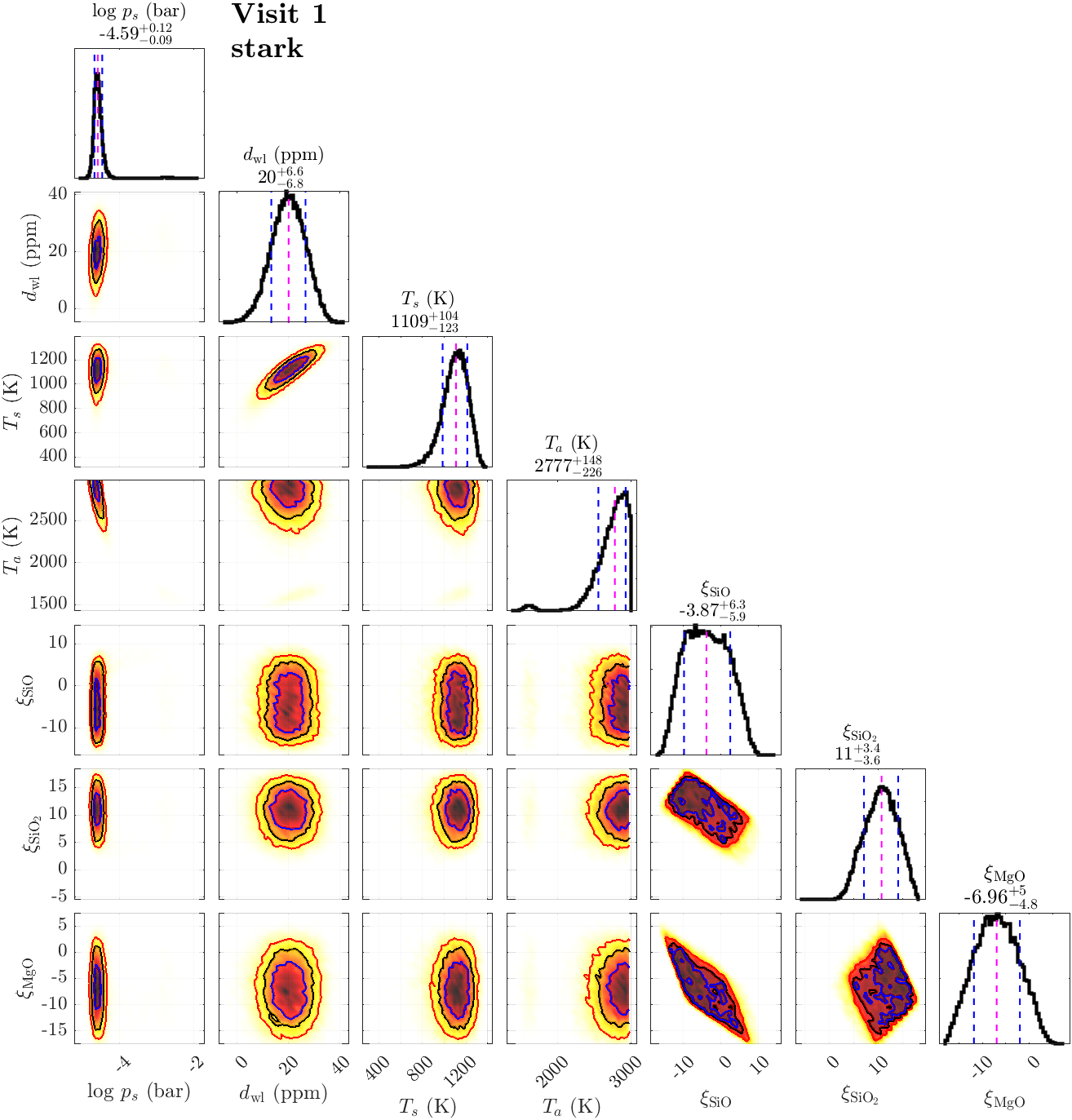} \includegraphics[width=0.6\textwidth]{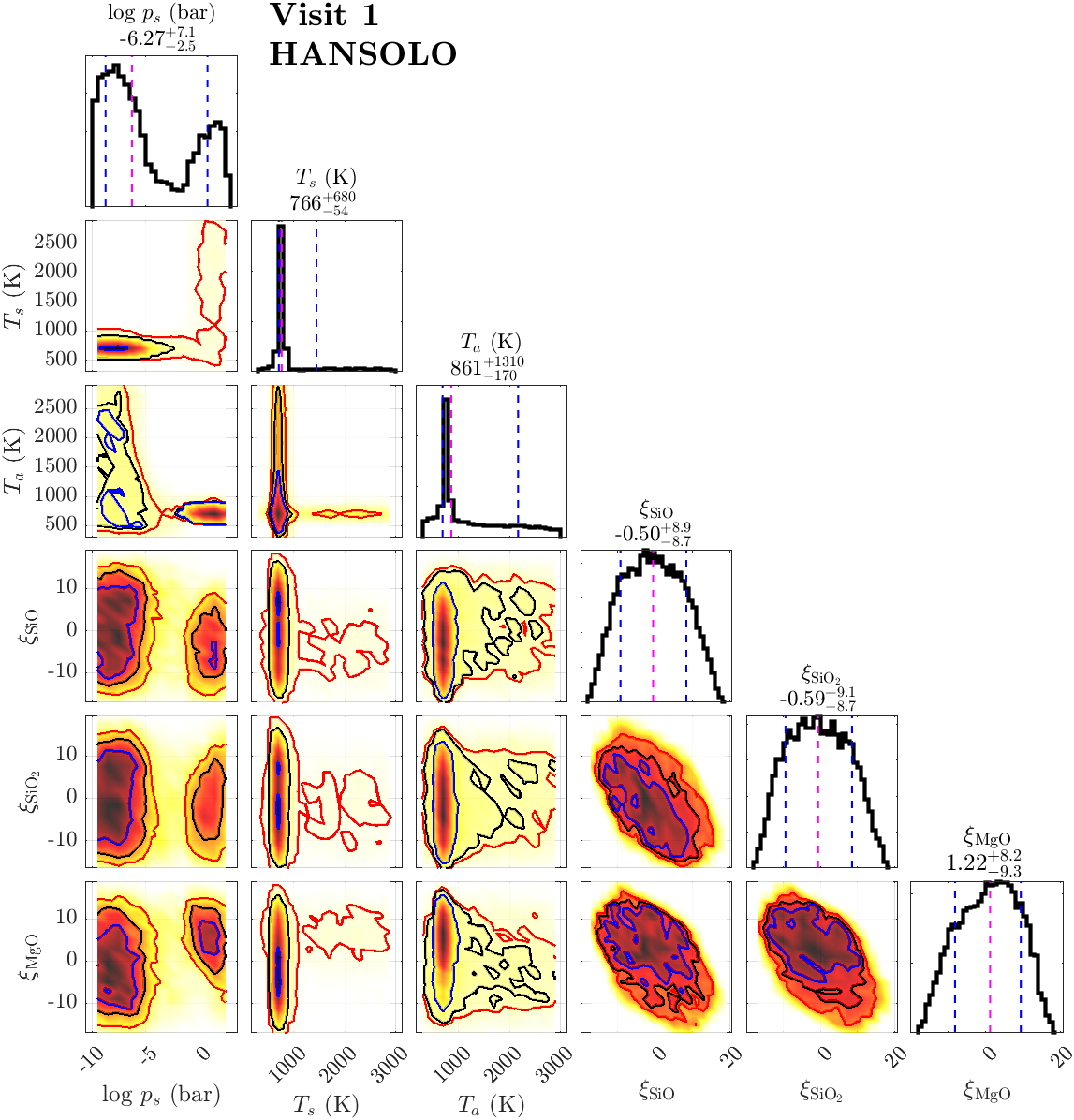}
    \caption{Posterior distributions of the free parameters for Visit 1, representing the SiO/\ch{SiO2}/MgO-atmosphere scenario. Results are shown for the \stark (left) and \hans (right) reductions.}
    \end{center}
    \label{fig:retrieval_posterior_sio_visit1}
\end{figure}
\end{landscape}

\begin{landscape}
\begin{figure}
    \begin{center}
    \includegraphics[width=0.7\textwidth]{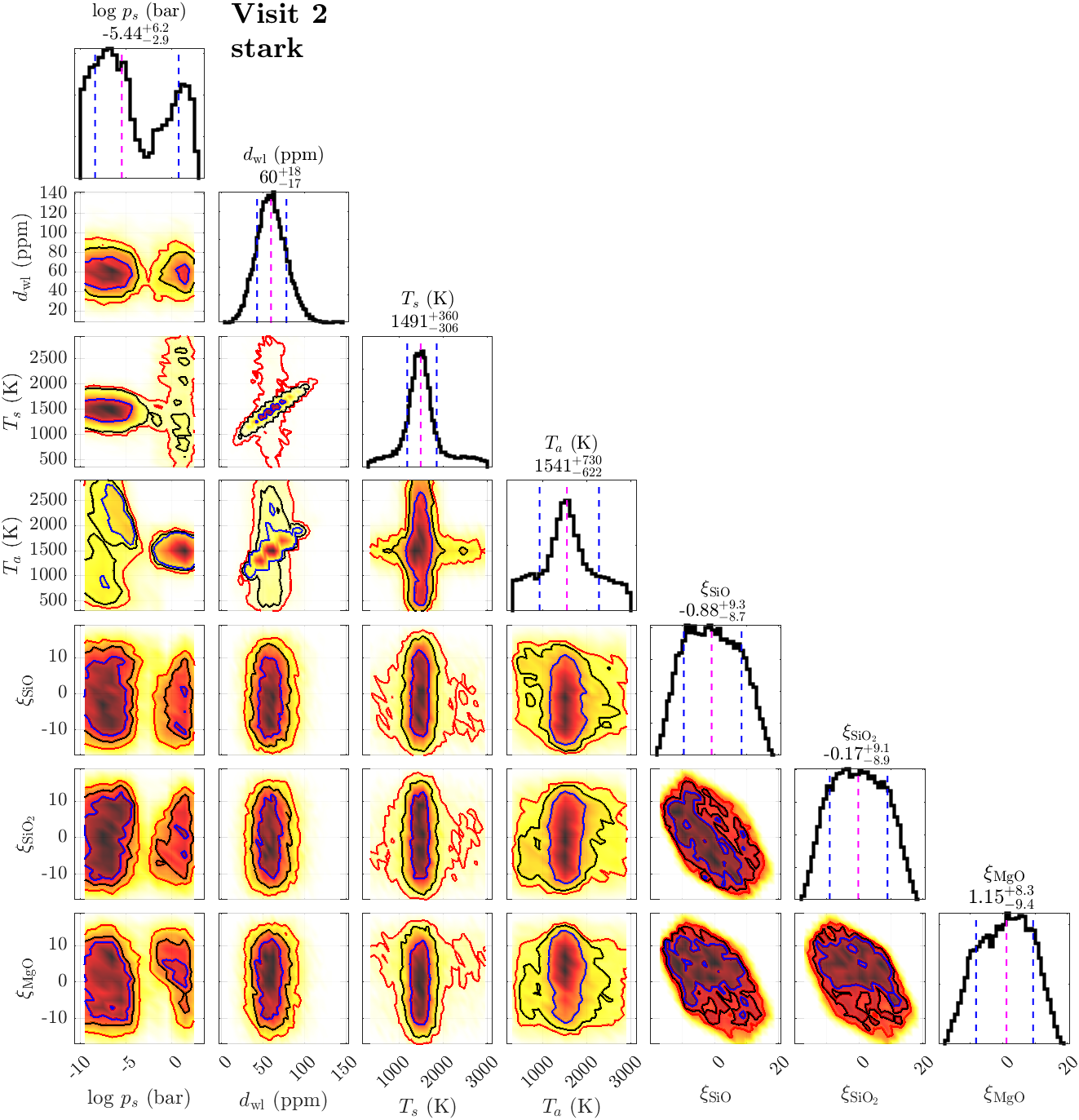} \includegraphics[width=0.6\textwidth]{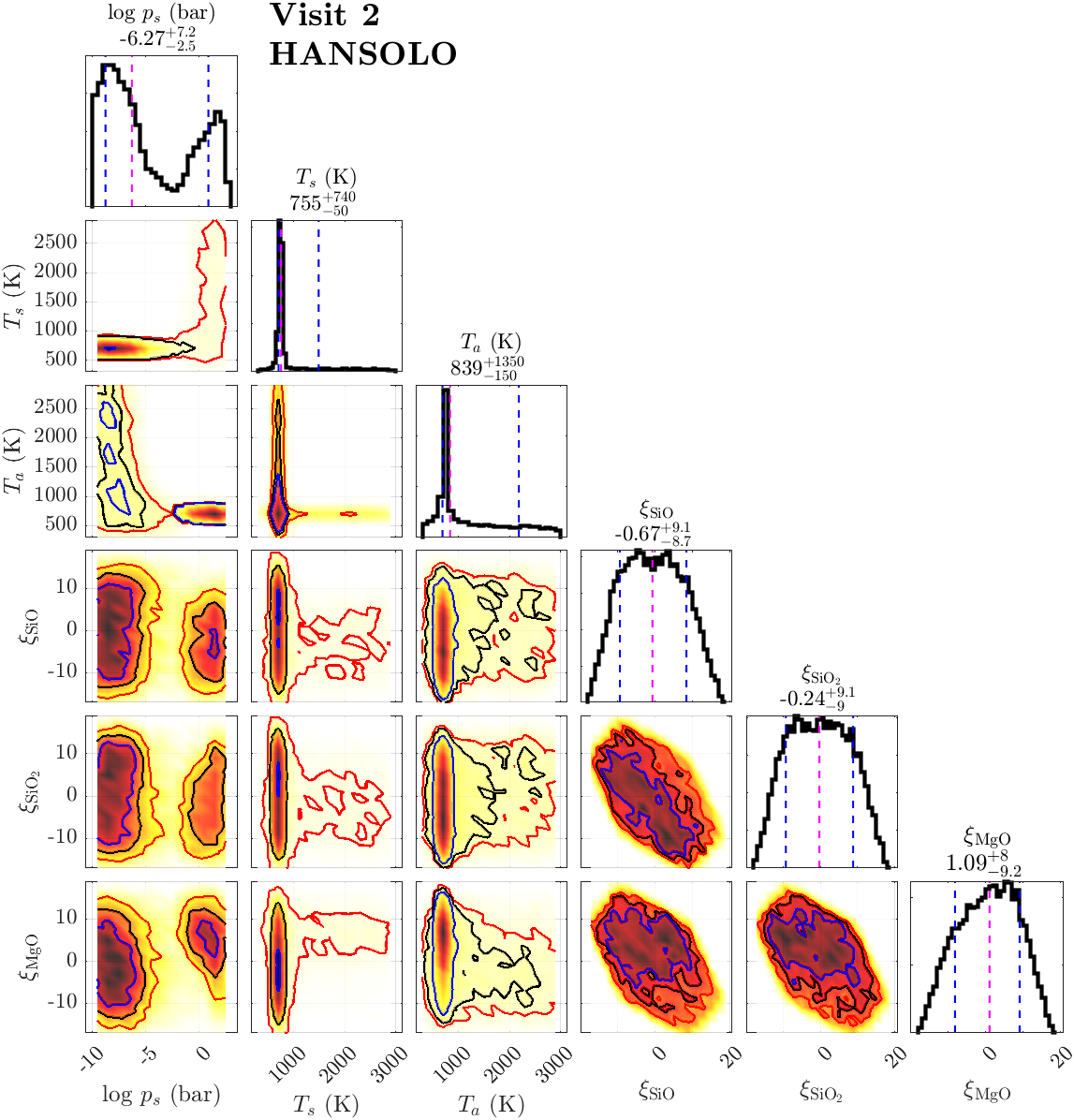}
    \caption{Posterior distributions of the free parameters for Visit 2, representing the SiO/\ch{SiO2}/MgO-atmosphere scenario. Results are shown for the \stark (left) and \hans (right) reductions.}
    \end{center}
    \label{fig:retrieval_posterior_sio_visit2}
\end{figure}
\end{landscape}

\begin{figure*}
    \centering
    \includegraphics[width=0.6\textwidth]{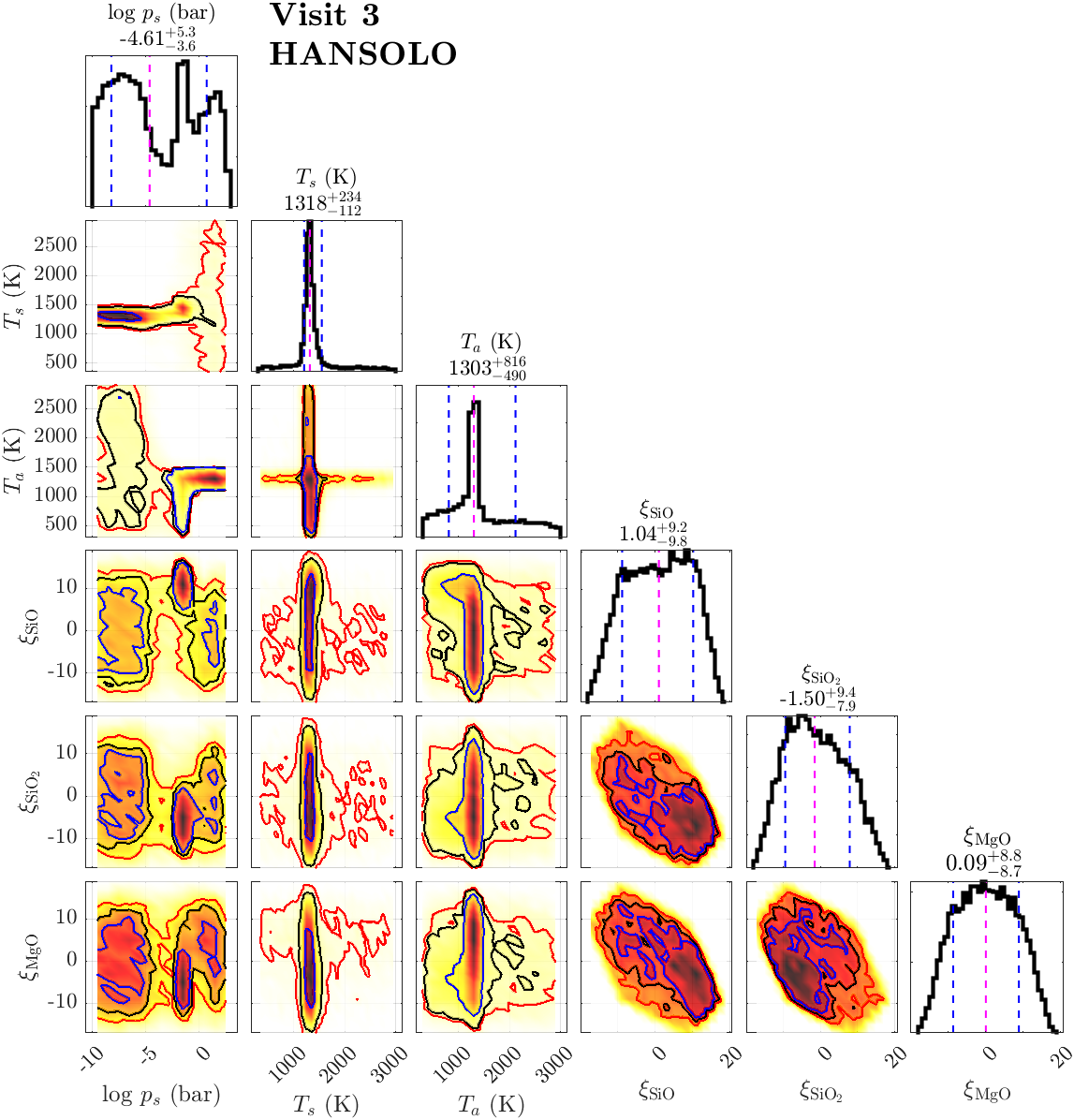}
    \caption{Posterior distributions of the free parameters for Visit 3, representing the SiO/\ch{SiO2}/MgO-atmosphere scenario. Results are shown for the \hans reduction. The corresponding distributions for the \stark reduction are depicted in Fig.~\ref{fig:retrieval_posterior_visit3_red1}.}
    \label{fig:retrieval_posterior_sio_visit3}
\end{figure*}

\begin{landscape}
\begin{figure}
    \begin{center}
    \includegraphics[width=0.7\textwidth]{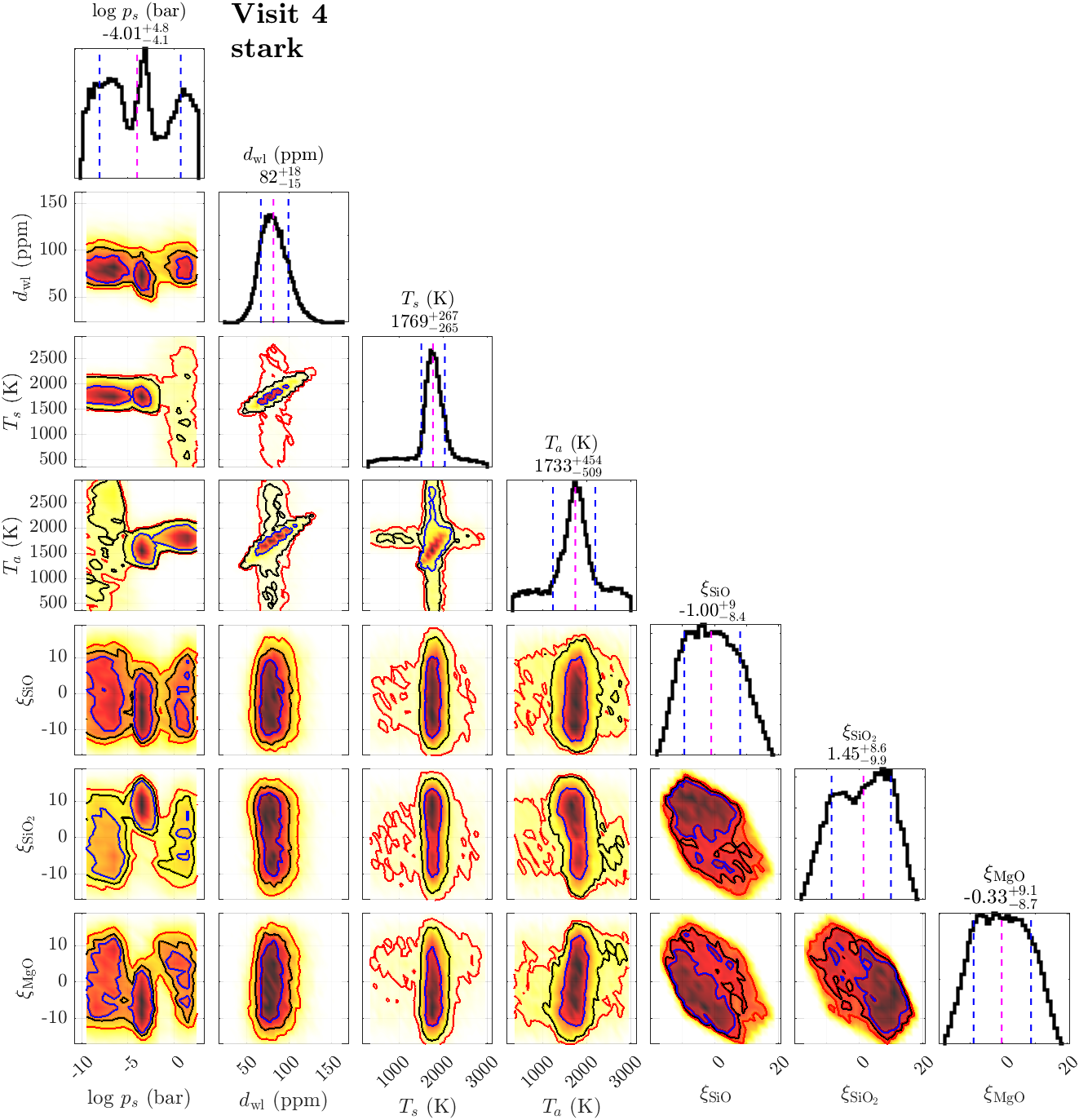} \includegraphics[width=0.6\textwidth]{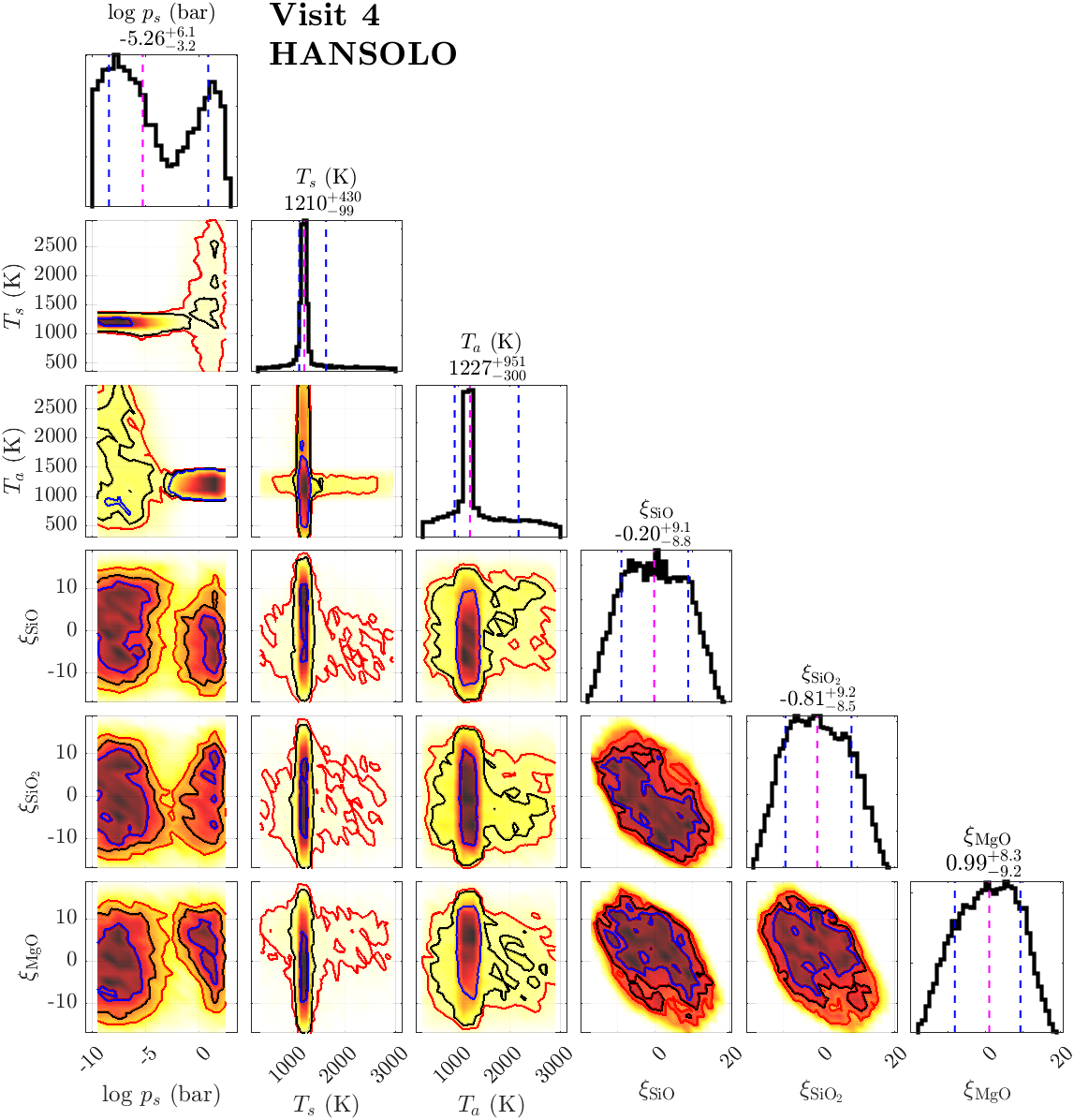}
    \caption{Posterior distributions of the free parameters for Visit 4, representing the SiO/\ch{SiO2}/MgO-atmosphere scenario. Results are shown for the \stark (left) and \hans (right) reductions.}
    \end{center}
    \label{fig:retrieval_posterior_sio_visit4}
\end{figure}
\end{landscape}

\begin{landscape}
\begin{figure}
    \begin{center}
    \includegraphics[width=0.7\textwidth]{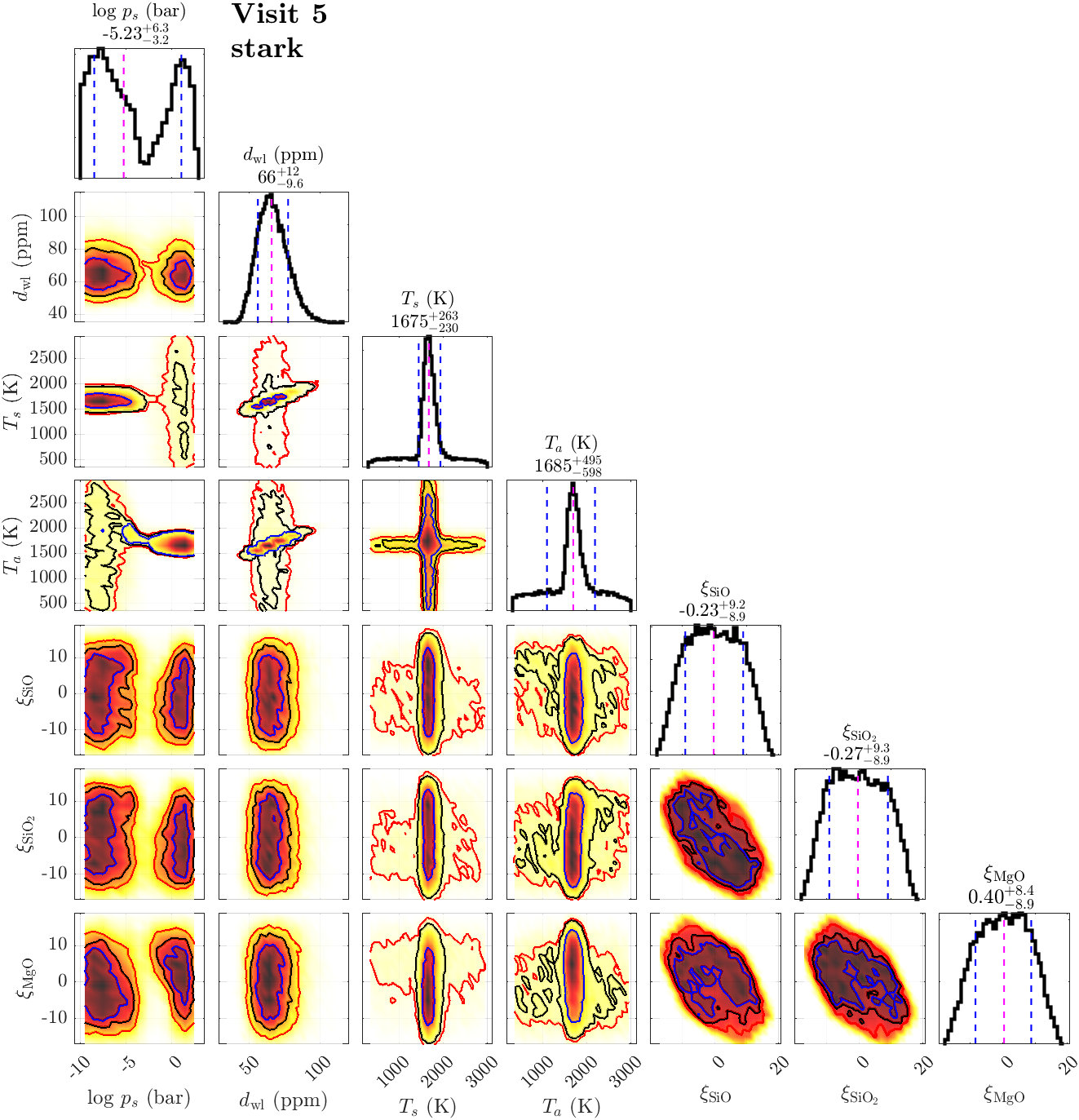} \includegraphics[width=0.6\textwidth]{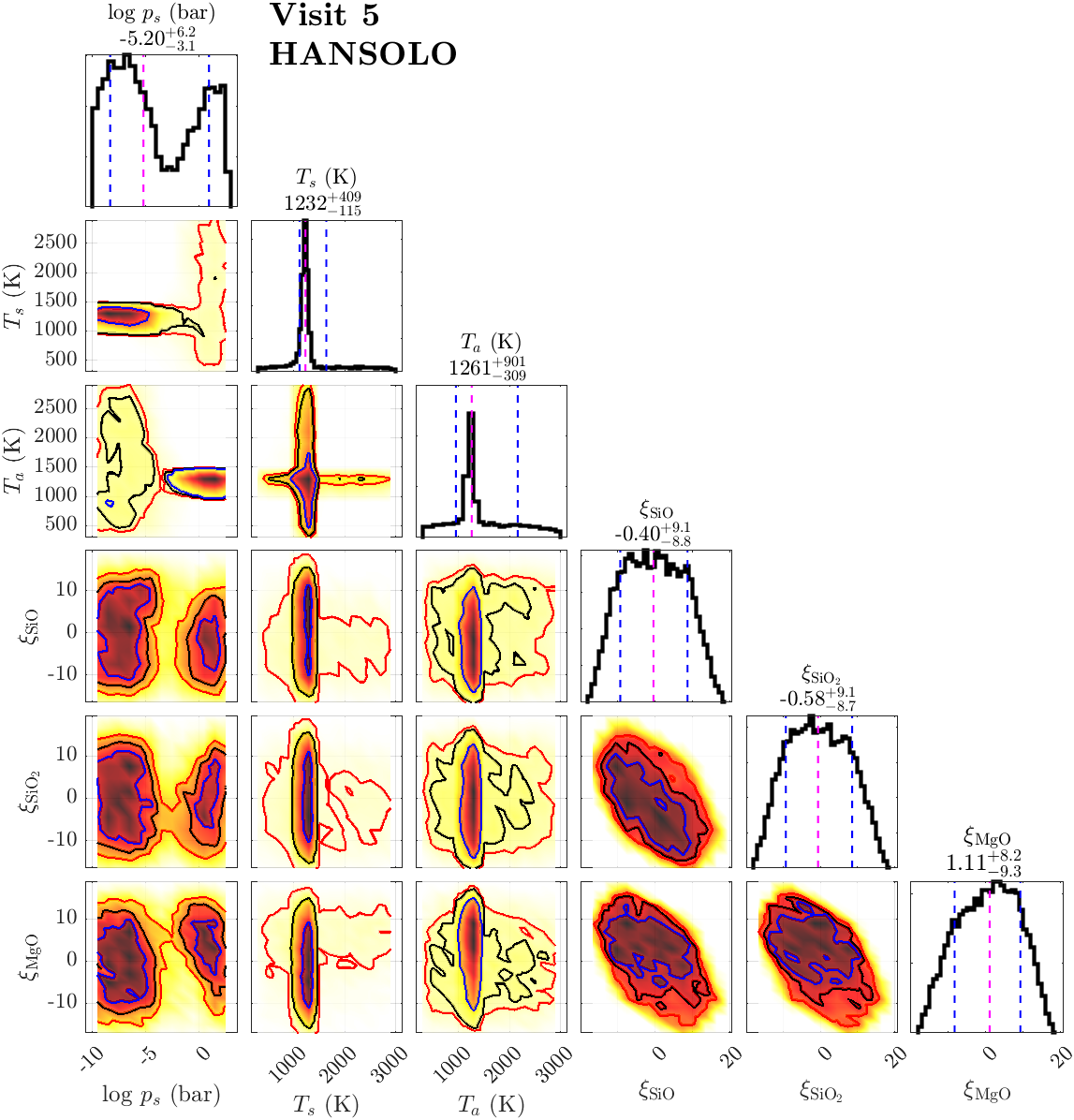}
    \caption{Posterior distributions of the free parameters for Visit 5, representing the SiO/\ch{SiO2}/MgO-atmosphere scenario. Results are shown for the \stark (left) and \hans (right) reductions.}
    \end{center}
    \label{fig:retrieval_posterior_sio_visit5}
\end{figure}
\end{landscape}

\end{appendix}

\end{document}